\documentclass[10pt]{article}

\usepackage{authblk}
\usepackage{natbib}
\usepackage{mathtools}
\usepackage{amsmath}
\usepackage{amsthm}
\usepackage{amssymb}
\usepackage{amsbsy}
\usepackage{amsfonts}
\usepackage{amscd}
\usepackage{mathrsfs}
\usepackage{bm}
\usepackage{breqn}
\usepackage{color, soul}
\usepackage{enumerate}
\usepackage{empheq}
\usepackage{rotating}
\usepackage{ftnxtra}
\usepackage{fnpos}
\usepackage[titletoc,toc,title]{appendix}
\usepackage{euscript}
\usepackage{graphicx}
\usepackage{epsfig}
\usepackage{epstopdf}
\DeclareGraphicsExtensions{.pdf,.png,.jpg,.eps}
\usepackage{pstool}
\usepackage{upgreek}
\usepackage{mathrsfs}
\usepackage[T1]{fontenc}
\usepackage[utf8]{inputenc}

\usepackage{psfrag}

\numberwithin{equation}{section}

\theoremstyle{plain}	
\newtheorem{thm}{Theorem}[section]

\newtheorem{prop}[thm]{Proposition}
\newtheorem*{prop*}{Proposition} 
\theoremstyle{definition}
\newtheorem{defi}[thm]{Definition}
\newtheorem{remark}{Remark}[section]

\usepackage{graphicx}
\usepackage{lipsum}

\usepackage{caption}
\usepackage{subcaption}

\usepackage{hyperref}
\hypersetup{colorlinks=true, linkcolor=blue}
\hypersetup{colorlinks=true,citecolor=blue}

\DeclareMathAlphabet{\mathpzc}{OT1}{pzc}{m}{it}

\usepackage{amsmath, amsthm, amssymb}

\usepackage{cleveref}

\DeclarePairedDelimiter\abs{\lvert}{\rvert}

\makeatletter
\newsavebox{\@brx}
\newcommand{\llangle}[1][]{\savebox{\@brx}{\(\m@th{#1\langle}\)}%
  \mathopen{\copy\@brx\mkern2mu\kern-0.9\wd\@brx\usebox{\@brx}}}
\newcommand{\rrangle}[1][]{\savebox{\@brx}{\(\m@th{#1\rangle}\)}%
  \mathclose{\copy\@brx\mkern2mu\kern-0.9\wd\@brx\usebox{\@brx}}}%
\let\oldabs\abs
\def\abs{\@ifstar{\oldabs}{\oldabs*}}
\makeatother

\usepackage{accents}

\newcommand{\Fe}{\accentset{e}{\boldsymbol{\mathbf{F}}}}
\newcommand{\Fa}{\accentset{a}{\boldsymbol{\mathbf{F}}}}
\newcommand{\Ce}{\accentset{e}{\boldsymbol{\mathbf{C}}}}

\newcommand{\cCs}{\accentset{s}{\mathsf{C}}}
\newcommand{\cCa}{\accentset{a}{\mathsf{C}}}		

\newcommand{\director}[1]{\underaccent{\mathfrak{a}}{#1}}

\newcommand{\As}{\accentset{s}{\boldsymbol{\mathsf{A}}}}
\newcommand{\Aa}{\accentset{a}{\boldsymbol{\mathsf{A}}}}
\newcommand{\Bs}{\accentset{s}{\boldsymbol{\mathsf{B}}}}
\newcommand{\Ba}{\accentset{a}{\boldsymbol{\mathsf{B}}}}

\newcommand{\N}{\accentset{1}{\boldsymbol{\mathsf{N}}}}
\newcommand{\NN}{\accentset{2}{\boldsymbol{\mathsf{N}}}}
\newcommand{\NNN}{\accentset{3}{\boldsymbol{\mathsf{N}}}}
\newcommand{\Nj}{\accentset{j}{\boldsymbol{\mathsf{N}}}}
\newcommand{\n}{\accentset{1}{\boldsymbol{\mathsf{n}}}}
\newcommand{\nn}{\accentset{2}{\boldsymbol{\mathsf{n}}}}
\newcommand{\nnn}{\accentset{3}{\boldsymbol{\mathsf{n}}}}
\newcommand{\nj}{\accentset{j}{\boldsymbol{\mathsf{n}}}}

    {\end{bmatrix}}%

\usepackage{enumitem}

\usepackage[utf8]{inputenc}
\usepackage{dutchcal}
\usepackage{multirow}

\usepackage{empheq} 

\usepackage[dvipsnames]{xcolor}
\usepackage{tikz}
\usetikzlibrary{fit}
\usetikzlibrary{positioning}
\tikzset{box1/.style={draw=black, thick, rectangle,rounded corners, minimum height=2cm, minimum width=2cm}}

\begin{document}


\title{\textbf{Nonlinear Cauchy Elasticity}}

\author[1,2]{Arash Yavari\thanks{Corresponding author, e-mail: arash.yavari@ce.gatech.edu}}
\author[3]{Alain Goriely}
\affil[1]{\small \textit{School of Civil and Environmental Engineering, Georgia Institute of Technology, Atlanta, USA}}
\affil[2]{\small \textit{The George W. Woodruff School of Mechanical Engineering, Georgia Institute of Technology, Atlanta,  USA}}
\affil[3]{\small \textit{Mathematical Institute, University of Oxford, Oxford,  UK}}

\maketitle

\begin{abstract} Most theories and applications of elasticity rely on an energy function that depends on the strains from which the stresses can be derived. This is the traditional setting of Green elasticity, also known as hyper-elasticity. However, in its original form the theory of elasticity does not assume the existence of an strain energy function. In this case, called Cauchy elasticity, stresses are directly related to the strains. Since the emergence of modern elasticity in the 1940s, research on Cauchy elasticity has been relatively limited. One possible reason is that for Cauchy materials, the net work performed by stress along a closed path in the strain space may be nonzero. Therefore, such materials may require  access to both energy sources and sinks. This characteristic has led some mechanicians to question the viability of Cauchy elasticity as a physically plausible theory of elasticity. In this paper,  motivated by its relevance to recent applications, such as the modeling of active solids, we revisit Cauchy elasticity in a modern form.
    First, we show that in the general theory of anisotropic Cauchy elasticity, stress can be expressed in terms of six functions, that we call \emph{Edelen-Darboux potentials}. For isotropic Cauchy materials, this number reduces to three, while for incompressible isotropic Cauchy elasticity, only two such potentials are required. 
    Second, we show that in Cauchy elasticity, the link between balance laws and symmetries is lost, in general, since Noether's theorem does not apply. In particular, we show that, unlike hyperleasticity, objectivity is not equivalent to the balance of angular momentum.
    Third, we formulate the balance laws of Cauchy elasticity covariantly and derive a generalized Doyle-Ericksen formula. 
    Fourth, the material symmetry and work theorems of Cauchy elasticity are revisited, based on  the \emph{stress-work $1$-form} that emerges as a fundamental quantity in Cauchy elasticity. The stress-work $1$-form allows for a classification via Darboux’s theorem that leads to  a classification of Cauchy elastic solids based on their generalized energy functions.
    Fifth, we discuss the relevance of Carath\'eodory's theorem on accessibility property of Pfaffian equations.
    Sixth, we show that Cauchy elasticity has an intrinsic geometric hystresis, which is the net work of stress in cyclic deformations. If the orientation of a cyclic deformation is reversed, the sign of the net work of stress changes, from which we conclude that stress in Cauchy elasticity is neither dissipative nor conservative.
    Seventh, we establish connections between Cauchy elasticity and the existing constitutive equations for active solids. 
    Eighth, linear anisotropic Cauchy elasticity is examined in detail, and simple displacement-control loadings are proposed for each symmetry class to characterize the corresponding antisymmetric elastic constants.
    Ninth, we discuss  both isotropic and anisotropic Cauchy anelasticity and show that the existing solutions for stress fields of distributed eigenstrains (and particularly defects) in hyperelastic solids can be readily extended to Cauchy elasticity.
    Tenth, we introduce  Cosserat-Cauchy materials and demonstrate that an anisotropic three-dimensional Cosserat-Cauchy elastic solid has at most twenty four generalized energy functions.
\end{abstract}

\begin{description}
\item[Keywords:] Cauchy elasticity, Green elasticity, hyper-elasticity, nonlinear elasticity, linear elasticity, odd elasticity.
\end{description}

\tableofcontents

\section{Introduction}

One of the greatest intellectual achievements of the last century was the formulation of a general mathematical theory of \emph{passive materials} that respond to external loads. This theory of rational mechanics has been fundamental for our understanding of  material behavior, and, as a consequence, extremely successful to design new structures and to  develop  new devices in mechanical and bio-medical engineering. However, over the last two decades, researchers from multiple fields of engineering and sciences have been increasingly interested in \emph{active materials} that are characterized by a solid matrix with embedded active components. For instance, many biological systems can  be modeled as active solids: growing plant stems change their shape to find light and growing neurons sense their environment to create operational neural networks, and the mammalian cell  adapts to its environment by changing the properties of its membrane. Whereas these systems may appear very different from technological devices, they share many common features and can be studied mathematically within the same general setup. We believe that Cauchy elasticity (and Cauchy anelasticity) may be a rational framework for modeling such material systems.
Another motivation for revisiting Cauchy elasticity stems from the recent interest in the physics literature regarding ``odd elasticity" \citep{Scheibner2020,Fruchart2023} which, as we will show, is the simplest version of Cauchy elasticity, namely linearized Cauchy elasticity.

Nonlinear elasticity is almost two hundred years old, with its theoretical foundations laid by Augustin-Louis Cauchy \citep{Truesdell1992-Cauchy}.
In Cauchy's original theory of elasticity in 1828, stress at any point depends explicitly on strain at that point without any history dependence \citep{Cauchy1828}. Hence, we follow Truesdell and refer to this theory  as \textit{Cauchy elasticity} \citep{Truesdell1952}. A \textit{particular} class of Cauchy elastic solids are those for which the stresses derive from a strain energy function and in this case the theory is called \textit{Green elasticity} \citep{Green1838,Green1839,Spencer2015} or, more commonly, \textit{hyperelasticity} \citep{Truesdell1952}.\footnote{Already in $1893$, \citet{Becker1893} proposed an isotropic constitutive equation expressed in terms of Biot stress as a function of logarithmic stretches. His constitutive equation turns out to be non-hyperelastic \citep{NeffEidelMartin2016}. Similarly, in $1928$, \citet{Hencky1928} considered another isotropic constitutive equation in terms of the Cauchy stress as a function of  logarithmic stretches and realized that it is not hyperelastic \citep{Neff2014}. These two constitutive equations are discussed in detail in \S\ref{Examples-NonHyperelastic}. \label{Becker-Hencky}} 
A natural question is whether all Cauchy materials must be Green materials based on fundamental thermodynamic principles.
\citet{GreenLaws1967} investigated the constitutive equations of Cauchy elasticity and their consistency with the second law of thermodynamics and concluded that ``\textit{Cauchy elasticity is consistent with the basic thermodynamical theory}."
Further, \citet{GreenNaghdi1971}  demonstrated the consistency of Cauchy elasticity with the first and second laws of thermodynamics. They showed that the net work of stress in a Cauchy elastic solid undergoing a cyclic motion, i.e., a closed path in the strain space, may be non-zero (also explicitly demonstrated later by \citet{Edelen1977}), and by changing the orientation of the cyclic motion, the sign of the net work of stress changes. This has been referred to as ``\textit{perpetual motion}" in the literature. \citet{GreenNaghdi1971} clearly stated that a non-vanishing net work of stress in cyclic motions does not violate the second law of thermodynamics. To refute the possibility of solids exhibiting this property, they proposed the following ``\textit{additional principle}": ``\textit{The total thermal energy and mechanical work per unit mass supplied to or extracted from any part of the body in a closed cycle of deformation of the type $\mathscr{L}$ is always zero}."\footnote{By "type $\mathscr{L}$" they meant a cyclic motion, i.e., both strain and velocities return to their initial values after one cycle, and hence, the total work done on the body is the work of stress, see \eqref{Total-Work}.}

Based on such discussions, Cauchy elasticity was mostly dismissed as an interesting, but ultimately irrelevant, theory of elasticity that is not viable for the modeling of elastic materials by most of the community \citep{Coleman1962,Rivlin1986,Casey2005,Carroll2009,Leonov2000,Rajagopal2011},\footnote{In the discussion of the paper \citep{Truesdell1964}, R.S. Rivlin questioned the physical validity of Cauchy elasticity as a material model. The following is Truesdell's response: ``While it is possible that the clastic material without a stored-energy function will not describe any physical material, before such a conclusion can be reached it will be necessary to know, by means of clearly stated and proved mathematical theorems, what undesirable properties such a material has."}  while a smaller community still persisted in using it \citep*{Truesdell1952,Truesdell1964,TruesdellNoll2004,Ogden1984,Bordiga2022}.\footnote{\citet{Bordiga2022} considered two-dimensional lattices preloaded by follower forces. They showed that the effective small-on-large response of such lattices is non-hyperelastic. More specifically, they showed that the acoustic tensor of the effective medium is non-symmetric. To the best of our knowledge, \citet{Bordiga2022}'s homogenized pre-stressed lattices are the first concrete example of linear non-hyperelastic Cauchy elastic solids.}

As a result, theoretical advancements in nonlinear elasticity and anelasticity, such as plasticity and mechanics of growth, have almost exclusively focused on hyperelasticity. Yet, emerging applications of active matter strongly motivate the revival of Cauchy elasticity and the mechanics of non-hyperelastic Cauchy solids. Indeed, many active materials  have access to external sources and sinks of energy. The net work of stress in cyclic motions is the key in the experimental determination of their non-hyperelastic  properties. In the case of linear elasticity, these materials are characterized by the presence of  antisymmetric elastic constants and in recent years have been referred to as \textit{odd elastic} (see \S\ref{Sec:Linear-Cauchy-elasticity}).

In hyperelasticity, one assumes the existence of an energy function that depends explicitly on strains. In the presence of inelastic deformations, strain energy would depend on the elastic part of strain. In Cauchy elasticity an energy function does not exist, in general. 
The fundamental object in Cauchy elasticity is the \emph{stress-work $1$-form} \citep{Edelen1977,Kadic1980,Cardin1995}. Using exterior calculus we will generalize the works of \citep{Ericksen1956} and \citep{Edelen1977} and classify Cauchy elastic solids via their generalized energy functions utilizing the Darboux classification of differential forms \citep{Darboux1882,Sternberg1999,Bryant2013,Suhubi2013}. As we will show, anisotropic Cauchy elastic solids have at most six generalized energy functions (explicitly depending on the right Cauchy-Green strain) while compressible and incompressible isotropic Cauchy elastic solids have at most three and two generalize energy functions, respectively. We will write the constitutive equations of Cauchy elasticity in terms of the generalized energy functions in the form of a generalized Doyle-Ericksen formula.
We will show that Cauchy elasticity has a natural geometric hystresis, namely the net work of stress in cyclic deformations. This geometric hystresis is key in experimental characterization of Cauchy elastic solids. 
We will investigate anisotropic non-hyperelastic Cauchy linear elastic solids, which is the correct general theory for linear elastic materials that are elastic but not hyperelastic. While in 2D, this non-hyperelastic behavior requires chirality, in $3$D anisotropic elastic solids, non-hyperelasticity can exist without it.
We will show that anisotropic linear Cauchy elasticity has $15$ extra elastic constant---the antisymmetric elastic constants  \citep{RogersPipkin1963,Podio1987,Yong1991,He1996,Ostrosablin2017}. These additional elastic constants  need to be characterized using the geometric hysteresis. Designing proper experiments in line with this theory will be essential in the experimental characterization of the mechanical properties of active solids. For each symmetry class, we will suggest simple experiments that can be used to characterize all the non-symmetric elastic constants.

In continuum mechanics, the second law of thermodynamics has been crucial in formulating the mathematical foundations of constitutive equations, going back to the seminal work of \citet{ColemanNoll1963}. The second law of thermodynamics has traditionally been used in the form of the Clausius-Duhem inequality \citep{Serrin2012,Silhavy2013}. However, in Cauchy elasticity, there is no underlying energy function and the fundamental object is the stress-work $1$-form, which is a differential $1$-form. For this more general class of elastic solids, the axiomatic or geometric formulation of thermodynamics as proposed by Carathéodory \citep{Caratheodory1909,Buchdahl2009,Frankel2011} is more convenient and natural.\footnote{It should be mentioned that Carathéodory's abstract formulation of thermodynamics has been used in formulating non-equilibrium thermodynamics of solids \citep{Valanis1971,Nemat-Nasser1975}.} This has been recognized by several researchers \citep{Ericksen1956,Edelen1977,Rivlin1986}. However, there is limited research on the constitutive equations of Cauchy elasticity and their consistency with the second law of thermodynamics.
It has been argued that Carathéodory’s accessibility principle excludes a large class of Cauchy elastic solids. 
For these solids, for any two nearby deformed configurations, there exists a path in the strain space that connects them without any work of stress.
At first glance, this may seem peculiar, but Carathéodory’s principle does not assert that every deformation path exhibits this zero work property. One can imagine that certain Cauchy elastic solids have soft modes that are the only ones activated in a specific deformation path, leading to zero work. These peculiar or anomalous cases should not be excluded from the outset, as they may represent exotic materials with unusual mechanical properties. 
Characterizing paths of zero-stress-work in the strain space is closely related to Pfaffian differential equations and exterior differential systems \citep{Sternberg1999,Bryant2013,Suhubi2013}---a branch of mathematics that, to the best of our knowledge, has not found systematic applications in elasticity and materials science to this date.

It should be emphasized that Cauchy elasticity does not encompass all elastic solids. In recent years, there has been research on what are known as implicit constitutive equations \citep{Morgan1966,Rajagopal2003,Rajagopal2007,Rajagopal2011,Bustamante2009,Bustamante2011,Yavari2024ImplicitElasticity}. More specifically, \citet{Morgan1966} and \citet{Rajagopal2003,Rajagopal2007} proposed constitutive equations of the form $\boldsymbol{\mathcal{F}}(\boldsymbol{\sigma},\mathbf{b})=\mathbf{0}$. Cauchy elasticity is a subset of this class of solids. 
\citet{RajagopalSrinivasa2007} defined an elastic body as one whose response is rate independent and non-dissipative. In this paper, we restrict ourselves to Cauchy elasticity.

\paragraph{Contributions of this paper.} Contribution of this paper can be summarized as follows.
\begin{itemize}[topsep=0pt,noitemsep, leftmargin=10pt]
\item We present a covariant formulation of Cauchy elasticity and derive a generalized Doyle-Ericksen formula.

\item We discuss constitutive equations and work theorems for Cauchy materials.

\item We describe Edelen-Darboux potentials in Cauchy elasticity and define Ericksen materials. We show that for incompressible isotropic Cauchy elastic solids, hyperelasticity and Ericksen elasticity are the only possibilities. We also show that compressible isotropic Cauchy elastic solids have at most three generalized energy functions.

\item We show that for anisotropic Cauchy elastic solids, in terms of the invariants of the right Cauchy-Green strain, the number of generalized energy functions explicitly depends on the symmetry class. For example, in the case of transversely isotropic Cauchy elastic solids this number is five.

\item We show that Cauchy elasticity has a natural vector bundle structure and a geometric hystresis. 

\item We  connect explicitly Cauchy elasticity and active stress in biological systems.

\item We  present a detailed analysis of two universal deformations (one homogeneous and one inhomogeneous) and provide examples of cyclic deformations that have non-vanishing net work of stress.

\item We study anisotropic linear Cauchy elasticity  in detail and show that only six out of the eight symmetric classes have non-vanishing antisymmetric elastic constants. For each symmetry class, simple cyclic deformations are suggested to characterize all the corresponding antisymmetric elastic constants.

\item We introduce Cauchy anelasticity and demonstrate that, in the presence of eigenstrains, the generalized energy functions depend on the principal invariants of the elastic distortions calculated using the Euclidean metric or the principal invariants of the total deformation calculated using the Riemannian material metric. It is also noted that all the existing exact solutions for distributed eigenstrains and defects in hyperelastic solids can be easily extended to Cauchy elasticity.

\item We introduce Cosserat-Cauchy elastic materials and  show that an anisotropic Cosserat-Cauchy elastic solid in three dimensions has at most twenty four generalized energy functions.

\end{itemize}

 \vskip 0.1in
This paper is organized as follows. 
We start by reviewing finite-dimensional classical mechanics in the presence of non-conservative forces in \S\ref{Finite-Dimensional}. This sets the stage for Cauchy elasticity in which stress is non-conservative and yet non-dissipative.
In \S\ref{Sec:ExteriorCalculus} exterior differential systems, and particularly, Pfaffian differential equations and Darboux's classification of differential forms are briefly reviewed. A concise review of Carathéodory’s formulation of thermodynamics is also provided. 
Several aspects of Cauchy elasticity are investigated in detail in \S\ref{Cauchy Elasticity}. 
Constitutive equations of Cauchy elasticity are discussed in \S\ref{Sec:Constitutive-Equations}.
In \S\ref{Sec:Geometric-Phase}, it is shown that Cauchy elasticity has a natural vector bundle structure and a geometric hystresis.
The connection between Cauchy elasticity and some of the existing constitutive equations proposed for active solids is investigated in \S\ref{Sec:Active-Matter}. 
Two examples of universal deformations are discussed in detail in \S\ref{Sec:Examples}.
\S\ref{Sec:Linear-Cauchy-elasticity} discusses anisotropic linear Cauchy elasticity and characterization of its antisymmetric elastic constants. 
Cauchy anelasticity is introduced and briefly discussed in \S\ref{Sec:CauchyAnelasticity}, as is Cosserat–Cauchy elasticity in \S\ref{Sec:GeneralizedCauchyElasticity}.
We conclude in \S\ref{Discussion} with a discussion of the implications of our results. Additional remarks and directions for future work are provided in \S\ref{FutureWork}.

\section{Non-Conservative Forces in Mechanical Systems} \label{Finite-Dimensional}

\begin{figure}[t!]
\centering
\includegraphics[width=0.7\textwidth]{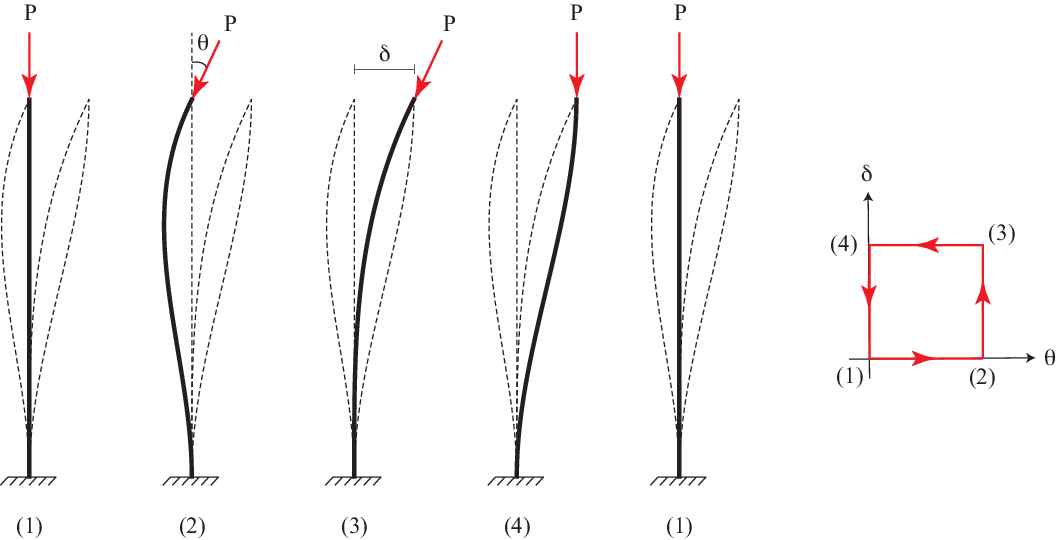}
\vspace*{-0.0in}
\caption{Follower forces are examples of non-conservative forces. Here, for instance a follower force acts along the tangent on a beam (column) while it is deflected. In the cyclic deformation $(1) \to (2) \to (3)\to (4)\to (1)$ the follower force does the net work $-P\delta \sin\theta$. In the reverse cyclic deformation $(1) \to (4) \to (3)\to (2)\to (1)$, it would do the net work $P\delta \sin\theta$.}
\label{Follower-Force}
\end{figure}

Before we discuss the general problem of non-conservative effects in a continuum, it is instructive to  briefly discuss non-conservative forces in finite-dimensional mechanical systems.
In classical mechanics, a \textit{conservative force} is a force that is derived from a potential field.\footnote{An alternative definition of a conservative force is  a force whose work is path independent. If the force only depends on position (or generalized coordinates) this would be equivalent to the existence of a potential. However, if a force is velocity dependent its work can be path independent even in the absence of a potential. Such forces are called gyroscopic forces \citep{Ziegler1977}.}
The usual examples of non-conservative forces, such as friction and viscous forces, are dissipative. However, a non-conservative force, i.e., a force field that does not have a corresponding potential, is not necessarily dissipative. For instance, {follower forces} are non-conservative but non-dissipative forces \citep{Ziegler1952,Ziegler1953,Bolotin1963,Bigoni2011}.
A \textit{follower force} is a force (or moment) that continuously changes its direction to keep a particular orientation with the deformed configuration of a body. 
A typical example of a follower force is one applied along the tangent of a deformed beam or column. As the column deforms, the force adjusts its direction to remain aligned with the tangent at the point of application, making it dependent on the evolving geometry of the column (see Fig.~\ref{Follower-Force}).
Since the 1950s, various engineering communities have been interested in the instability of systems subjected to follower forces \citep{Pfluger1950,Pfluger1955,Beck1952,Ziegler1952,Ziegler1953,Ziegler1977,Bolotin1963}. Follower forces find applications in structural mechanics, aeroelasticity, fluid-structure interactions, rotordynamics, etc. 
Follower forces are notoriously difficult to realize in experiments (see \cite{Elishakoff2005} for an exhaustive review), and their physical relevance has been strongly questioned \citep{Koiter1996}.\
However, in recent years follower forces have been realized experimentally  by Bigoni and his co-workers \citep{Bigoni2011,Bigoni2018,Cazzolli2020}.

In the literature, there are plenty of other instances of non-conservative forces.  For instance, the 
force fields $\boldsymbol{\mathsf{F}}=\boldsymbol{\mathsf{F}}(\mathbf{x})$ that depend on position $\mathbf{x}\in\mathbb{R}^n$ but not on velocity and such that $\operatorname{curl}\boldsymbol{\mathsf{F}}(\mathbf{x})\neq \mathbf{0}$ have been called \emph{curl forces} \citep{Berry2012}. 
 Non-conservative position-dependent forces have also been called \emph{positional forces}, \emph{pseudo- gyroscopic forces}, and \emph{circulatory forces} \citep{Ziegler1977,Kirillov2021}. 
These are all examples of non-conservative yet non-dissipative forces, i.e., under such force fields the phase space volume is preserved. Particle dynamics under such force fields is fundamentally different from that under conservative forces. For example, Noether's theorem, which is the link between conservation laws and symmetries is broken \citep{Berry2012,Berry2013,Berry2015}. 


A single particle under a curl force field is the zero-dimensional analogue of Cauchy elasticity \citep{Berry2013,Berry2015,YavariGoriely2025CurlForces}. Indeed, we will show that stress in hyperelasticity is non-dissipative and conservative. However, the class of non-dissipative solids is significantly broader than that of hyperelastic materials \citep{RajagopalSrinivasa2007} and in Cauchy elasticity, stress is non-dissipative but non-conservative.

Let us consider a finite-dimensional mechanical system with $n$ degrees of freedom $\mathbf{q}(t)=\{q_1(t),\hdots,q_n(t)\}\in\mathbb{R}^n$. We  assume that this system is under the action of both conservative and non-conservative forces. Then, Lagrange's equations read
\begin{equation}
	\frac{d}{dt}\frac{\partial\mathsf{L}}{\partial \dot{q}_i}-\frac{\partial\mathsf{L}}{\partial q_i}=\mathsf{F}_i
	\,,\qquad i=1,\hdots,n\,,
\end{equation}
where $\mathsf{L}=\mathsf{T}-\mathsf{V}$ is the Lagrangian, $\mathsf{T}=\frac{1}{2}\sum_{i,j}\mathsf{g}_{ij}\dot{q}_i\dot{q}_j$ is the kinetic energy, $\mathsf{g}_{ij}=\mathsf{g}_{ij}(\mathbf{q})$, $\mathsf{V}(\mathbf{q})$ is the potential of conservative forces, and $\mathsf{F}_i$ is the $i$th component of the non-conservative force $\boldsymbol{\mathsf{F}}=\boldsymbol{\mathsf{F}}(t,\mathbf{q},\dot{\mathbf{q}})$. 
Lagrange's equations can then equivalently be written as \citep{Krechetnikov2006,Zhuravlev2008}
\begin{equation} \label{Lagrange-Equations}
	\frac{d}{dt}\frac{\partial\mathsf{T}}{\partial \dot{q}_i}-\frac{\partial\mathsf{T}}{\partial q_i}
	=\mathsf{F}_i-\frac{\partial\mathsf{V}}{\partial q_i}\,,\qquad i=1,\hdots,n\,.
\end{equation}
Note that
\begin{equation}
\begin{aligned}
	\frac{\partial\mathsf{T}}{\partial \dot{q}_i} &=\sum_{j=1}^n\mathsf{g}_{ij} \dot{q}_j\,,\\
	\frac{d}{dt}\frac{\partial\mathsf{T}}{\partial \dot{q}_i}
	&=\sum_{j=1}^n \left( \sum_{k=1}^n \frac{\mathsf{g}_{ij}}{\partial g_k} \dot{q}_j\dot{q}_k
	+\mathsf{g}_{ij} \ddot{q}_j \right)\,,\\
	\frac{\partial\mathsf{T}}{\partial q_i} &=\frac{1}{2}\sum_{j,k=1}^n \frac{\partial \mathsf{g}_{jk}}{\partial q_i} 
	\,\dot{q}_j\dot{q}_k
	\,.
\end{aligned}
\end{equation}
Multiplying both sides of \eqref{Lagrange-Equations} by $q_i$ and summing over $i$, one obtains
\begin{equation} \label{Identity}
	\sum_{i=1}^n \left[ \frac{1}{2}\sum_{j,k=1}^n 
	\frac{\partial \mathsf{g}_{jk} }{\partial q_i}\,\dot{q}_i\dot{q}_j\dot{q}_k
	+\sum_{j=1}^n \mathsf{g}_{ij} q_i \ddot{q}_j
	\right]
	=\sum_{i=1}^n \left(\mathsf{F}_i \dot{q}_i-\frac{\partial\mathsf{V}}{\partial q_i} \dot{q}_i\right)
	\,.
\end{equation}
For this system, we define the \textit{energy} as $\mathsf{H}=\mathsf{T}+\mathsf{V}$, and hence
\begin{equation}
\begin{aligned}
	\frac{d \mathsf{H}}{dt}
	& =\frac{d \mathsf{T}}{dt}+\frac{d \mathsf{V}}{dt}
	=\sum_{i=1}^n \left(\frac{\partial \mathsf{T}}{\partial q_i}\dot{q}_i
	+\frac{\partial \mathsf{T}}{\partial \dot{q}_i}\ddot{q}_i 
	+\frac{\partial \mathsf{V}}{\partial q_i}\dot{q}_i\right) \\
	& = \sum_{i=1}^n \left[ \frac{1}{2}\sum_{j,k=1}^n 
	\frac{\partial \mathsf{g}_{jk} }{\partial q_i}\,\dot{q}_i\dot{q}_j\dot{q}_k
	+\sum_{j=1}^n \mathsf{g}_{ij} q_i \ddot{q}_j
	\right]
	+\sum_{i=1}^n \frac{\partial \mathsf{V}}{\partial q_i}\dot{q}_i \,.
\end{aligned}
\end{equation}
Using the identity \eqref{Identity} one obtains the rate of change of the energy due to the external forces
\begin{equation} \label{Rate-Of-Energy}
	\frac{d \mathsf{H}}{dt}= \sum_{i=1}^n \mathsf{F}_i \dot{q}_i\,.
\end{equation}
If the potential has an explicit time dependence, i.e., $\mathsf{V}=\mathsf{V}(t,\mathbf{q})$, then instead of \eqref{Rate-Of-Energy} one has \citep{Krechetnikov2006}
\begin{equation}
	\frac{d \mathsf{H}}{dt}= \sum_{i=1}^n \mathsf{F}_i \dot{q}_i+\frac{\partial \mathsf{V}}{\partial t}\,.
\end{equation}

If the dependence of a force on generalized coordinates and velocities is linear, $\boldsymbol{\mathsf{F}}=\boldsymbol{\mathsf{A}}\,\mathbf{q}+\boldsymbol{\mathsf{B}}\,\dot{\mathbf{q}}$, the force can be decomposed into symmetric $(s)$ and antisymmetric $(a)$ parts as follows:
\begin{equation} \label{LinearForce-Decomposition}
\boldsymbol{\mathsf{F}}=\As\,\mathbf{q}+\Aa\,\mathbf{q}+\Bs\,\dot{\mathbf{q}}+\Ba\,\dot{\mathbf{q}}\,,
\end{equation}
where $\As=\frac{1}{2}(\boldsymbol{\mathsf{A}}+\boldsymbol{\mathsf{A}}^{\mathsf{T}})$, $\Aa=\frac{1}{2}(\boldsymbol{\mathsf{A}}-\boldsymbol{\mathsf{A}}^{\mathsf{T}})$, $\Bs=\frac{1}{2}(\boldsymbol{\mathsf{B}}+\boldsymbol{\mathsf{B}}^{\mathsf{T}})$, and $\Ba=\frac{1}{2}(\boldsymbol{\mathsf{B}}-\boldsymbol{\mathsf{B}}^{\mathsf{T}})$. The forces $\As\,\mathbf{q}$, $\Aa\,\mathbf{q}$, $\Bs\,\dot{\mathbf{q}}$, and $\Ba\,\dot{\mathbf{q}}$ are called \textit{potential}, \textit{positional}, \textit{dissipative}, and \textit{gyroscopic}, respectively.

A central theme of this paper is that the properties of non-conservative systems are best explored by considering the work done during a motion. In the simpler case of rigid systems, we can compute the work done on a closed path in the configuration space. We start with a finite-dimensional mechanical system under the action of a force $\boldsymbol{\mathsf{F}}$ going from a point $A_1$ at time $t=t_1$ to point $A_2$ and time $t_2$ along a path $\gamma$ parametrized by $\mathbf{q}:[t_1,t_2]\to\mathbb{R}^n$. The work done on the system is 
\begin{equation}
	W(\gamma)=\int_{\gamma} \boldsymbol{\mathsf{F}}\cdot d\mathbf{q}\,.
\end{equation}
We note that for a conservative force, this integral does not depend on  $\gamma$, but just on the end points, since on a closed path $\int_{\gamma} \boldsymbol{\mathsf{F}}\cdot d\mathbf{q}=\int_{\gamma} d\mathsf{V}=0$. 
However, for a non-conservative force, $W(\gamma)$ depends on $\gamma$. Indeed, when $\gamma$ is a closed path enclosing a surface with a non-vanishing area, $W(\gamma)$ may be nonzero. 
For instance, consider the follower force illustrated in Fig.~\ref{Follower-Force} (motivated by a similar discussion in \citep{Bolotin1963}). This force acts along the tangent to the deformed column at all times. In the cyclic deformation shown,  the work done by the follower force is negative and is due only to the segment $(2)\to(3)$. In the reverse cyclic deformation, the work of the follower force would be positive, illustrating the crucial difference with dissipative forces (for which the work would have the same negative sign regardless of the path direction as we show next).

We can gain further insight by looking at  a force that linearly depends on generalized velocities:
\begin{equation}
	W(\gamma)=\int_{\gamma} \boldsymbol{\mathsf{B}}\,\dot{\mathbf{q}}\cdot d\mathbf{q}
	=\int_{t_1}^{t_2} \boldsymbol{\mathsf{B}}\,\dot{\mathbf{q}}\cdot \dot{\mathbf{q}}\,dt
	=\int_{t_1}^{t_2} \Bs\,\dot{\mathbf{q}}\cdot \dot{\mathbf{q}}\,dt
	\,.
\end{equation}
First, we note that the antisymmetric part $\Ba$ of $\boldsymbol{\mathsf{B}}$ does not contribute to work, i.e., gyroscopic forces are workless. Second, if $\Bs$ is negative-semi-definite $\Bs\,\dot{\mathbf{q}}$ is a \emph{dissipative force} while if $\Bs$ is positive-semi-definite $\Bs\,\dot{\mathbf{q}}$ is an \emph{accelerating force}.

Next we show that for a general non-conservative and non-dissipative force $\boldsymbol{\mathsf{F}}=\boldsymbol{\mathsf{F}}(\mathbf{q})$, $W(-\gamma)=-W(\gamma)$, while for a dissipative (or accelerating force), $W(-\gamma)=W(\gamma)$. The path $\gamma$ is parametrized as $\mathbf{q}:[t_1,t_2]\to\mathbb{R}^n$ while $-\gamma$ is parametrized by $\mathbf{q}\circ \psi:[t_1,t_2]\to\mathbb{R}^n$\ where $\psi:[t_1,t_2]\to[t_1,t_2]$ is a smooth function such that $\psi(t_1)=t_2$ and $\psi(t_2)=t_1$. One choice would be $\psi(t)=t_1+t_2-t$. For $\boldsymbol{\mathsf{F}}=\boldsymbol{\mathsf{F}}(\mathbf{q})$,
\begin{equation}
\begin{aligned}
	W(-\gamma)
	&=\int_{t_1}^{t_2} \boldsymbol{\mathsf{F}}(\mathbf{q}(\psi(t)))\cdot d\mathbf{q}(\psi(t)) \\
	& =\int_{t_1}^{t_2} \boldsymbol{\mathsf{F}}(\mathbf{q}(\psi(t)))\cdot \mathbf{q}(\psi(t)))\,\dot{\psi}(t)\,dt \\
	&=-\int_{t_1}^{t_2} \boldsymbol{\mathsf{F}}(\mathbf{q}(\psi(t)))\cdot \mathbf{q}(\psi(t))\,dt \\
	&=-\int_{t_2}^{t_1} \boldsymbol{\mathsf{F}}(\mathbf{q}(\tau))\cdot \mathbf{q}(\tau))\,(-d\tau) \\
	& =-\int_{t_1}^{t_2} \boldsymbol{\mathsf{F}}(\mathbf{q}(\tau))\cdot \mathbf{q}(\tau))\,d\tau
	 =-W(\gamma)
	\,.
\end{aligned}
\end{equation}
Therefore, if a non-conservative, non-dissipative force does positive work in a cycle, it does the opposite work when the orientation of the cycle is reversed.
For a dissipative (or accelerating) force of the form  $\boldsymbol{\mathsf{F}}=\boldsymbol{\mathsf{B}}(\mathbf{q})\dot{\mathbf{q}}$,
\begin{equation}
\begin{aligned}
	W(-\gamma)
	&=\int_{t_1}^{t_2} \boldsymbol{\mathsf{B}}(\mathbf{q}(\psi(t)))\frac{d}{dt}\left[\mathbf{q}(\psi(t))\right]
	\cdot d\mathbf{q}(\psi(t)) \\
	& =\int_{t_1}^{t_2} \boldsymbol{\mathsf{B}}(\mathbf{q}(\psi(t)))\,\dot{\mathbf{q}}(\psi(t))\,\dot{\psi}(t)
	\cdot \dot{\mathbf{q}}(\psi(t))\,\dot{\psi}(t)\,dt \\
	& =\int_{t_1}^{t_2} \boldsymbol{\mathsf{B}}(\mathbf{q}(\psi(t)))\,\dot{\mathbf{q}}(\psi(t))
	\cdot \dot{\mathbf{q}}(\psi(t))\,dt \\
	& =\int_{t_2}^{t_1} \boldsymbol{\mathsf{B}}(\mathbf{q}(\tau))\,\dot{\mathbf{q}}(\tau)
	\cdot \dot{\mathbf{q}}(\tau)\,(-d\tau)\\
	& =\int_{t_1}^{t_2} \boldsymbol{\mathsf{B}}(\mathbf{q}(\tau))\,\dot{\mathbf{q}}(\tau)
	\cdot \dot{\mathbf{q}}(\tau)\,d\tau
	=W(\gamma)\,.
\end{aligned}
\end{equation}
As expected, the work done by a dissipative force is always negative regardless of the orientation of the path.

\begin{figure}[t!]
\centering
\scalebox{0.7}{
\begin{minipage}[t]{0.48\textwidth}
\centering
\begin{tikzpicture}
  \begin{scope}[blend group = soft light]
    \fill[red!20!white]   ( 0:2.50) circle (2.50);
    \fill[blue!20!white]  (0:5.50) circle (2.50);
  \end{scope}
  \node at (0:6.50)   {\textsf{Conservative (C)}};
  \node at (0:1.5)    {\textsf{Dissipative (D)}};   
  \draw (1.9,2.8)     node {\textsf{Non-Conservative (NC)}};
  \draw (6.1,2.8)     node {\textsf{Non-Dissipative (ND)}};
  \draw[->, thick, dashed] (4,0) node[below]{} -- (4,3.5) node[above, scale = 1.0] {\textsf{Non-Conservative (NC)} \& \textsf{~Non-Dissipative (ND)}};
\end{tikzpicture}
\vskip 0.2cm
\caption{General forces $\boldsymbol{\mathsf{F}}=\boldsymbol{\mathsf{F}}(t,\mathbf{q},\dot{\mathbf{q}})$}
\label{Conservative-Non-Conservative}
\end{minipage}
\hspace{0.02\textwidth} 
\begin{minipage}[t]{0.48\textwidth}
\centering
\begin{tikzpicture}
\node[box1, fill=Lavender] (c1) {$\begin{aligned} & \textsf{Dissipative} \\ &\Bs^{\mathsf{T}}=\Bs\\ & \quad\textsf{NC}/\text{D} \end{aligned}$};
\node[box1, fill=CornflowerBlue, right=.1cm of c1] (c2) {$\begin{aligned} & \textsf{Potential} \\ &\As^{\mathsf{T}}=\As\\ & ~\textsf{C}/\text{ND} \end{aligned}$};
\node[box1, fill=SkyBlue, below=.1cm of c2] (c3) {$\begin{aligned} & \quad \textsf{Curl} \\ &\Aa^{\mathsf{T}}=-\Aa\\ & ~\text{NC}/\textsf{ND} \end{aligned}$};
\node[box1, fill=Salmon, left=.1cm of c3] (c3) {$\begin{aligned} & \textsf{Gyroscopic} \\ &\Ba^{\mathsf{T}}=-\Ba\\ & ~\,\textsf{NC}/\text{ND} \end{aligned}$};
\end{tikzpicture}
\vskip 0.3cm
\caption{Linear forces $\boldsymbol{\mathsf{F}}=\boldsymbol{\mathsf{A}}\,\mathbf{q}+\boldsymbol{\mathsf{B}}\,\dot{\mathbf{q}}$}
\label{Linear-Forces}
\end{minipage}
}
\vspace{0.3cm}
\caption{(a) Dissipative forces (stresses) are non-conservative, while conservative forces (stresses) are non-dissipative. However, there are forces (stresses) that are non-dissipative yet still not conservative. In Cauchy elasticity,  stress is non-conservative yet non-dissipative. (b) A force that depends on the generalized coordinates and velocities linearly can be decomposed as in \eqref{LinearForce-Decomposition}.}
\label{fig:Conservative-Nonconservative}
\end{figure}

\begin{remark}
A force field is either conservative (i.e., it is derived from a potential field, and conserves energy in all motions), or is non-conservative (i.e. it does not conserve energy in all motions and does not derive from a potential). It is either dissipative (i.e. it dissipates energy in all motions) or non-dissipative (i.e. it does not dissipate energy in any motion). 
A conservative force is non-dissipative but the converse is not necessarily true.
A non-conservative force is not necessarily dissipative. A dissipative force has no potential, and hence, it is always non-conservative. 
The set-theoretic relationship among these four types of forces is illustrated schematically in Figure \ref{Conservative-Non-Conservative}, with the particular linear case shown in Figure \ref{Linear-Forces}.
We will show that, similarly, a stress field can be either conservative or non-conservative. Additionally, it can be classified as either dissipative or non-dissipative. In particular, the stress field in Cauchy elastic solids is, in general, non-conservative and non-dissipative.
\end{remark}

We can now generalize these concepts to the general problem of a continuum and study the role and effect of non-conservative forces and stresses in a general 
elastic material.

\section{An Axiomatic Formulation of Thermodynamics} \label{Sec:ExteriorCalculus}

In hyperelasticity the constitutive equations are fully specified by a strain energy function. In Cauchy elasticity, such a function does not exist, in general. The only specification is that at any material point, stress  depends exclusively and explicitly on the strain at that point. This information is enough to calculate the \emph{stress-work $1$-form} $\boldsymbol{\Omega}$, which replaces the strain energy function as  the fundamental object in Cauchy elasticity \citep{Edelen1977,Kadic1980,Cardin1995}, see Definition~\ref{Stress-Work-Form}. At any material point, the stress-work $1$-form is defined on the manifold of strains. The partial differential equation $\boldsymbol{\Omega}=0$ is called a \textit{Pfaffian equation}, the solutions of which are submanifolds (curves, surfaces, or higher-dimensional manifolds) called  \textit{integral manifolds}. Such integral manifolds are physically important because in a motion (a curve in the manifold of strains) completely lying on an integral manifold stress does zero work. The study of integral manifolds and Pfaffian equations require the theory of exterior differential systems that we briefly review next.

\subsection{Exterior Differential Systems} \label{ExteriorDifferentialSystems}

Let $\boldsymbol{\alpha}$ be a differential form on an $n$-manifold $\mathcal{M}$. Let us define 
\begin{equation}
    (\boldsymbol{\alpha})^{k}=\overbrace{\boldsymbol{\alpha} \wedge \hdots \wedge \boldsymbol{\alpha}}^{k-\text{factors}}
    \,.
\end{equation}
We define $\Lambda^1(\mathcal{M})$ to be the set of differential $1$-forms on the $n$-dimensional manifold $\mathcal{M}$. Let  $\boldsymbol{\Omega}\in \Lambda^1(\mathcal{M})$, then the positive integer $k$ is  the \textit{rank of} $\boldsymbol{\Omega}$ if  $(d\boldsymbol{\Omega})^k\neq 0$ and $(d\boldsymbol{\Omega})^{k+1}= 0$. 
Note that $(d\boldsymbol{\Omega})^k$ is a $2k$-form, and the condition $(d\boldsymbol{\Omega})^k \neq 0$ implies that $n > 2k$ or $n - k > k$.

\begin{thm}[Darboux \citep{Darboux1882,Slebodzinski1970,Sternberg1999,Bryant2013,Suhubi2013}]
 Let $\boldsymbol{\Omega}\in \Lambda^1(\mathcal{M})$ be of rank $k$. If everywhere on $\mathcal{M}$, $\boldsymbol{\Omega}\wedge(d\boldsymbol{\Omega})^k= 0$, then in a neighborhood of any point there are coordinates $\{y^1,\hdots,y^{n-k},z^1,\hdots,z^k\}$ such that
\begin{equation}
    \boldsymbol{\Omega}=y^1dz^1+\hdots+y^k dz^k
    \,.
\end{equation}
If everywhere on $\mathcal{M}$, $\boldsymbol{\Omega}\wedge(d\boldsymbol{\Omega})^k\neq 0$, then in a neighbourhood of every point there are coordinates $\{y^1,\hdots,y^{n-k},$ $z^1,\hdots,z^k\}$ such that
\begin{equation}
    \boldsymbol{\Omega}=y^1dz^1+\hdots+y^kdz^k+dy^{k+1}
    \,.
\end{equation}
\end{thm}
\noindent These are called the \textit{canonical (normal) forms} of $\boldsymbol{\Omega}$.

We can use the notion of normal forms for the study of Cauchy elasticity by considering the normal forms of a $1$-form in dimensions one, two, three, and six, corresponding to incompressible and compressible isotropic Cauchy elasticity,  as well as anisotropic Cauchy elasticity, respectively.

If  $\mathcal{M}$ is one-dimensional, then $d\boldsymbol{\Omega}=0$, and hence $k=0$. This means that there exists a   scalar field $\psi$ such that $\boldsymbol{\Omega}=d\psi$. 
For a two-dimensional manifold one has either $k=0$ or $1$, and hence one has the following possibilities (note that $\boldsymbol{\Omega}\wedge d\boldsymbol{\Omega}$ is a $3$-form and identically vanishes on any $2$-manifold):
\begin{equation} \label{Darboux-2D}
\begin{aligned}
    k=0:& \quad d\boldsymbol{\Omega}=0\,,~ \boldsymbol{\Omega}\wedge (d\boldsymbol{\Omega})^0=\boldsymbol{\Omega}\neq 0 && \Rightarrow ~ \boldsymbol{\Omega}=d\psi_1\,, \\
    k=1:& \quad d\boldsymbol{\Omega}\wedge d\boldsymbol{\Omega}=0, ~ 
    \boldsymbol{\Omega}\wedge d\boldsymbol{\Omega} =0  && \Rightarrow ~ \boldsymbol{\Omega}=\phi^1 d\psi_1\,.
\end{aligned}
\end{equation}
For a three-dimensional manifold still either $k=0$ or $1$, and hence one has the following possibilities:
\begin{equation}
\begin{aligned}
    k=0:& \quad d\boldsymbol{\Omega}=0\,,~ \boldsymbol{\Omega}\wedge (d\boldsymbol{\Omega})^0=\boldsymbol{\Omega}\neq 0 ~ \Rightarrow ~ \boldsymbol{\Omega}=d\psi_1\,,\\
    k=1:& \quad d\boldsymbol{\Omega}\wedge d\boldsymbol{\Omega}=0, ~ 
    \begin{dcases}
    \boldsymbol{\Omega}\wedge d\boldsymbol{\Omega} =0    ~ \Rightarrow ~ \boldsymbol{\Omega}=\phi^1 d\psi_1\,,\\
    \boldsymbol{\Omega}\wedge d\boldsymbol{\Omega} \neq 0    ~ \Rightarrow ~ \boldsymbol{\Omega}=\phi^1 d\psi_1+d\psi_2 \,.
    \end{dcases}
\end{aligned}
\end{equation}
Finally, for a six-dimensional manifold, the rank of a $1$-form can take any of the values $0,1,2,3$. Thus, one has the following possibilities (note that $\boldsymbol{\Omega}\wedge d\boldsymbol{\Omega}\wedge d\boldsymbol{\Omega}\wedge d\boldsymbol{\Omega} $ is a $7$-form and identically vanishes on any $6$-manifold):
\begin{equation}
\begin{aligned}
    k=0:& \quad d\boldsymbol{\Omega}=0\,,~ \boldsymbol{\Omega}\wedge (d\boldsymbol{\Omega})^0=\boldsymbol{\Omega}\neq 0 ~ \Rightarrow ~ \boldsymbol{\Omega}=d\psi_1\,, \\
    k=1:& \quad d\boldsymbol{\Omega}\wedge d\boldsymbol{\Omega}=0, ~ 
    \begin{dcases}
    \boldsymbol{\Omega}\wedge d\boldsymbol{\Omega} =0    ~ \Rightarrow ~ \boldsymbol{\Omega}=\phi^1 d\psi_1\,, \\
    \boldsymbol{\Omega}\wedge d\boldsymbol{\Omega} \neq 0    ~ \Rightarrow ~ \boldsymbol{\Omega}=\phi^1 d\psi_1+d\psi_2\,,
    \end{dcases}\\
    k=2:& \quad d\boldsymbol{\Omega}\wedge d\boldsymbol{\Omega}\wedge d\boldsymbol{\Omega}=0, ~ 
    \begin{dcases}
    \boldsymbol{\Omega}\wedge d\boldsymbol{\Omega}\wedge d\boldsymbol{\Omega} =0    ~ \Rightarrow ~ \boldsymbol{\Omega}=\phi^1 d\psi_1+\phi^2 d\psi_2\,,\\
    \boldsymbol{\Omega}\wedge d\boldsymbol{\Omega}\wedge d\boldsymbol{\Omega} \neq 0    ~ \Rightarrow ~ \boldsymbol{\Omega}=\phi^1 d\psi_1+\phi^2 d\psi_2+d\psi_3\,,
    \end{dcases}\\ 
    k=3:& \quad d\boldsymbol{\Omega} \wedge d\boldsymbol{\Omega}\wedge d\boldsymbol{\Omega}\wedge d\boldsymbol{\Omega}=0, ~ 
    \boldsymbol{\Omega}\wedge d\boldsymbol{\Omega}\wedge d\boldsymbol{\Omega}\wedge d\boldsymbol{\Omega} =0    ~ \Rightarrow ~ \boldsymbol{\Omega}=\phi^1 d\psi_1+\phi^2 d\psi_2+\phi^3 d\psi_3\,.
\end{aligned}
\end{equation}

Let $\boldsymbol{\Omega}\in\Lambda^1(\mathcal{M})$, then the first-order partial differential equation 
\begin{equation}\label{Pfaff}
	\boldsymbol{\Omega}=0\,,
\end{equation}
is called a \textit{Pfaffian equation}. This is the simplest example of an exterior differential system.
A $p$-dimensional integral manifold of \eqref{Pfaff} is an immersion (not necessarily an embedding) $f:\mathcal{W}\to\mathcal{M}$ such that $f^*\boldsymbol{\Omega}=0$ ($f^*$ denotes the pullback by $f$).
An integral curve of $\boldsymbol{\Omega}$ is a curve $c:I\to\mathcal{M}$, where $I$ is an interval on the real line, such that $\boldsymbol{\Omega}(c(t))=0$, $\forall t\in I$. The Pfaffian equation \eqref{Pfaff} is said to have the \emph{local accessibility} property if, for every point $x\in\mathcal{M}$, there exists a neighborhood $U$ such that for every point $y\in U$, there is an integral curve of \eqref{Pfaff} that connects $x$ and $y$.
The Pfaffian equation \eqref{Pfaff} has the \emph{local inaccessibility} property if, in every neighborhood $U$ of every point $x \in \mathcal{M}$, there is at least one point $y \in$ U that cannot be connected to $x$ by any integral curve of \eqref{Pfaff}.

The rank of the Pfaffian equation \eqref{Pfaff} is the integer $r$ such that $\boldsymbol{\Omega}\wedge(d\boldsymbol{\Omega})^r\neq 0$ and $\boldsymbol{\Omega}\wedge(d\boldsymbol{\Omega})^{r+1}= 0$. Carath\'eodory's theorem tells us that if the rank of a Pfaffian equation is constant, the Pfaffian equation has the local accessibility property if and only if $r\geq 2$. Therefore, a Pfaffian equation has the local inaccessibility property only when $r=0$ or $r=1$, which correspond to the normal forms $\boldsymbol{\Omega}=d\psi_1$ and $\boldsymbol{\Omega}=\phi^1 d\psi_1$, respectively. In either case, $\boldsymbol{\Omega}=0$ implies that $\psi_1=\text{constant}$, which are are hypersurfaces---the integral manifolds of maximum possible dimension $n-1$.

\subsection{Carath\'eodory's Abstract Formulation of Thermodynamics}\label{Thermodynamics}

Carath\'eodory \citep{Caratheodory1909} reformulated thermodynamics as follows: The first law is described by the vanishing of a differential $1$-form on a manifold of thermodynamic states, i.e., a Pfaffian equation on a manifold. The second law is stated as an axiom asserting that this differential $1$-form has the property of inaccessibility.

The set of all states of the system is denoted by $\mathcal{M}$ and is assumed to be a connected manifold. 
Carath\'eodory states the first law of thermodynamics as 
\begin{equation}
    \boldsymbol{\theta}=du-\boldsymbol{\omega}_h-\boldsymbol{\omega}_w=0\,,
\end{equation}
where $u$ is the internal energy of the system, $\boldsymbol{\omega}_h$ is the heat $1$-form and $\boldsymbol{\omega}_w$ is the work $1$-form all defined on $\mathcal{M}$ \citep{Mrugala1978}.
An \textit{adiabatic process} (or change of state) of a thermodynamic system is one in which there is no exchange of heat with its surroundings. An adiabatic process is a curve in $\mathcal{M}$, i.e., $c:I\to\mathcal{M}$, where $I$ is an interval on the real line, such that $\boldsymbol{\omega}_h(c(t))=0$, $\forall t\in I$.

Carath\'eodory's statement of the second law of thermodynamics is: \textit{For any thermodynamic system, in every neighborhood of a given state, there exist states that are inaccessible via adiabatic changes of state} \citep{Pogliani2000,Frankel2011}.  In other words, the heat $1$-form has the inaccessibility property. In simple terms, inaccessibility means that one cannot move between all possible states of the system without violating the laws of thermodynamics. For instance, while some changes in a system’s energy or state might be possible, others would require additional input, like heat, that the isolated system cannot provide.  Carath\'eodory's theorem then tells us that the heat $1$-form is integrable, i.e., $\boldsymbol{\omega}_h=T ds$, where $T$ is absolute temperature and $s$ is entropy \citep{Cooper1967,Boyling1972,Buchdahl2009}.

\section{Covariant Cauchy Elasticity: Kinematics and Balance Laws} \label{Cauchy Elasticity}

We now turn our attention to a general theory of Cauchy elasticity. First, we introduce the general theoretical background. We consider a body that is deforming in the Euclidean ambient space $(\mathcal{S},\mathbf{g})$, where $\mathbf{g}$ is a fixed flat background metric (a metric lets one measure distances). We identify the body with a Riemannian manifold $(\mathcal{B},\mathbf{G})$, where $\mathcal{B}$ is a $3$-manifold embeddable in the Euclidean ambient space and $\mathbf{G}=\mathbf{g}\big|_{\mathcal{B}}$ is the induced flat material metric.

We assume that the body $\mathcal{B}$ is made of a Cauchy elastic solid. Hence at any $X\in\mathcal{B}$, the stress depends only the strain at $X$ without any history dependence. Importantly, we do not assume the existence of an underlying energy function.

\subsection{Kinematics} 

In nonlinear elasticity, kinematics is unaffected by existence or lack thereof an energy function. In the following we briefly review the kinematics of nonlinear elasticity.

A \textit{motion} (or deformation) is a one-parameter family of mappings $\varphi_t:\mathcal{B}\rightarrow \mathcal{C}_t\subset\mathcal{S}$, where $\mathcal{C}_t=\varphi_t(\mathcal{B})$ is the current configuration. A material point $X\in\mathcal{B}$ is mapped to $x_t=\varphi_t(X)=\varphi(X,t)\in\mathcal{S}$.
The \textit{material velocity} of motion $\mathbf{V}_t:\mathcal{B}\rightarrow T_{\varphi_t(X)}\mathcal{C}_t$ is defined as $\mathbf{V}_t(X)=\mathbf{V}(X,t)=\frac{\partial \varphi(X,t)}{\partial t}$. Here, $T_{\varphi_t(X)}\mathcal{C}_t$ is the tangent space of the current configuration at $\varphi_t(X)\in\mathcal{C}_t$.
One defines the \textit{spatial velocity} as $
\mathbf{v}=\mathbf{V}\circ\varphi_t^{-1}$.

In order to be able to measure local distances the ambient space is equipped with a background (Euclidean) metric $\mathbf{g}$ making it a Riemannian manifold $(\mathcal{S},\mathbf{g})$. The body is an embedded submanifold of the ambient space and the inclusion map $\iota:\mathcal{B}\to \mathcal{S}$ induces a metric $\mathbf{G}$ on $\mathcal{B}$ defined as $\mathbf{G}=\mathbf{g}\big |_{\mathcal{B}}=\iota^*\mathbf{g}$. The Levi-Civita connections of $\mathbf{G}$ and $\mathbf{g}$ are denoted by $\nabla^{\mathbf{G}}$ and $\nabla^{\mathbf{g}}$, respectively. With respect to the coordinate charts $\{X^A\}$ and $\{x^a\}$ for $\mathcal{B}$ and $\mathcal{S}$, respectively, the connection coefficients (Christoffel symbols) are denoted by $\Gamma^A{}_{BC}$ and $\gamma^a{}_{bc}$, respectively. They are defined such that 
\begin{equation}
	\nabla^{\mathbf{G}}_{\frac{\partial}{\partial X^B}}\frac{\partial}{\partial X^A}
	=\Gamma^A{}_{BC}\,\frac{\partial}{\partial X^C}\,,\qquad
	\nabla^{\mathbf{g}}_{\frac{\partial}{\partial x^b}}\frac{\partial}{\partial x^a}
	=\gamma^a{}_{bc}\,\frac{\partial}{\partial x^c}
	\,.
\end{equation}
The connection coefficients have the following expressions in terms of the metric components
\begin{equation}
	\gamma^a{}_{bc}=\frac{1}{2}g^{ak}\left(g_{kb,c}+g_{kc,b}-g_{bc,k}\right)\,,\qquad \Gamma^A{}_{BC}
	=\frac{1}{2}G^{AK}\left(G_{KB,C}+G_{KC,B}-G_{BC,K}\right)\,.
\end{equation}
For a vector field on $T\mathcal{S}$ but defined with respect to $\mathcal{B}$, e.g., the material velocity $\mathbf{V}$, covariant derivative is defined as follows
\begin{equation}
	\tilde{\nabla}^{\mathbf{g}}_{\mathbf{Y}} (\mathbf{V}\circ\varphi)
	= \nabla^{\mathbf{g}}_{\varphi_*\mathbf{Y}}\mathbf{V}\,,\quad \forall \mathbf{Y}\in T\mathcal{B}\,.
\end{equation}
In components, $V^a{}_{|A}=V^a{}_{|b}\,F^b{}_A=V^a{}_{,A}+\gamma^a{}_{bc} V^b \,F^c{}_A$.

The \textit{deformation gradient} denoted by $\mathbf{F}=T\varphi$ is the tangent map (or derivative) of $\varphi_t$ (a metric-independent two-point tensor). 
At each material point $X \in \mathcal{B}$, the deformation gradient is a linear map $\mathbf{F}(X):T_{X}\mathcal{B}\rightarrow T_{\varphi_t(X)}\mathcal{C}_t$, where $T_{X}\mathcal{B}$ is the tangent space of the reference configuration at $X\in\mathcal{B}$. 
We introduce local coordinate charts $\{x^a\}$ and $\{X^A\}$ on $\mathcal{S}$ and $\mathcal{B}$, respectively, which in general may be curvilinear. We denote the cotangent space of the reference configuration at $X\in\mathcal{B}$ by $T^*_{X}\mathcal{B}$, which is the space of covectors or $1$-form (a $1$-form when paired with a vector yields a scalar). One can show that $\{dX^A\}$ and $\{\frac{\partial}{\partial x^a}\}$ are bases for $T^*_{X}\mathcal{B}$ and $T_{\varphi_t(X)}\mathcal{C}_t$, respectively \citep{MarsdenHughes1983}.
With respect to these coordinate charts the deformation gradient has the following representation 
\begin{equation}
    \mathbf{F}=F^a{}_A\,\frac{\partial}{\partial x^a}\otimes\,dX^A=
	\frac{\partial \varphi^a}{\partial X^A}\,\frac{\partial}{\partial x^a}\otimes\,dX^A\,.
\end{equation}
The transpose of deformation gradient (a metric-dependent tensor) is defined as
\begin{equation}
	\mathbf{F}^{\textsf{T}}:T_{x_t}\mathcal{C}_t 
	\rightarrow T_{X}\mathcal{B},\quad \llangle \mathbf{FV},\mathbf{v} \rrangle_{\mathbf{g}}    
	=\llangle\mathbf{V},\mathbf{F}^{\textsf{T}}\mathbf{v}\rrangle_{\mathbf{G}}, \quad
	\forall \mathbf{V} \in T_{X}\mathcal{B},~\mathbf{v} \in T_{x_t} \mathcal{C}_t\,,
\end{equation}
which has components $(F^{\textsf{T}})^A{}_{a}=G^{AB} \,F^b{}_{B}\,g_{ab}$. Thus, $\mathbf{F}^{\textsf{T}}=\mathbf{G}^\sharp \mathbf{F}^\star \mathbf{g}$.

The \textit{right Cauchy-Green deformation tensor} is defined as $\mathbf{C}(X)=\mathbf{F}^{\textsf{T}}(X) \mathbf{F}(X):T_{X} \mathcal{B}\rightarrow T_{X} \mathcal{B}$. It has components $C^A{}_{B}=(F^{\textsf{T}})^A{}_{a}\,F^a{}_{B}$. 
Notice that $C_{AB}=(g_{ab}\circ \varphi)F^a{}_{A}\,F^b{}_{B}$, which implies that $\mathbf{C}^\flat=\varphi^*\mathbf{g}=\mathbf{F}^\star \mathbf{g}\,\mathbf{F}$ is the pull-back of the spatial metric to the current configuration, where $\flat$ is the flat operator induced by the metric $\mathbf{g}$ and $\mathbf{F}^\star(X): T_{\varphi_t(X)}\mathcal{C}_t\to T_{X}\mathcal{B}$ is the dual of $\mathbf{F}$ and is defined as 
\begin{equation}
    \mathbf{F}^\star=F^a{}_A\,dX^A \otimes \frac{\partial}{\partial x^a} \,.
\end{equation}
The \textit{left Cauchy-Green deformation tensor} is defined as $\mathbf{B}^{\sharp}=\varphi^*(\mathbf{g}^{\sharp})=\mathbf{F}^{-1}\mathbf{g}^\sharp \mathbf{F}^{-\star}$, i.e., the pull back of the inverse of the spatial metric to the current configuration.  It has components $B^{AB}=F^{-A}{}_a\,F^{-B}{}_b\,g^{ab}$, where $F^{-A}{}_a$ are components of $\mathbf{F}^{-1}$. 
The spatial analogs of $\mathbf{C}^\flat$ and $\mathbf{B}^{\sharp}$ are denoted as $\mathbf{c}^\flat$ and $\mathbf{b}^{\sharp}$, respectively, and are defined as (push forwards of the material metric and its inverse) $\mathbf{c}^\flat=\varphi_*\mathbf{G}$, with components $c_{ab}=F^{-A}{}_a\,F^{-B}{}_b\,G_{AB}$, and $\mathbf{b}^{\sharp}=\varphi_*\mathbf{G}^{\sharp}=\mathbf{F}\mathbf{G}^\sharp\mathbf{F}^\star$, with components $b^{ab}=F^a{}_{A}\,F^b{}_{B}\,G^{AB}$.
Note that $\mathbf{F}\mathbf{F}^{\textsf{T}} = \mathbf{F}\mathbf{G}^\sharp \mathbf{F}^\star \mathbf{g}=\mathbf{b}^\sharp \mathbf{g}=\mathbf{b}$.

The tensors $\mathbf{C}$ and $\mathbf{b}$ have identical \textit{principal invariants} $I_1$, $I_2$, and $I_3$, which are defined as \citep{Ogden1984}\footnote{To see this, let us assume that $\mathbf{W}$ is an eigenvector of $\mathbf{C}$ with its corresponding eigenvalue $\lambda$, i.e., $\mathbf{C}\mathbf{W}=\lambda\mathbf{W}$. Thus, $\mathbf{F}\mathbf{C}\mathbf{W}=\lambda\mathbf{F}\mathbf{W}$. This can be rewritten as $\mathbf{F}\mathbf{G}^\sharp \mathbf{F}^\star \mathbf{g}\,\mathbf{F}\mathbf{W}=\mathbf{b}^\sharp \mathbf{g}\,\mathbf{F}\mathbf{W}=\mathbf{b}\,\mathbf{F}\mathbf{W}=\lambda\mathbf{F}\mathbf{W}$. This shows that $\mathbf{b}$ has the same eigenvalue $\lambda$ with a corresponding eigenvector $\mathbf{F}\mathbf{W}$.}
\begin{equation}\label{Principal-Invariants}
    I_1 =\operatorname{tr}\mathbf{b}\,,\qquad
    I_2 =\frac{1}{2}\left(I_1^2-\operatorname{tr}\mathbf{b}^2\right)
    \,,\qquad  I_3 =\det \mathbf{b}\,.
\end{equation}
These three invariants are related to changes of lengths, areas, and volumes in a deformation, respectively \cite{kuhl2024too}. 

In the geometric setting, the \textit{polar decomposition} of the deformation gradient can be stated as 
\begin{equation}\label{Polar-Decomposition}
    \mathbf{F}=\mathbf{R}\mathbf{U}=\mathbf{V}\mathbf{R} \,,
\end{equation}
where $\mathbf{U}$ and $\mathbf{V}$ (not to be confused with the material velocity) are the material and the spatial stretch tensors, respectively, and $\mathbf{R}:T\mathcal{B}\to\mathcal{C}$ is a $(\mathbf{G},\mathbf{g})$-orthogonal tensor field in the sense that \citep{SimoMarsden1984}:
\begin{equation} \label{G-g-Orthogonal}
	\mathbf{R}^\star (\mathbf{g}\circ\varphi)\, \mathbf{R}=\mathbf{G}    \,.
\end{equation}
In components, $R^a{}_A\,(g_{ab}\circ\varphi)\,R^b{}_B=G_{AB}$. 
The polar decomposition in components reads
\begin{equation}
	F^a{}_A= R^a{}_B\, U^B{}_A =V^a{}_b\,R^b{}_A    \,.
\end{equation}
Eq.~\eqref{G-g-Orthogonal} implies that $(\det\mathbf{R})^2\det\mathbf{g}=\det\mathbf{G}$, and from \eqref{Polar-Decomposition} we observe that $\det\mathbf{U}=\det\mathbf{V}$. Recall that Jacobian of deformation is defined by the relation $dv=J\,dV$, and it is expressed as
\begin{equation}
	J=\sqrt{\frac{\det\mathbf{g}}{\det\mathbf{G}}}\,\det\mathbf{F}=\det\mathbf{U}=\det\mathbf{V}  \,.
\end{equation}

The material stretch tensor $\mathbf{U}:T_X\mathcal{B}\to T\mathcal{B}$ and spatial stretch tensor $\mathbf{V}:T_x\mathcal{C}\to T\mathcal{C}$ are related to the right and left Cauchy-Green deformation tensors as
\begin{equation} \label{C-U-b_v}
\begin{aligned}
	\mathbf{C} &=\mathbf{F}^{\textsf{T}} \mathbf{F}
	=(\mathbf{R}\mathbf{U})^{\textsf{T}}\mathbf{R}\mathbf{U}
	=\mathbf{G}^\sharp(\mathbf{R}\mathbf{U})^\star \mathbf{g}\mathbf{R}\mathbf{U}
	=\mathbf{G}^\sharp \mathbf{U}^\star \mathbf{R}^\star \mathbf{g}\mathbf{R}\mathbf{U}
	=\mathbf{G}^\sharp \mathbf{U}^\star \mathbf{G} \mathbf{U}=\mathbf{U}^2 \,,\\
	\mathbf{b} &=\mathbf{F}\mathbf{F}^{\textsf{T}}
	=\mathbf{V}\mathbf{R} (\mathbf{V}\mathbf{R})^{\textsf{T}}
	=\mathbf{V}\mathbf{R}\mathbf{G}^\sharp(\mathbf{V}\mathbf{R})^\star\mathbf{g}
	=\mathbf{V}\mathbf{R}\mathbf{G}^\sharp \mathbf{R}^\star\mathbf{V}^\star \mathbf{g}
	=\mathbf{V} \mathbf{g}^\sharp \mathbf{V}^\star \mathbf{g}
	=\mathbf{V}^2	    \,.
\end{aligned}
\end{equation}
Or equivalently
\begin{equation}
	\mathbf{C}^\flat=\mathbf{U}^\star \mathbf{G} \mathbf{U}\,,\qquad 
	\mathbf{b}^\sharp = \mathbf{V} \mathbf{g}^\sharp \mathbf{V}^\star
	\,,
\end{equation}
which in components read $C_{AB} = U^M{}_A\,G_{MN}\,U^N{}_B$ and $b^{ab}=V^a{}_m\,g^{mn}\,V^b{}_n$. 
The relations \eqref{C-U-b_v} are usually written as $\mathbf{U}=\sqrt{\mathbf{C}}$ and $\mathbf{V}=\sqrt{\mathbf{b}}$.

The right Cauchy-Green deformation and the material stretch tensors have the following spectral representations \citep{Ogden1984}:
\begin{equation} \label{C-Spectral}
	\mathbf{C}^\sharp= \lambda_1^2 \,\N \otimes\N	+\lambda_2^2 \,\NN \otimes\NN
	+\lambda_3^2 \,\NNN \otimes\NNN
	\,,\qquad
	\mathbf{U}^\sharp= \lambda_1 \,\N \otimes\N+\lambda_2 \,\NN \otimes\NN
	+\lambda_3 \,\NNN \otimes\NNN\,,
\end{equation}
where $\lambda_1, \lambda_2, \lambda_3$ and $\N, \NN, \NNN$ are the principal stretches and their corresponding principal directions. 
Note that $\N\otimes\N+ \NN\otimes\NN+ \NNN\otimes\NNN=\mathbf{G}^\sharp$.
The representation \eqref{C-Spectral}$_2$ is equivalent to $\mathbf{U}= \lambda_1 \,\N \otimes\N^\flat+\lambda_2 \,\NN \otimes\NN^\flat+\lambda_3 \,\NNN \otimes\NNN^\flat$, and hence $\mathbf{U}\Nj=\lambda_j\Nj$ (no summation). Note that $\mathbf{F}\Nj=\mathbf{R}\mathbf{U}\Nj=\lambda_j \mathbf{R}\Nj=\mathbf{V}\mathbf{R}\Nj$. This implies that denoting the eigenbasis of $\mathbf{V}$ by $\nj$, $j=1,2,3$ one has \citep{Ogden1984}: $\nj=\mathbf{R}\Nj$, $j=1,2,3$.
Therefore, the Finger and spatial stretch tensors have the following spectral representations:
\begin{equation}
	\mathbf{b}^\sharp= 
	\lambda_1^2 \,\n \otimes\n+\lambda_2^2 \,\nn \otimes\nn+\lambda_3^2 \,\nnn \otimes\nnn\,,\qquad
	\mathbf{V}^\sharp= 
	\lambda_1 \,\n \otimes\n+\lambda_2 \,\nn \otimes\nn+\lambda_3 \,\nnn \otimes\nnn
	\,.
\end{equation}
Note that
\begin{equation} 
	\sum_{j=1}^3 \nj\otimes \nj = \sum_{j=1}^3 \mathbf{R}\Nj\otimes \mathbf{R}\Nj
	= \mathbf{R} \left(\sum_{j=1}^3 \Nj\otimes \Nj\right) \mathbf{R}^\star
	= \mathbf{R} \mathbf{G}^\sharp \mathbf{R}^\star =\mathbf{g}^\sharp \,.
\end{equation}

Suppose $f:\mathbb{R}\to\mathbb{R}$ is a smooth monotone function such that $f(1)=0$ and $f'(1)=1$. Hill's strain measures are defined as \citep{Hill1968,Hill1970,Hill1978}
\begin{equation}
	f(\mathbf{V}^\sharp)= 
	f(\lambda_1) \,\n \otimes\n+ f(\lambda_2) \,\nn \otimes\nn+f(\lambda_3) \,\nnn \otimes\nnn
	\,.
\end{equation}
In particular, Hencky's logarithmic strain is defined as 
\begin{equation}
	\mathbf{h}^\sharp= \log\mathbf{V}^\sharp= 
	\log \lambda_1 \,\n \otimes\n+\log\lambda_2 \,\nn \otimes\nn+\log\lambda_3 \,\nnn \otimes\nnn
	\,.
\end{equation}

\subsection{Stress measures and the stress-work $1$-form} 

In continuum mechanics, several stress measures are commonly used.
Here, we consider four of them, namely, the Cauchy $\boldsymbol{\sigma}$, the first Piola-Kirchhoff $\mathbf{P}$, the second Piola-Kirchhoff $\mathbf{S}$, and the Kirchhoff $\boldsymbol{\tau}$ stress tensors.

In the reference configuration, a surface element $dA$ with unit normal vector $\mathbf{N}$ is mapped to its corresponding deformed surface element $da$, which has the unit normal vector $\mathbf{n}$.
The force acting on the deformed surface element is given by $\mathbf{f} = \mathbf{t} \, da$, where $\mathbf{t}$ is the traction vector.
This is related to the Cauchy stress as $\mathbf{t} = \boldsymbol{\sigma} \cdot \mathbf{n} = \llangle \boldsymbol{\sigma}, \mathbf{n} \rrangle_{\mathbf{g}}$, where $\llangle \cdot, \cdot \rrangle_{\mathbf{g}}$ denotes the inner product with respect to the metric $\mathbf{g}$. In component form, the traction vector is expressed as $t^a = \sigma^{ab} g_{bc}\, n^c$.

The Piola identity tells us that $\mathbf{n}^{\flat}da=J\mathbf{F}^{-\star}\mathbf{N}^{\flat}dA$, which in component form is expressed as  $n_ada=J\,F^{-A}{}_aN_AdA$. In the continuum mechanics literature this is known as Nanson’s formula.
The first Piola-Kirchhoff stress is defined such that $\mathbf{f}=\mathbf{t}_0dA=\mathbf{P}\cdot\mathbf{N}dA= \llangle \mathbf{P}, \mathbf{N} \rrangle_{\mathbf{G}}\, dA$. Thus, $\mathbf{P} = J \boldsymbol{\sigma} \mathbf{F}^{-\star}$.
The second Piola-Kirchhoff $\mathbf{S}$ stress is defined such that $ \llangle \mathbf{S}, \mathbf{N} \rrangle_{\mathbf{G}}\, dA=\varphi^*\mathbf{f}=\mathbf{F}^{-1}\mathbf{f}$, and hence, $\mathbf{S} = \mathbf{F}^{-1} \mathbf{P}=J \mathbf{F}^{-1} \boldsymbol{\sigma} \mathbf{F}^{-\star}$.

\begin{defi}\label{Stress-Work-Form}
The \emph{stress work $1$-form} is defined as
\begin{equation}
	\boldsymbol{\Omega}(\mathbf{C}(X,t))
	=\frac{1}{2}S^{AB}(X,t)\,\dot{C}_{AB}(X,t) \,dt
	=\frac{1}{2}S^{AB}(X,t)\,dC_{AB}(X,t)
	=\frac{1}{2} \mathbf{S}\!:\!d\mathbf{C}^\flat
	\,.
\end{equation}
This is a $1$-form in the space of symmetric second-order tensors (a six dimensional space).
\end{defi}

One can define $\boldsymbol{\Omega}$ using the deformation gradient and the first Piola-Kirchhoff stress as well. However, caution is required when defining a time derivative for the deformation gradient, as $\mathbf{F}$ maps a fixed vector in $T_X\mathcal{B}$ to different tangent spaces over time: $\mathbf{F}(X,t): T_X\mathcal{B} \to T_{\varphi_t(X)}\mathcal{C}$. To account for this, one can use a covariant time derivative, defined as $D_t \mathbf{F} := \nabla^{\mathbf{g}}_{\frac{\partial}{\partial t}} \mathbf{F}$.
Using the following identity \citep{Nishikawa2002}
\begin{equation}
	\nabla^{\mathbf{g}}_{\frac{\partial}{\partial t}} \frac{\partial \varphi^a}{\partial X^A}
	=\tilde{\nabla}^{\mathbf{g}}_{\frac{\partial}{\partial X^A}} \frac{\partial \varphi^a}{\partial t}
	=\tilde{\nabla}^{\mathbf{g}}_{\frac{\partial}{\partial X^A}} V^a
	\,,
\end{equation}
one concludes that
\begin{equation}
	D_t\mathbf{F}=\tilde{\nabla}^{\mathbf{g}}\mathbf{V}
	\,.
\end{equation}
Therefore, the stress-work $1$-form is written in terms of the deformation gradient as
\begin{equation} \label{Stress-Work-Form-First-Piola}
\begin{aligned}
	\boldsymbol{\Omega}(\mathbf{F}(X,t))
	&=P_a{}^A(X,t)\,V^a{}_{|A}(X,t) \,dt
	=P_a{}^A(X,t)\,dF^a{}_A(X,t) \\
	&=\mathbf{g}\mathbf{P}\!:\!d\mathbf{F}\,,
\end{aligned}
\end{equation}
which is a $1$-form on the space of deformation gradients.

Note that from \eqref{Stress-Work-Form-First-Piola} one can write
\begin{equation}
\begin{aligned}
	\boldsymbol{\Omega} &=J \sigma^{ab} F^{-A}{}_b\,V_{a|A} \,dt 
	=J \sigma^{ab} \,V_{a|b} \,dt =J \sigma^{ab} \,\frac{1}{2} \left[V_{a|b}+V_{b|a}\right] \,dt \\
	&= \frac{1}{2} J \sigma^{ab} \, (\mathfrak{L}_{\mathbf{V}}\mathbf{g}\circ\varphi)_{ab} \,dt \,.
\end{aligned}
\end{equation}
Thus
\begin{equation}
	\boldsymbol{\Omega} = \frac{1}{2} \tau^{ab} \, 
	(\mathfrak{L}_{\mathbf{V}}\mathbf{g}\circ\varphi)_{ab} \,dt
	=\frac{1}{2} \boldsymbol{\tau}\!:\! \mathfrak{L}_{\mathbf{V}}\mathbf{g}\circ\varphi\,dt
	=\boldsymbol{\tau}\!:\! (\mathbf{d}\circ\varphi)\,dt
	\,,
\end{equation}
where $\boldsymbol{\tau}=J \boldsymbol{\sigma}$ is the Kirchhoff stress and $\mathbf{d}=\frac{1}{2}\mathfrak{L}_{\mathbf{V}}\mathbf{g}$ is the spatial rate of deformation tensor.
The strain measure conjugate to the Kirchhoff stress is defined as one whose objective rate coincides exactly with the spatial rate of deformation tensor. Under specific conditions, this strain measure had been shown to correspond to Hencky’s logarithmic strain \citep{Gurtin1983, Hoger1986}. This was later proved for arbitrary deformations by \citet{Xiao1997}.
This, in particular, implies that the stress-work $1$-form can be written in terms of the Kirchhoff stress as\footnote{A similar expression appears in \citep[Eq.~(3.8*)]{Richter1948}. For a hyperelastic solid \citet{Vallee1978} writes the constitutive equations (in our notation) as $\boldsymbol{\tau}(\mathbf{h})=D_{\mathbf{h}^\flat}\psi(\mathbf{h})=\frac{\partial \psi}{\partial \mathbf{h}^\flat}$.}
\begin{equation} \label{Stress-Work-Form-Kirchhoff}
	\boldsymbol{\Omega} =\boldsymbol{\tau}\!:\!d\mathbf{h}^\flat
	\,.
\end{equation}

\subsection{Balance laws}

One method for deriving the governing equations of a field theory, such as elasticity, is to use the Lagrangian field theory approach; an action is defined, and extremizing this action leads to the Euler-Lagrange equations. In the case of hyperelasticity, these equations correspond to the balance of linear momentum. 

Noether’s theorem states that invariance of the Lagrangian density under a group of symmetries corresponds to a conserved quantity. For instance, in hyperelasticity, invariance of the Lagrangian density under rigid-body rotations of the Euclidean ambient space leads to the balance of angular momentum. 
In the case of Cauchy elasticity, there is no energy function, and as a result, a Lagrangian density cannot be defined. This implies that the connection between conserved quantities and symmetries via Noether's theorem is lost.

Another approach to derive  balance laws  is to use thermodynamics and invariance arguments. This idea goes back to the work of \citet{Green64} who postulated an energy balance (the first law of thermodynamics) and its invariance under superposed rigid body motions of the ambient Euclidean space. Using these two assumptions they obtained the balance of linear and angular momenta as well as conservation of mass, embodied in the Green-Naghdi-Rivlin theorem. 
A different version of this theorem is due to \citet{Noll1963}, who interpreted superposed motions passively, viewing them as time-dependent coordinate charts on the Euclidean ambient space.
Recently, \citet{SadikYavari2025} generalized the Coleman-Noll procedure by deriving the balance laws of hyperelasticity and hyper-anelasticity directly from the first and second laws of thermodynamics, without assuming invariance. 

In some applications the ambient space may be curved, e.g., when modeling dynamics of fluid membranes \citep{Arroyo2009,Yavari2016}. \citet{HuMa1977} pointed out that in such problems postulating integral balance laws does not make sense as on a manifold a vector field cannot be integrated intrinsically. Instead, they postulated the first law of thermodynamics and its invariance under arbitrary diffeomorphisms of the Riemannian ambient space---this is called \textit{covariance of the energy balance}. Using these two assumptions they derived conservation of mass, the balance of linear momentum, and the Doyle-Ericksen formula \citep{DoyleEricksen1956}, which implies the balance of angular momentum (see also \citep{MarsdenHughes1983,SimoMarsden1983,YaMaOr2006,YavariOzakin2008,Yavari2008,YavariMarsden2009a,YavariMarsden2009b,Yavari2010,YavariGolgoon2019}).

In the following, we adopt this approach by first introducing the first law of thermodynamics and then imposing invariance under diffeomorphisms.
\subsubsection{The first law of thermodynamics}

In Section~\ref{Thermodynamics}, we discussed Carath\'eodory’s abstract formulation of thermodynamics, and particularly, the first law written as vanishing of a differential $1$-form. 
The classical formulation of the first law in solid mechanics is usually presented as an integral form over an arbitrary subbody. 
In this section we start with this classical approach and derive the governing equations of Cauchy elasticity postulating covariance of the balance of energy. 
At the end of Section~\ref{Second-Law-Continuum} we establish a connection with Carath\'eodory’s formulation of the first law (see Remark \ref{Work-Heat-Forms}).

For any sub-body $\mathcal{U}\subset\mathcal{B}$ of a hyperelastic body $\mathcal{B}$, the balance of energy is written as
\begin{equation}\label{Hyperelastic-Energy-Balance}
	\frac{d}{dt}
	\int_{\mathcal{U}} \rho_0 \left(\mathscr{E}+\frac{1}{2} \Vert\mathbf V\Vert^2_{\mathbf g}\right)
	dV
	=\int_{\mathcal{U}}\rho_0 \left(\llangle\mathbf{B},\mathbf{V}\rrangle_{\mathbf{g}}
	+R\right)dV+\int_{\partial \mathcal{U}}
	\left(\llangle\mathbf{T},\mathbf{V}\rrangle_{\mathbf{g}}+H\right)dA\,,
\end{equation}
where $\rho_0(X,t)$ is the material mass density, $\mathscr{E}$ is the internal energy (or energy function), 
$\mathbf{B}$ is the body force, $\mathbf{T}$ is the boundary traction vector field per unit material area, ${R=R(X,t)}$ is the specific heat supply, and $H=-\llangle \mathbf{Q},\mathbf{N}\rrangle_{\mathbf{G}}$ is the material heat flux, ${\mathbf Q=\mathbf Q(X,T,dT,\mathbf C^\flat,\mathbf G)}$ is the external heat flux per unit material area, $T$ is temperature, $dT$ is its exterior derivative, and $\mathbf{N}$ is the $\mathbf{G}$-unit normal to the boundary $\partial\mathcal{B}$.

Despite the fact that  Cauchy elastic solids do not necessarily have an energy function,  one can still write an energy balance \citep{GreenNaghdi1971}.
For any sub-body $\mathcal{U}\subset\mathcal{B}$ of a Cauchy elastic body $\mathcal{B}$, we write the energy balance  as\footnote{This implies that the rate of change of energy is written as 
\begin{equation}\label{Energy-Rate}
	\rho_0 \dot{\mathscr{E}}
	=\rho_0 ( \tilde{\mathbf{P}}\!:\!\tilde{\nabla}^{\mathbf{g}}\mathbf{V}+\mathsf{f} \dot{\mathscr{N}} ) \,.
\end{equation}
It might seem natural to assume that $\rho_0 \tilde{\mathbf{P}}  = \mathbf{P}$ is the first Piola-Kirchhoff stress and $\mathsf{f} = T$ is the absolute temperature. However, we show in the following that one can prove these using invariance arguments. It should be emphasized that $\mathscr{E}$ is not necessarily a function of $\mathbf{F}$.
}
\begin{equation}\label{Cauchy-Energy-Balance}
\begin{aligned}
	\frac{d}{dt}
	\int_{\mathcal{U}} \frac{1}{2} \rho_0 \Vert\mathbf V\Vert^2_{\mathbf g} ~dV
	+\int_{\mathcal{U}}\rho_0 \left( \tilde{\mathbf{P}}\!:\!\tilde{\nabla}^{\mathbf{g}}\mathbf{V}
	+\mathsf{f} \dot{\mathscr{N}} 
	\right)dV
	&=\int_{\mathcal{U}}\rho_0 \left(\llangle\mathbf{B},\mathbf{V}\rrangle_{\mathbf{g}}+R\right)dV \\
	&\quad +\int_{\partial \mathcal{U}}  \left(\llangle\mathbf{T},\mathbf{V}\rrangle_{\mathbf{g}}+H\right)dA\,,
\end{aligned}
\end{equation}
where $\tilde{\mathbf{P}}$ is a two-point tensor to be determined, $\tilde{\nabla}^{\mathbf{g}}\mathbf{V}$ is the \textit{velocity gradient} with components $V^a{}_{|A}$ (note that the covariant derivative is with respect to the Levi-Civita connection of the metric $\mathbf{g}$), $\mathscr{N}$ is the \textit{specific entropy}, and $\mathsf{f}$ is a scalar field to be determined.
Note that the Cauchy stress theorem does not rely on the existence of an energy function, and hence $\llangle\mathbf{T},\mathbf{V}\rrangle_{\mathbf{g}}=\mathbf{P}\mathbf{N}^\flat$, where $\mathbf{P}$ is the \textit{first Piola-Kirchhoff stress tensor}.\footnote{It should be emphasized that Cauchy stress theorem can be proved only using the balance law \eqref{Cauchy-Energy-Balance} without relying on the balance of linear momentum \citep[\S2.1(1.9)]{MarsdenHughes1983}.}
Thus
\begin{equation}
	\int_{\partial \mathcal{U}} \llangle\mathbf{T},\mathbf{V}\rrangle_{\mathbf{g}} ~dA
	=\int_{ \mathcal{U}}  \left(\llangle \operatorname{Div}\mathbf{P},\mathbf{V}\rrangle_{\mathbf{g}}
	+ \mathbf{P}\!:\!\nabla\mathbf{V} \right)dV
	\,.
\end{equation}
Therefore, the balance of energy is simplified to read
\begin{equation} \label{Cauchy-Energy-Balance1}
\begin{aligned}
	& \int_{ \mathcal{U}}  \left(\llangle \operatorname{Div}\mathbf{P}+\rho_0 (\mathbf{B}-\mathbf{A}),
	\mathbf{V}\rrangle_{\mathbf{g}}
	+ (\mathbf{P}-\rho_0 \tilde{\mathbf{P}})\!:\!\nabla\mathbf{V}
	- \frac{1}{2} \dot{\rho}_0 \Vert\mathbf V\Vert^2_{\mathbf g}+\rho_0 R-\rho_0 \mathsf{f} \dot{\mathscr{N}}   
	\right)dV \\
	& \quad +\int_{\partial \mathcal{U}} H dA=0
	\,.
\end{aligned}
\end{equation}

\subsubsection{Covariance of the energy balance}

We consider an arbitrary time-dependent spatial diffeomorphism $\xi_t:\mathcal{S}\to\mathcal{S}$ such that $\xi_{t=0}=\operatorname{id}_{\mathcal{S}}$, the identity map.
We define $\mathbf{w}=\frac{\partial \xi_t}{\partial t}$, and $\mathbf{W}=\mathbf{w}\circ\varphi^{-1}$. Under this change of frame $\mathbf{g}'=\xi_{t*}\mathbf{g}=(T\xi_t)^{-\star}\,\mathbf{g}\,(T\xi_t)^{-1}$ and the deformation map is transformed to $\varphi'_t=\xi_t\circ\varphi_t$, and hence \citep{MarsdenHughes1983,Yavari2006}
\begin{equation} 
	\mathbf{V}'=\xi_{t*}\mathbf{V}+\mathbf{W}=T\xi_{t}\mathbf{V}+\mathbf{W}
	\,.
\end{equation}
Thus, at $t=0$, $\mathbf{V}'=\mathbf{V}+\mathbf{W}$.
The \textit{covariance of energy balance} is defined as the invariance of the form of the energy balance \eqref{Cauchy-Energy-Balance1}  under $\xi_t$, i.e.,
\begin{equation} \label{Cauchy-Energy-Balance2}
\begin{aligned}
	& \int_{ \mathcal{U}}  \left(\llangle \operatorname{Div}'\mathbf{P}'+\rho'_0 (\mathbf{B}'-\mathbf{A}'),
	\mathbf{V}'\rrangle_{\mathbf{g}'}
	+ (\mathbf{P}'-\rho'_0 \tilde{\mathbf{P}}')\!:\!\tilde{\nabla}^{\mathbf{g}'}\mathbf{V}'
	- \frac{1}{2} \dot{\rho}'_0 \Vert\mathbf{V}'\Vert^2_{\mathbf{g}'}+\rho'_0R'
	-\rho'_0 \mathsf{f}' \dot{\mathscr{N}}'  \right)dV' \\
	& \quad +\int_{\partial \mathcal{U}} H' dA'=0
	\,.
\end{aligned}
\end{equation}
Under the change of frame the following transformations are assumed (notice that material fields and operators are unaffected by a spatial change of frame) \citep{MarsdenHughes1983}
\begin{equation} \label{Transformed-Fields}
\begin{aligned}
	& \rho'_0=\rho_0\,,\quad 
	\mathsf{f}' = \mathsf{f} \,,\quad
	\mathscr{N}' = \mathscr{N} \,,\quad
	H'=H\,,\quad R'=R\,,\\
	& \mathbf{B}'-\mathbf{A}'=\xi_{t*}(\mathbf{B}-\mathbf{A})=T\xi_{t}\cdot(\mathbf{B}-\mathbf{A})\,,\\
	& \tilde{\mathbf{P}}' = \xi_{t*}\mathbf{P} = T\xi_t\cdot\mathbf{P}\,,\quad
	\tilde{\mathbf{P}}' = \xi_{t*}\tilde{\mathbf{P}} = T\xi_t\cdot\tilde{\mathbf{P}}\,,\\
    & \tilde{\nabla}^{\mathbf{g}'}=\tilde{\nabla}^{\mathbf{g}}\,,\quad dV'=dV\,,\quad dA'=dA
	\,.
\end{aligned}
\end{equation}
Substituting \eqref{Transformed-Fields} into \eqref{Cauchy-Energy-Balance2} and evaluating at $t=0$, one obtains
\begin{equation} \label{Cauchy-Energy-Balance3}
\begin{aligned}
	& \int_{ \mathcal{U}}  \Bigg(\llangle \operatorname{Div}\mathbf{P}+\rho_0 (\mathbf{B}-\mathbf{A}),
	\mathbf{V}+\mathbf{W}\rrangle_{\mathbf{g}}
	+ (\mathbf{P}-\rho_0 \tilde{\mathbf{P}})\!:\!(\tilde{\nabla}^{\mathbf{g}}\mathbf{V}
	+\tilde{\nabla}^{\mathbf{g}}\mathbf{W}) \\
	& \qquad - \frac{1}{2} \dot{\rho}_0 \Vert \mathbf{V}+\mathbf{W} \Vert^2_{\mathbf{g}}
	+\rho_0 R-\rho_0 \mathsf{f} \dot{\mathscr{N}} \Bigg)dV 
	+\int_{\partial \mathcal{U}} H dA=0
	\,.
\end{aligned}
\end{equation}
Subtracting \eqref{Cauchy-Energy-Balance1} from \eqref{Cauchy-Energy-Balance3} one gets
\begin{equation} \label{Cauchy-Energy-Balance4}
	 \int_{ \mathcal{U}}  \left[\llangle \operatorname{Div}\mathbf{P}+\rho_0 (\mathbf{B}-\mathbf{A}),
	\mathbf{W}\rrangle_{\mathbf{g}}
	+ (\mathbf{P}-\rho_0 \tilde{\mathbf{P}})\!:\!\tilde{\nabla}^{\mathbf{g}}\mathbf{W}
	-  \dot{\rho}_0 \left(\frac{1}{2}\Vert \mathbf{W} \Vert^2_{\mathbf{g}}
	+\llangle \mathbf{V},\mathbf{W}\rrangle_{\mathbf{g}} \right)   \right]dV =0
	\,.
\end{equation}
The above identity must hold for arbitrary vectors $\mathbf{W}$. We choose $\mathbf{W}=\beta \hat{\mathbf{W}}$ for some $\mathbf{g}$-unit vector $\hat{\mathbf{W}}$. Taking derivatives of both side with respect to $\beta$ twice one obtains
\begin{equation} 
	 \int_{ \mathcal{U}}  \dot{\rho}_0\, dV =0
	\,,
\end{equation}
which must hold for arbitrary $\mathcal{U}\subset\mathcal{B}$. Therefore, $\dot{\rho}_0=0$, which we recognize as the \textit{conservation of mass}. Now \eqref{Cauchy-Energy-Balance4} is simplified to read
\begin{equation} \label{Cauchy-Energy-Balance5}
	 \int_{ \mathcal{U}}  \left[\llangle \operatorname{Div}\mathbf{P}+\rho_0 (\mathbf{B}-\mathbf{A}),
	\mathbf{W}\rrangle_{\mathbf{g}}
	+ (\mathbf{P}-\rho_0 \tilde{\mathbf{P}})\!:\!\tilde{\nabla}^{\mathbf{g}}\mathbf{W} \right]dV =0
	\,.
\end{equation}
As $\mathbf{W}$ and $\nabla\mathbf{W}$ are independent one concludes that
\begin{equation} \label{Covariance-Results}
	\operatorname{Div}\mathbf{P}+\rho_0 \mathbf{B} =\rho_0\mathbf{A}\,,\qquad
	\rho_0 \tilde{\mathbf{P}}  = \mathbf{P}
	\,.
\end{equation}
We conclude that the covariance of the energy balance leads to both the conservation of mass and the balance of linear momentum. We can use thes two relations to simplify the first law of thermodynamics:
\begin{equation} \label{Cauchy-Energy-Balance6}
	\int_{ \mathcal{U}} \rho_0 \left(R- \mathsf{f} \dot{\mathscr{N}} \right)dV
	+\int_{\partial \mathcal{U}} H dA=0
	\,.
\end{equation}
Recall that $H=-\llangle \mathbf{Q},\mathbf{N}\rrangle_{\mathbf{G}}$, where ${\mathbf{Q}=\mathbf Q(X,T,dT,\mathbf C^\flat,\mathbf G)}$ is the external heat flux per unit material area, $T$ is temperature, and $\mathbf{N}$ is the $\mathbf{G}$-unit normal to the boundary $\partial\mathcal{U}$.
Thus, the energy balance in local form reads
\begin{equation} \label{Cauchy-Energy-Balance-Local}
	 \rho_0 R -\operatorname{Div}\mathbf{Q}
	  -\rho_0 \mathsf{f} \dot{\mathscr{N}} =0
	\,.
\end{equation}
Eq.~\eqref{Energy-Rate} is now rewritten as
\begin{equation}
	 \dot{\mathscr{E}}= \frac{1}{\rho_0}\mathbf{P}\!:\!\nabla\mathbf{V}+\mathsf{f} \dot{\mathscr{N}} 
	=\frac{1}{2\rho_0} \mathbf{S}\!:\!\dot{\mathbf{C}}^\flat+\mathsf{f} \dot{\mathscr{N}}
	=\frac{1}{2\rho_0} \mathbf{S}\!:\!\dot{\mathbf{C}}^\flat+\mathsf{f} \dot{\mathscr{N}} \,,
\end{equation}
where $\mathbf{S}$ is the \textit{second Piola-Kirchhoff stress tensor}.
This can be rewritten as
\begin{equation} \label{Energy-1-Form}
	 \dot{\mathscr{E}} dt= \frac{1}{2\rho_0} \mathbf{S}\!:\!d\mathbf{C}^\flat+\mathsf{f} \dot{\mathscr{N}}dt \,.
\end{equation}

\subsubsection{The second law of thermodynamics} \label{Second-Law-Continuum}

In \S\ref{Thermodynamics}, our discussion of Carathéodory’s abstract formulation of thermodynamics was motivated by its relevance to certain work theorems in elasticity, which will be examined in \S\ref{WorkTheorems}.
There have been careful studies of the relation between Carath\'eodory's postulate and the principle of increase of entropy \citep{Boyling1972}. 
Here, we adopt the standard formulation of the second law of thermodynamics for Cauchy elasticity.

For any sub-body $\mathcal{U}\subset\mathcal{B}$, the second law of thermodynamics can be expressed in terms of the Clausius-Duhem inequality as
\citep{Truesdell1952,gurtin1974modern,MarsdenHughes1983}
\begin{equation}\label{Second-Law}
	\gamma=\frac{d}{dt} \int_{\mathcal{U}} \rho_0 \mathscr{N}dV
	-\int_{\mathcal{U}}\rho_0 \frac{R}{T}dV-\int_{\partial\mathcal{U}}\frac{H}{T}dA \geq 0\,,
\end{equation}
where $\gamma$ is the entropy production.
In localized form, the material Clausius-Duhem inequality is written as
\begin{equation} \label{Second-Law-Local}
	\dot\eta = \rho_0 T \dot{\mathscr{N}} 
	+ \operatorname{Div}\mathbf{Q}  - \rho_0 R -\frac{1}{T}\langle dT,\mathbf{Q}\rangle
	\geq 0\,,
\end{equation}
where $\dot\eta$ denotes the material rate of energy dissipation density, and $\langle .,. \rangle$ is the natural pairing of $1$-forms and vectors.
Substituting \eqref{Cauchy-Energy-Balance-Local} into \eqref{Second-Law-Local} one obtains
\begin{equation} \label{Second-Law-Local1}
	\dot\eta = \rho_0 (T-\mathsf{f} ) \dot{\mathscr{N}} 
	-\frac{1}{T}\langle dT,\mathbf{Q}\rangle \geq 0\,.
\end{equation}
Suppose that the  temperature is uniform, i.e., $dT=0$, then $\dot\eta = \rho_0 (T-\mathsf{f} ) \dot{\mathscr{N}} \geq 0$, and hence $\mathsf{f}=T$.
Therefore, the material rate of energy dissipation density is written as
\begin{equation} \label{Rate-Energy-Dissipation}
	\dot\eta =   -\frac{1}{T}\langle dT,\mathbf{Q}\rangle \geq 0\,.
\end{equation}
This, in particular, implies that
\begin{equation} \label{Entropy-Rate}
	\rho_0 T \dot{\mathscr{N}} = \rho_0 R  -\operatorname{Div}\mathbf{Q} \,.
\end{equation}
From \eqref{Energy-Rate}, \eqref{Covariance-Results}$_2$, and \eqref{Entropy-Rate} the rate of change of energy is written as 
\begin{equation} \label{Energy-Rate2}
	\rho_0\dot{\mathscr{E}}= \mathbf{P}\!:\!\tilde{\nabla}^{\mathbf{g}}\mathbf{V}+\rho_0 T \dot{\mathscr{N}}
	= \mathbf{P}\!:\!\tilde{\nabla}^{\mathbf{g}}\mathbf{V}+\rho_0 R  -\operatorname{Div}\mathbf{Q}
	\,.
\end{equation}
This also implies that the integral form of the balance of energy in Cauchy elasticity reads
\begin{equation}\label{Cauchy-Energy-Balance-Final}
\begin{aligned}
	\frac{d}{dt}
	\int_{\mathcal{U}} \frac{1}{2} \rho_0 \Vert\mathbf V\Vert^2_{\mathbf g} ~dV
	+\int_{\mathcal{U}} \left(\mathbf{P}\!:\!\nabla\mathbf{V}+\rho_0T \dot{\mathscr{N}} \right)dV
	& =\int_{\mathcal{U}}\rho_0 \left(\llangle\mathbf{B},\mathbf{V}\rrangle_{\mathbf{g}}+R\right)dV\\
	& \quad +\int_{\partial \mathcal{U}}  \left(\llangle\mathbf{T},\mathbf{V}\rrangle_{\mathbf{g}}+H\right)dA\,.
\end{aligned}
\end{equation}

In summary, the first law of thermodynamics and its covariance give us the conservation of mass and the balance of linear momentum. Instead of the rate of change of internal energy, in the expression of the first law there are two a priori unknown terms conjugate to the velocity gradient $\tilde{\nabla}^{\mathbf{g}}\mathbf{V}$ and the rate of entropy $\dot{\mathscr{N}}$. Together, covariance of the first law and the second law tell us that the term conjugate to $\tilde{\nabla}^{\mathbf{g}}\mathbf{V}$ is the first Piola-Kirchhoff stress and that conjugate to $\dot{\mathscr{N}}$ is the temperature.

\begin{remark}\label{Work-Heat-Forms}
From \eqref{Energy-1-Form} one concludes that
\begin{equation}
	\dot{\mathscr{E}} dt= \frac{1}{2\rho_0} \mathbf{S}\!:\!d\mathbf{C}^\flat+T \dot{\mathscr{N}}dt  \,.
\end{equation}
This implies that the work and heat $1$-forms are written as $\boldsymbol{\omega}_w=\frac{1}{2\rho_0} \mathbf{S}\!:\!d\mathbf{C}^\flat=\frac{1}{\rho_0}\boldsymbol{\Omega}$ and $\boldsymbol{\omega}_h=T \dot{\mathscr{N}}dt $.
\end{remark}

\begin{remark}
Suppose there is no heat source or flux in a Cauchy elastic solid, i.e., from \eqref{Energy-Rate2}, we obtain 
\begin{equation}\rho_0\dot{\mathscr{E}} dt= \mathbf{P}\!:\!\tilde{\nabla}^{\mathbf{g}}\mathbf{V} dt=\boldsymbol{\Omega}.\end{equation} Therefore, for a path $\Gamma$ in the strain space, one has
\begin{equation}
	\int_{\Gamma} \boldsymbol{\Omega} = \int_{t_1}^{t_2} \rho_0\,\dot{\mathscr{E}} dt
	\,.
\end{equation}
As $\mathscr{E}$ is not necessarily a function of $\mathbf{F}$, the above integral is path dependent, in general. 
In particular, along closed paths, the net work of stress is equal to the total energy either supplied to or extracted from the body.
Although it may seem non-intuitive, this is entirely consistent with the first and second laws of thermodynamics.
Recall that Cauchy elasticity is characterized by the possibility of having non-zero stress work along closed paths in  strain space that enclose regions with non-zero area. However, for closed paths in  strain space that do not enclose a surface, non-hyperelastic Cauchy elasticity is indistinguishable from hyperelasticity. 
Importantly, in Cauchy elasticity, the net work of stress along a closed path $\Gamma$, followed by its reverse $-\Gamma$, is always zero. In contrast, for a dissipative solid this work is always negative.
\end{remark}

\subsubsection{Balance of angular momentum and the generalized Doyle-Ericksen formula}

We will show in \S\ref{Edelen-Darboux-Potentials} that the most general Cauchy elastic solid has six Darboux-Edelen potentials in terms of $\mathbf{C}^\flat$. In this case, the stress-work $1$-form can be written as \begin{equation}
    \boldsymbol{\Omega} =  \phi_1 d\psi_1+\phi_2 d\psi_2+\phi_3 d\psi_3,
\end{equation} where $\phi_i=\phi_i(X,\mathbf{F},\mathbf{G},\mathbf{g})$ and $\psi_i=\psi_i(X,\mathbf{F},\mathbf{G},\mathbf{g})$. The term $\mathbf{P}\!:\!\tilde{\nabla}^{\mathbf{g}}\mathbf{V}$ in the balance of energy \eqref{Cauchy-Energy-Balance-Final} is $\dot{\boldsymbol{\Omega}}$, which is written as
\begin{equation}
	\dot{\boldsymbol{\Omega}}
	=\sum_{i=1}^3 \phi_i \frac{d\psi_i}{dt}
	=\sum_{i=1}^3 \phi_i \frac{\partial \psi_i}{\partial \mathbf{F}}\!:\!\tilde{\nabla}^{\mathbf{g}}\mathbf{V}
	\,.
\end{equation}
Therefore, we can re-write the balance of energy as
\begin{equation}\label{Cauchy-Energy-Balance-Edelen}
\begin{aligned}
	& \frac{d}{dt}
	\int_{\mathcal{U}} \frac{1}{2} \rho_0 \Vert\mathbf V\Vert^2_{\mathbf g} ~dV
	+\int_{\mathcal{U}} \left[\sum_{i=1}^3 \phi_i \frac{\partial \psi_i}{\partial \mathbf{F}}\!:\!
	\tilde{\nabla}^{\mathbf{g}}\mathbf{V}
	+\rho_0T \dot{\mathscr{N}} \right]dV\\
	&\quad=\int_{\mathcal{U}}\rho_0 \left[\llangle\mathbf{B},\mathbf{V}\rrangle_{\mathbf{g}}+R\right]dV
	 +\int_{\partial \mathcal{U}}  \left[\llangle\mathbf{T},\mathbf{V}\rrangle_{\mathbf{g}}+H\right]dA\,.
\end{aligned}
\end{equation}
Covariance of the energy balance implies two things: First, the six potentials are covariant, i.e.,
\begin{equation}\label{Covariance-Ptentials}
	\phi_i=\hat{\phi}_i(X,\mathbf{C}^\flat,\mathbf{G})\,,\qquad
	\psi_i=\hat{\psi}_i(X,\mathbf{C}^\flat,\mathbf{G},)\,,\qquad i=1,2,3
	\,,
\end{equation}
and second,
\begin{equation} 
	 \int_{ \mathcal{U}}  
	 \left[\mathbf{P}-\sum_{i=1}^3 \phi_i \frac{\partial \psi_i}{\partial \mathbf{F}}\right]\!:\!
	 \tilde{\nabla}^{\mathbf{g}}\mathbf{W} dV =0
	\,,
\end{equation}
which implies, through localization,
\begin{equation} \label{Generalized-DE}
	 \mathbf{P}=\sum_{i=1}^3 \phi_i \frac{\partial \psi_i}{\partial \mathbf{F}}
	\,.
\end{equation}
This last expression is a \textit{generalized Doyle-Ericksen formula}. Indeed when restricted to hyperlelastic materials ($\phi_1=1$, $\phi_2=\phi_3=\psi_2=\psi_3=0$) one recovers the classical Doyle-Ericksen formula of hyperelasticity. Using \eqref{Covariance-Ptentials} the formula \eqref{Generalized-DE} can be rewritten as
\begin{equation} 
	 \mathbf{P}=2\sum_{i=1}^3 \phi_i \,\mathbf{F}\,\frac{\partial \psi_i}{\partial \mathbf{C}^\flat}
	 \,, \quad \text{or}\qquad
	 \mathbf{S}=2\sum_{i=1}^3 \phi_i \,\frac{\partial \psi_i}{\partial \mathbf{C}^\flat}	\,.
\end{equation}
These relations have the important property that $\mathbf{P}\mathbf{F}^{\star}=\mathbf{F}\mathbf{P}^{\star}$ (or equivalently, $\mathbf{S}^{\star}=\mathbf{S}$), which by analogy with hyperelasticity, can be interpreted as the  balance of angular momentum.

Hence, we have proved the following result:

\begin{prop}
In a Cauchy elastic solid, the balance of energy and its covariance, together with the second law of thermodynamics, imply the conservation of mass, the balance of linear momentum, and the balance of angular momentum through a generalized Doyle-Ericksen formula.
\end{prop}

\subsubsection{Balance of angular momentum and objectivity in Cauchy elasticity}

In the absence of an energy function, the connection between conservation laws and symmetries via Noether’s theorem is broken. 
A key assumption to obtain $\mathbf{S}=\mathbf{S}^\star$ in the previous section was  objectivity implying that $\mathbf{S}=\hat{\mathbf{S}}(X,\mathbf{C}^\flat,\mathbf{G})$.
Therefore, in Cauchy elasticity objectivity implies the balance of angular momentum. It turns out that unlike the case of hyper-elasticity the converse is not true.

To show the balance of angular momentum implies objectivity for hyperelastic material, we follow \citet{Kadic1980} who writes the stress-work $1$-form in terms of the deformation gradient:
\begin{equation} \label{Liouville-1-form}
	\boldsymbol{\Omega}(\mathbf{F}(X,t),\mathbf{G}(X,t),\mathbf{g}(\varphi(X,t)))
	=g_{ab}(\varphi(X,t))\,P^{bA}(X,t)\,dF^a{}_{A}(X,t)\,.
\end{equation}
In the case of a hyperelastic solid 
\begin{equation} 
	\boldsymbol{\Omega}(\mathbf{F},\mathbf{G},\mathbf{g})=d\psi_1(\mathbf{F},\mathbf{G},\mathbf{g})
	=\frac{\partial \psi_1}{\partial F^a{}_A}\, dF^a{}_A
	\,,
\end{equation}
and hence
\begin{equation}
	g_{ab}\,P^{bA}=\frac{\partial \psi_1}{\partial F^a{}_A}\,,\quad \text{or} \qquad
	P^{aA}=g^{ab}\,\frac{\partial \psi_1}{\partial F^b{}_A}
	\,.
\end{equation}
Note that
\begin{equation}
	0=d\circ d\psi_1(\mathbf{F},\mathbf{G},\mathbf{g})
	=\frac{\partial P_a{}^A}{\partial F^b{}_B}\, dF^b{}_B\wedge d F^a{}_A
	=\frac{1}{2}\left[\frac{\partial P_a{}^A}{\partial F^b{}_B}
	-\frac{\partial P_b{}^B}{\partial F^a{}_A}\right] dF^b{}_B\wedge d F^a{}_A
	\,.
\end{equation}
Thus
\begin{equation}
	\frac{\partial P_a{}^A}{\partial F^b{}_B} = \frac{\partial P_b{}^B}{\partial F^a{}_A} \,.
\end{equation}
Balance of angular momentum is written as $\mathbf{P}\mathbf{F}^\star=\mathbf{F}\mathbf{P}^\star$ or in components $P^{aA}\,F^b{}_A = P^{bA}\,F^a{}_A$. Thus
\begin{equation} \label{Angular-Momentum-Green}
	g^{ac}\,\frac{\partial \psi_1}{\partial F^c{}_A}\,F^b{}_A 
	= g^{bc}\,\frac{\partial \psi_1}{\partial F^c{}_A}\,F^a{}_A
	\,,\quad\text{or}\qquad
	\frac{\partial \psi_1}{\partial F^a{}_A}\,F^b{}_A 
	- \frac{\partial \psi_1}{\partial F^b{}_A}\,F^a{}_A=0
	 \,.
\end{equation}
These relations are satisfied if and only if $\psi_1=\hat{\psi}_1(\mathbf{C}^\flat,\mathbf{G})$ \citep{Kadic1980,Duff1956}, i.e., the balance of angular momentum implies objectivity.

Conversely, let us start with objectivity, which implies that $\psi_1=\hat{\psi}_1(\mathbf{C}^\flat,\mathbf{G})$, and hence
\begin{equation}
	\mathbf{P}=2\mathbf{F}\,\frac{\partial \hat{\psi}_1}{\partial \mathbf{C}^\flat}\,,
	\quad\text{or~in~components}\quad
	P^a{}^A= 2F^a{}_N\,\frac{\partial \hat{\psi}_1}{\partial C_{NA}} \,.
\end{equation}
Note that symmetry of $\mathbf{C}^\flat$ implies that
\begin{equation}
	\mathbf{P}\mathbf{F}^\star=2\mathbf{F}\,\frac{\partial \hat{\psi}_1}{\partial \mathbf{C}^\flat}\mathbf{F}^\star
	=\left[2\mathbf{F}\,\frac{\partial \hat{\psi}_1}{\partial \mathbf{C}^\flat}\mathbf{F}^\star\right]^\star
	=\mathbf{F}\mathbf{P}^\star
	 \,,
\end{equation}
i.e., the balance of angular momentum. Therefore, in hyperelasticity objectivity and balance of angular momentum are equivalent \citep{Noll1955}, which is a consequence of Noether's theorem.

Now consider the simplest Cauchy material, namely  an Ericksen elastic solid (see \S\ref{Edelen-Darboux-Potentials}), with two potentials
\begin{equation}
	\boldsymbol{\Omega}(\mathbf{F},\mathbf{G},\mathbf{g})
	=\phi_1(\mathbf{F},\mathbf{G},\mathbf{g})\,d\psi_1(\mathbf{F})
	=\phi_1(\mathbf{F},\mathbf{G},\mathbf{g})\,\frac{\partial \psi_1}{\partial F^a{}_A}\, dF^a{}_A
	=g_{ab}\,P^{bA}\,dF^a{}_{A}
	\,,
\end{equation}
and hence
\begin{equation}
	P^{aA}=g^{ab}\,\phi_1(\mathbf{F},\mathbf{G},\mathbf{g})\,\frac{\partial \psi_1}{\partial F^b{}_A}
	\,.
\end{equation}
Balance of angular momentum is written as
\begin{equation}
	g^{ac}\,\phi_1(\mathbf{F},\mathbf{G},\mathbf{g})\,\frac{\partial \psi_1}{\partial F^c{}_A}\,F^b{}_B 
	= g^{bc}\,\phi_1(\mathbf{F},\mathbf{G},\mathbf{g})\,\frac{\partial \psi_1}{\partial F^c{}_A}\,F^a{}_B
	\,,
\end{equation}
or
\begin{equation}
	\phi_1(\mathbf{F},\mathbf{G},\mathbf{g})\left[\frac{\partial \psi_1}{\partial F^a{}_A}\,F^b{}_B 
	- \frac{\partial \psi_1}{\partial F^b{}_A}\,F^a{}_B\right]=0
	 \,.
\end{equation}
As $\phi_1\neq 0$, this gives us \eqref{Angular-Momentum-Green}. Therefore, one concludes that one of the potentials, namely $\psi_1$, is objective, i.e., $\psi_1=\psi_1(\mathbf{C}^\flat,\mathbf{G})$. However, there is no condition on $\phi_1(\mathbf{F},\mathbf{G},\mathbf{g})$ \citep{Kadic1980}. Therefore, in Ericksen elasticity objectivity is a stronger condition that implies the balance of angular momentum but the converse is not necessarily true.

For an Edelen solid of type III, the first Piola-Kirchhoff stress is written as
\begin{equation}
	P^{aA}=g^{ab}\,\phi_1(\mathbf{F},\mathbf{G},\mathbf{g})\,\frac{\partial \psi_1}{\partial F^b{}_A}
	+g^{ab}\,\frac{\partial \psi_2}{\partial F^b{}_A}
	\,.
\end{equation}
The balance of angular momentum tells us that
\begin{equation}
	\phi_1(\mathbf{F},\mathbf{G},\mathbf{g})\left[\frac{\partial \psi_1}{\partial F^a{}_A}\,F^b{}_B 
	- \frac{\partial \psi_1}{\partial F^b{}_A}\,F^a{}_B\right]
	+\frac{\partial \psi_2}{\partial F^a{}_A}\,F^b{}_B 
	- \frac{\partial \psi_2}{\partial F^b{}_A}\,F^a{}_B=0
	 \,.
\end{equation}
In this case none of the three potentials needs to be a function of the right Cauchy-Green strain for the balance of angular momentum to hold. However, objectivity would again imply the balance of angular momentum. The same conclusion can be reached for Edelen solids of types IV, V, and VI.

\begin{prop}
In Cauchy elasticity, objectivity implies the balance of angular momentum. However, the converse is not necessarily true.
\end{prop}

\section{Constitutive Equations for Cauchy Elastic Solids} \label{Sec:Constitutive-Equations}

In an inhomogeneous Cauchy elastic solid one has the constitutive equation $\mathbf{P}=\hat{\mathbf{P}}(X,\mathbf{F},\mathbf{G},\mathbf{g})$. In terms of Cauchy stress one has $\boldsymbol{\sigma}=\hat{\boldsymbol{\sigma}}(x,\mathbf{c}^\flat,\mathbf{g})$ \citep{YavariSozio2023}.

\subsection{Objectivity in Cauchy elasticity}

Objectivity, also known as material-frame indifference, implies that for all deformation gradients $\mathbf{F}$, under the left action of the orthogonal group one has
\begin{equation}
	\hat{\mathbf{P}}(X,\mathbf{Q}\mathbf{F},\mathbf{G},\mathbf{g})
	=\mathbf{Q}\hat{\mathbf{P}}(X,\mathbf{F},\mathbf{G},\mathbf{g})\,,
	\qquad \forall \mathbf{Q}\in \text{Orth}(\mathbf{g}),
\end{equation}
where Orth$(\mathbf{g})=\{\mathbf Q:T_{\varphi(X)}\mathcal{C}\to T_{\varphi(X)}\mathcal{C} ~|~ \mathbf{Q}^*\mathbf{g}=\mathbf{Q}^\star\mathbf{g}\mathbf{Q}=\mathbf{g} \}$ is the group of $\mathbf{g}$-orthogonal transformations.

\begin{prop}[\citep{TruesdellNoll2004}]
Objectivity implies that the constitutive equation of a Cauchy elastic solid is written in the following form 
\begin{equation} \label{C-S-ConstitituveEquation}
	\mathbf{S}=\hat{\mathbf{S}}(X,\mathbf{C}^\flat,\mathbf{G})\,.
\end{equation}
\end{prop}
\begin{proof}
Consider two deformation gradients $\mathbf{F}_1$ and $\mathbf{F}_2$ such that $\mathbf{F}_1^{\mathsf{T}}\mathbf{F}_1=\mathbf{F}_2^{\mathsf{T}}\mathbf{F}_2=\mathbf{C}$, or equivalently, $\mathbf{F}_1^*\mathbf{g}=\mathbf{F}_2^*\mathbf{g}=\mathbf{C}^\flat$. 
Let us define $\mathbf{R}=\mathbf{F}_2\mathbf{F}_1^{-1}:T_x\mathcal{C}\to T_x\mathcal{C}$.
For arbitrary vectors $\mathbf{u}_1, \mathbf{u}_2\in T_x\mathcal{C}$, as $\mathbf{F}_1$ is invertible, one has $\mathbf{u}_1=\mathbf{F}_1\mathbf{U}_1$ and $\mathbf{u}_2=\mathbf{F}_1\mathbf{U}_2$, where $\mathbf{U}_1, \mathbf{U}_2\in T_X\mathcal{B}$.
Thus 
\begin{equation}
\begin{aligned}
	\llangle \mathbf{R}\mathbf{u}_1 , \mathbf{R}\mathbf{u}_2 \rrangle_{\mathbf{g}}
	&=\llangle \mathbf{R}\mathbf{F}_1\mathbf{U}_1 , 
	\mathbf{R}\mathbf{F}_1\mathbf{U}_2\rrangle_{\mathbf{g}} 
	=\llangle \mathbf{F}_2\mathbf{U}_1 , \mathbf{F}_2\mathbf{U}_2\rrangle_{\mathbf{g}} 
	=\llangle \mathbf{U}_1 , \mathbf{U}_2\rrangle_{\mathbf{F}_2^*\mathbf{g}}\\
	&=\llangle \mathbf{U}_1 , \mathbf{U}_2\rrangle_{\mathbf{F}_1^*\mathbf{g}} 
	=\llangle \mathbf{F}_1\mathbf{U}_1 , \mathbf{F}_1\mathbf{U}_2\rrangle_{\mathbf{g}}=\llangle \mathbf{u}_1 , 
	\mathbf{u}_2 \rrangle_{\mathbf{g}}		
	\,.
\end{aligned}
\end{equation}
This shows that $\mathbf{R}$ is a local isometry, and hence material frame-indifference implies that 
\begin{equation}
	\hat{\mathbf{P}}(X,\mathbf{R}\mathbf{F}_1,\mathbf{G},\mathbf{g})
	=\mathbf{R}\hat{\mathbf{P}}(X,\mathbf{F}_1,\mathbf{G},\mathbf{g})
	\,.
\end{equation}
Thus
\begin{equation}
	\hat{\mathbf{P}}(X,\mathbf{F}_2,\mathbf{G},\mathbf{g})
	=\mathbf{F}_2\mathbf{F}_1^{-1} \hat{\mathbf{P}}(X,\mathbf{F}_1,\mathbf{G},\mathbf{g})\,,\quad
	\text{or} \qquad
	\mathbf{F}_2^{-1}\hat{\mathbf{P}}(X,\mathbf{F}_2,\mathbf{G},\mathbf{g})
	=\mathbf{F}_1^{-1} \hat{\mathbf{P}}(X,\mathbf{F}_1,\mathbf{G},\mathbf{g})
	\,.
\end{equation}
This proves that $\mathbf{S}=\hat{\mathbf{S}}(X,\mathbf{C}^\flat,\mathbf{G})$ is well defined and is equal to the common value of $\mathbf{F}^{-1}\hat{\mathbf{P}}(X,\mathbf{F},\mathbf{G},\mathbf{g})$ for any $\mathbf{F}$ such that $\mathbf{C}^\flat=\mathbf{F}^*\mathbf{g}=\mathbf{F}^\star\mathbf{g}\mathbf{F}$.
\end{proof}

\subsection{Material symmetry in Cauchy elasticity}

The material symmetry group $\mathcal{G}_X$ of a Cauchy elastic solid at a point $X\in\mathcal{B}$ with respect to the Euclidean reference configuration $(\mathcal{B},\mathbf{G})$ is defined as 
\begin{equation} 
	\hat{\mathbf{S}}(X,\mathbf{K}^*\mathbf{C}^\flat,\mathbf{G})
	=\mathbf{K}^*\hat{\mathbf{S}}(X,\mathbf{C}^\flat,\mathbf{G})\,, \qquad \forall\,\,
	\mathbf{K}\in \mathcal{G}_X\leqslant \mathrm{Orth}(\mathbf{G})\,,
\end{equation}
or
\begin{equation} 
	\hat{\mathbf{S}}(X,\mathbf{K}^\star\mathbf{C}^\flat \mathbf{K},\mathbf{G})
	=\mathbf{K}^\star \hat{\mathbf{S}}(X,\mathbf{C}^\flat,\mathbf{G})\mathbf{K}\,, \qquad \forall\,\,
	\mathbf{K}\in \mathcal{G}_X\leqslant \mathrm{Orth}(\mathbf{G})\,,
\end{equation}
for all deformation gradients $\mathbf{F}$, where
Orth$(\mathbf{G})=\{\mathbf{Q}: T_X\mathcal{B}\to T_X\mathcal{B}~|~\mathbf{K}^*\mathbf{G}= \mathbf{Q}^{\star}\mathbf{G}\mathbf{Q}=\mathbf{G} \}$.

\subsubsection{Isotropic Cauchy elasticity}

For an isotropic Cauchy solid, $\mathcal{G}_X=\mathrm{Orth}(\mathbf{G})$, and the second Piola-Kirchhoff stress has the following classic representation \citep{Reiner1948,Richter1948,RivlinEricksen1955,Wang1969,Boehler1977}\footnote{See \citep{Graban2019} for a translation and discussion of \citep{Richter1948}.} 
\begin{equation}
	\mathbf{S}=\alpha_0 \mathbf{G}^\sharp+\alpha_1 \mathbf{C}^\sharp+\alpha_2 \mathbf{C}^{2\sharp}\,,
\end{equation}
where $\alpha_i=\alpha_i(I_1,I_2,I_3)$, $i=0,1,2$. Equivalently, one can write
\begin{equation}
	\mathbf{S}=\beta_0 \mathbf{G}^\sharp+\beta_1 \mathbf{C}^\sharp+\beta_{-1} \mathbf{C}^{-\sharp}\,,
\end{equation}
where $\beta_i=\beta_i(I_1,I_2,I_3)$, $i=-1,0,1$. For incompressible isotropic Cauchy solids $I_3=1$, and one has
\begin{equation}
	\mathbf{S}=-p\,\mathbf{C}^{-\sharp}+\beta_0 \mathbf{G}^\sharp+\beta_1 \mathbf{C}^\sharp\,,
\end{equation}
where $p=p(X,t)$ is the Lagrange multiplier associated with the incompressibility constraint $J=\sqrt{I_3}=1$, and $\beta_i=\beta_i(I_1,I_2)$, $i=0,1$.  

In terms of the Cauchy stress, the constitutive equations of compressible and incompressible isotropic Cauchy elastic solids are written as
\begin{equation} \label{Isotropic-Constitutive-Equations-Cauchy}
\begin{aligned}
	\boldsymbol{\sigma} &=\gamma_0\, \mathbf{g}^\sharp+\gamma_1 \mathbf{b}^\sharp
	+\gamma_2\mathbf{c}^\sharp\,, && \gamma_i=\gamma_i(I_1,I_2,I_3)\,,&& i=0,1,2\,, \\
	\boldsymbol{\sigma} &= -p\, \mathbf{g}^\sharp+\gamma_1 \mathbf{b}^\sharp
	+\gamma_2\mathbf{c}^\sharp\,, && \gamma_i=\gamma_i(I_1,I_2)\,,&& i=1,2\,.
\end{aligned}
\end{equation}
In components, they read $\sigma^{ab}=\gamma_0\,g^{ab}+\gamma_1b^{ab}+\gamma_2\,c^{ab}$, and $\sigma^{ab}=-p\,g^{ab}+\gamma_1b^{ab}+\gamma_2\,c^{ab}$, respectively.

\subsubsection{Anisotropic Cauchy elasticity}

For anisotropic solids, the material symmetry group is characterized using a finite collection of \emph{structural tensors} $\boldsymbol{\zeta}_i$, $i=1,\hdots,N$ \citep{liu1982,boehler1987,zheng1993,zheng1994theory,Lu2000,MazzucatoRachele2006}. 
These structural tensors serve as a basis for the space of tensors that remain invariant under the right action of the symmetry group $\mathcal{G}$. 

When structural tensors are used as arguments in a tensor function, e.g., the second Piola-Kirchhoff stress $\mathbf{S}$, the function becomes isotropic with respect to its augmented arguments. This is known as the \emph{principle of isotropy of space} \citep{Boehler1979}. 

Instead of using the set of tensors $\{\mathbf{C}^\flat,\mathbf{G},\boldsymbol{\zeta}_1,\dots,\boldsymbol{\zeta}_N\}$, one can consider a corresponding set of isotropic invariants.
According to a theorem by Hilbert, for any finite collection of tensors, there exists a finite set of isotropic invariants, known as the \emph{integrity basis}, for the set of isotropic invariants of the collection \citep{Spencer1971}. We denote the integrity basis by \(I_j, ~j=1,\dots,m\). Thus, one can express $\mathbf{S}$ as $\mathbf{S} = \mathbf{S}(X, I_1, \dots, I_m)$.

As an example, we consider transversely isotropic Cauchy elastic solids. Other symmetry classes can be treated in a similar manner. In a transversely isotropic solid, there is a single preferred material direction at every point, which is perpendicular to the plane of isotropy at that point. At $X \in\mathcal{B}$, this preferred material direction is identified by a unit vector $\mathbf{N}(X)$.
The second Piola-Kirchhoff stress $\mathbf{S}$ of an inhomogeneous transversely isotropic solid is expressed as $\mathbf{S}=\mathbf{S}(X,\mathbf{G},\mathbf{C}^\flat, \mathbf{A})$, where $\mathbf{A}=\mathbf{N}\otimes\mathbf{N}$ is a structural tensor \citep{DoyleEricksen1956,spencer1982formulation,Lu2000}. Equivalently, the stress depends on five independent invariants $I_1,\hdots,I_5$. The extra invariants $I_4$ and $I_5$ are defined as
\begin{equation}
	I_4=\mathbf{N}\cdot\mathbf{C}\cdot\mathbf{N}=N^AN^B\,C_{AB}\,,\qquad 
	I_5=\mathbf{N}\cdot\mathbf{C}^2\cdot\mathbf{N}=N^AN^B\,C_{BM}\,C^M{}_A\,.
\end{equation}
The second Piola-Kirchhoff stress has the following represetation \citep{Boehler1979}
\begin{equation} 
\begin{aligned}
	\mathbf{S} = \alpha_0\mathbf{G}^\sharp 
	&+\alpha_1 \mathbf{N}\otimes\mathbf{N}
	+ \alpha_2 \mathbf{C}^\sharp+\alpha_3  \mathbf{C}^{2\sharp}
	+\alpha_4 \left[\mathbf{N}\otimes(\mathbf{C}\cdot\mathbf{N})
	+(\mathbf{C}\cdot\mathbf{N})\otimes\mathbf{N} \right] \\
	&  +\alpha_5 \left[\mathbf{N}\otimes(\mathbf{C}^2\cdot\mathbf{N})
	+(\mathbf{C}^2\cdot\mathbf{N})\otimes\mathbf{N} \right]  \,,
\end{aligned}
\end{equation} 
where $\alpha_i=\alpha_i(I_1,\hdots,I_5)$, $i=0,\hdots,5$ are the response functions.

\subsection{Work theorems of elasticity} \label{WorkTheorems}

It has been suggested in the literature that not all thermodynamically-admissible constitutive equations, i.e., those that respect the laws of thermodynamics, are necessarily physically viable. 
In order to exclude material behaviors considered unphysical, additional constitutive assumptions are necessary.
A subset of such constitutive assumptions has been formulated based on the work done by stress along a path in strain space, with a lower bound assumed for this work. 

At a point $X\in\mathcal{B}$ the\textit{ stress power} is defined as $\frac{1}{2}S^{AB}(X,t)\,\dot{C}_{AB}$. The \textit{total stress work} in a subset $\mathcal{U}\subset\mathcal{B}$ over a time interval $[t_1,t_2]$ is found by integrating the stress power in time and space as
\begin{equation}
	W_{\text{s}}(\mathcal{U},[t_1,t_2])
	= \int_{\mathcal{U}}\int_{t_1}^{t_2} \frac{1}{2}S^{AB}(X,t)\,\dot{C}_{AB}(X,t) \,dt \,dV\,.
\end{equation}
Equivalently, we have
\begin{equation}
	W_{\text{s}}(\mathcal{U},[t_1,t_2]) = \int_{\mathcal{U}}
	\int_{\mathbf{C}_1(X)}^{\mathbf{C}_2(X)} \frac{1}{2}S^{AB}(X,t)\,dC_{AB}(X,t) \,dV
	=\int_{\mathcal{U}}\int_{\mathbf{C}_1(X)}^{\mathbf{C}_2(X)} 
	\boldsymbol{\Omega}(\mathbf{C}(X,t)) \,dV\,,
\end{equation}
where $\mathbf{C}_1(X)=\mathbf{C}(X,t_1)$ and $\mathbf{C}_2(X)=\mathbf{C}(X,t_2)$.

The first attempt to use {stress work $1$-form} as a constitutive restriction in nonlinear elasticity is due to \citet{Caprioli1955}. He postulates that for any closed path $\Gamma:I\to \mathbb{S}$, where $I$ is a time interval and $\mathbb{S}$ is the strain space, a material should satisfy
\begin{equation}
	\int_{\Gamma}\boldsymbol{\Omega}\geq 0\,.
\end{equation}
A consequence of this postulate is that the elastic solid must be hyperelastic. 
\citet{Bernstein1960} demonstrated that those hypoelastic materials that satisfy this work postulate are, in fact, hyperelastic.\footnote{For a hypoelastic material, the stress rate is a function of the current stress and strain rate \citep{Truesdell1955}.}
In plasticity, a similar postulate was introduced by \citet{Ilyushin1961}, where a positive net work of stress indicates plastic deformation during a deformation cycle, while zero work indicates that only elastic deformations have occurred.
More detailed discussions of these postulates can be found in \citep{TruesdellNoll2004,Hill1968}.

The total work done on the sub-body $\mathcal{U}\subset\mathcal{B}$ over a time interval $[t_1,t_2]$ is written as
\begin{equation}
\begin{aligned}
	W(\mathcal{U},\Gamma)
	&= \int_{t_1}^{t_2} 	\left[ \int_{\partial\mathcal{U}} \mathbf{T}\cdot \mathbf{V}
	+\int_{\mathcal{U}} \rho_0\mathbf{B}\cdot \mathbf{V}	\right] \,dV \,dt \\
	& = \int_{t_1}^{t_2}  \int_{\mathcal{U}} 
	\left[\left(\operatorname{Div}\mathbf{P}+\rho_0\mathbf{B}\right)\cdot \mathbf{V}
	+ \mathbf{P}:\nabla\mathbf{V}	\right]
	\,dV \,dt \\
	& =  \int_{\mathcal{U}} \int_{t_1}^{t_2}  \left[\rho_0\mathbf{A}\cdot \mathbf{V}
	+ \mathbf{P}:\nabla\mathbf{V}	\right] \,dt \,dV \\
	& =  \int_{\mathcal{U}} \left[\int_{t_1}^{t_2}  
	\frac{1}{2}\left( \frac{1}{2}\rho_0\mathbf{V}\cdot \mathbf{V}\right)\,dt
	+ \int_{\Gamma}\boldsymbol{\Omega} \right]  dV
	\,.
\end{aligned}
\end{equation}
Therefore
\begin{equation} \label{Total-Work}
	W(\mathcal{U},\Gamma)
	=  \int_{\mathcal{U}} \left[ \mathcal{K}_2-\mathcal{K}_1
	+ \int_{\Gamma}\boldsymbol{\Omega} \right]  dV
	\,,
\end{equation}
where $\mathcal{K}=\mathcal{K}(X,t)= \frac{1}{2}\rho_0\mathbf{V}\cdot \mathbf{V}$ is the kinetic energy density, and $\mathcal{K}_i=\mathcal{K}(X,t_i)$, $i=1,2$. 
A \emph{cyclic deformation} is a closed curve in $\mathbb{S}$, i.e., $\mathbf{C}(X,t_1)=\mathbf{C}(X,t_2)$ (without conditions on velocity at the two endpoints). A \emph{cyclic motion} is a cyclic deformation for which $\mathbf{V}(X,t_1)=\mathbf{V}(X,t_2)$. Obviously, in a cyclic motion the total work is equal to the work of stress.

\begin{remark}
Suppose one reverses the orientation of the path $\Gamma$, then
\begin{equation} 
	W(\mathcal{U},-\Gamma)
	=  \int_{\mathcal{U}} \left[ \mathcal{K}_1-\mathcal{K}_2
	+ \int_{-\Gamma}\boldsymbol{\Omega} \right]  dV
	=-W(\mathcal{U},\Gamma)
	\,.
\end{equation}
In other words, if energy is lost during a cyclic deformation, it is regained in the reverse cyclic deformation. In this sense, deformations in Cauchy elasticity are non-dissipative, yet still non-conservative.
It should be noted that in a dissipative system, energy is lost irrespective of the orientation of the cyclic deformation.
\end{remark}

\begin{remark}\label{Cyclic-Deformation-Motion}
Obviously, $\mathbf{V}(X,t_1)=\mathbf{V}(X,t_2)$ is a sufficient condition for $\mathcal{K}(X,t_1)=\mathcal{K}(X,t_2)$. However, it is not necessary. For example, if $\mathbf{V}(X,t_1)=-\mathbf{V}(X,t_2)$, one still has $\mathcal{K}(X,t_1)=\mathcal{K}(X,t_2)$. In other words, for a cyclic deformation, as long as the initial and final kinetic energies are equal, the total work done on the body is equal to the work of stress.
\end{remark}

\begin{remark}
\citet{Truesdell1966} pointed out that $\int_{\mathcal{U}} \int_{\Gamma}\boldsymbol{\Omega} \,dV$ ``is generally not the actual work done"; one would need to consider the change in the kinetic energy as can be seen in \eqref{Total-Work}.
\citet{Truesdell1966} considered a homogeneous deformation of a homogeneous body for which $\operatorname{Div}\mathbf{P}=\mathbf{0}$, and hence, body force must be equal to acceleration, i.e., $\mathbf{B}=\mathbf{A}$. If $\mathbf{B}=\mathbf{0}$, the work of body force vanishes, and hence, $\int_{\mathcal{U}} \int_{\Gamma}\boldsymbol{\Omega} \,dV$ is the total work done on the body. Truesdell then concluded that the acceleration must vanish, which implies that the displacement field is linear in time, and as a result, the deformation gradient is also linear in time. 
For such motions, the velocity is time-independent, and hence, $\mathcal{K}_2=\mathcal{K}_1$. However, the assumption of vanishing body force is unnecessary, and the total work done on a sub-body is still simply given by \eqref{Total-Work}. Assuming that kinetic energies match at the endpoints, we recover the classical result.  
\end{remark}

\subsubsection{Ericksen's work theorem} 

\citet{Ericksen1956} started with the stress-work $1$-form $\boldsymbol{\Omega}=P^{aA}\,g_{ab}\,dF^b{}_A$ and considered the differential equation $P^{aA}\,g_{ab}\,dF^b{}_A=0$, which he recognized to be a Pfaffian equation. He considered a general Cauchy elastic solid and required the dependence of $\mathbf{P}$ on $\mathbf{F}$ be such that: ``...arbitrarily close to each deformation, there are other deformations which cannot be attained in a motion in which the stresses do no work. It seems to us unlikely that relations not satisfying this condition can describe real materials." This postulate is equivalent to $\boldsymbol{\Omega}$ being completely integrable. He then used \citet{Caratheodory1909}'s theorem that implies that there exist functions $\phi$ and $\psi$ such that $\boldsymbol{\Omega}=\phi\,d\psi$ and showed that 
\begin{equation}
	P^{aA}=g^{ab}\,\phi\,\frac{\partial \psi}{\partial F^b{}_B}\,.
\end{equation}
Note that $d\boldsymbol{\Omega}=d\phi\,\wedge d\psi=0$ if and only if $\phi=\phi(\psi)$, which corresponds to hyper-elasticity as \citet{Ericksen1956} explicitly pointed out.
It should be emphasized that Ericksen’s postulate is merely a constitutive assumption, and a material that does not adhere to this postulate is not necessarily in violation of the second law of thermodynamics.  Additionally, it is important to define what constitutes a ``real material", as this concept may evolve over time with the emergence of new applications. Yet, Ericksen identified the most physically plausible class of non-hyperelastic materials and, as a consequence and in his honor, we refer to such Cauchy materials with two potentials as \textit{Ericksen materials} (see the next section).

\subsection{Characterization of the constitutive equations of Cauchy elasticity using exterior calculus: Green, Ericksen, and Edelen elastic solids} \label{Edelen-Darboux-Potentials}

Twenty years after the work of \citet{Ericksen1956}, and apparently being unaware of it, \citet{Edelen1977} used  Darboux's theorem \citep{Darboux1882} of exterior differential systems \citep{Bryant2013} and classified Cauchy elastic response functions.
More specifically, \citet{Edelen1977} pointed out that the stress-work $1$-form $\boldsymbol{\Omega}$ can only assume one of the following six canonical forms:
\begin{empheq}[left={\empheqlbrace }]{align}
	\label{Cauchy-1} 
	\boldsymbol{\Omega} & =  d\psi_1\,, \\
	\label{Cauchy-2} 
	\boldsymbol{\Omega} & =  \phi_1 d\psi_1\,, \\
	\label{Cauchy-3} 
	\boldsymbol{\Omega} & =  \phi_1 d\psi_1+d\psi_2\,, \\
	\label{Cauchy-4} 
	\boldsymbol{\Omega} & =  \phi_1 d\psi_1+\phi_2 d\psi_2\,, \\
	\label{Cauchy-5} 
	\boldsymbol{\Omega} & = \phi_1 d\psi_1+\phi_2 d\psi_2+d\psi_3\,,  \\
	\label{Cauchy-6} 
	\boldsymbol{\Omega} & =  \phi_1 d\psi_1+\phi_2 d\psi_2+\phi_3 d\psi_3
	\,,
\end{empheq}
where $\psi_i=\psi_i(X,\mathbf{C}^\flat,\mathbf{G})$ and $\phi_i=\phi_i(X,\mathbf{C}^\flat,\mathbf{G})$, $i=1,2,3$.
We call these potentials \emph{Edelen-Darboux potentials}.
Note that \eqref{Cauchy-1} corresponds to hyperelasticity and only in this case $d\boldsymbol{\Omega} =  d\circ d\psi_1=0$.
In a hyperelastic body
\begin{equation}
	W_{\text{s}}(\mathcal{U},[t_1,t_2])
	= \int_{\mathcal{U}}\int_{t_1}^{t_2} d\psi_1(\mathbf{C}(X,t)) \,dV
	= \int_{\mathcal{U}} \left[\psi_1(\mathbf{C}(X,t_2))-\psi_1(\mathbf{C}(X,t_1))\right]dV
	\,,
\end{equation}
which is path independent and vanishes in a cyclic deformation---a motion in which $\mathbf{C}(X,t_1)=\mathbf{C}(X,t_2)$ \citep{Sternberg1979}, i.e., a closed path in the space of strains.

We call solids with stress-work $1$-forms given in \eqref{Cauchy-1}-\eqref{Cauchy-6}, \emph{Edelen elastic solids} of types I, II, III, IV, V, and VI, respectively. In particular, Edelen elastic solids of type II, given by \eqref{Cauchy-2}, are the \emph{Ericksen elastic solids}.
Also, Edelen elastic solids of type I, given by \eqref{Cauchy-1}, are the Green elastic (hyperelastic) solids.

\begin{remark}
	It should be emphasized that Edelen-Darboux potentials are not unique. In particular, we note that in \eqref{Cauchy-6}, the stress-work $1$-form $\boldsymbol{\Omega}$ is invariant under the transformations
\begin{equation}
	(\psi_1,\phi_1,\psi_2,\phi_2,\psi_3,\phi_3) \mapsto
	\left(f_1(\psi_1)\,,\frac{\phi_1}{f_1'(\psi_1)}\,,f_2(\psi_2)\,,\frac{\phi_2}{f_2'(\psi_2)}
	\,,f_3(\psi_3)\,,\frac{\phi_3}{f_3'(\psi_3)} \right)
	\,,
\end{equation}
where $f_1$, $f_2$, and $f_3$ are arbitrary differentiable functions with non-vanishing derivatives. This invariance reflects the functional freedom inherent in the choice of Edelen-Darboux potentials and confirms that the physical content of the constitutive equations is encoded in the stress-work $1$-form $\boldsymbol{\Omega}$, rather than in any particular choice of potentials.
\end{remark}

\begin{remark}
In a recent work \citep{YavariGoriely2025CurlForces}, we showed that  Darboux's theorem can  be used in a similar way to decompose physical forces for rigid body mechanics into no more than two or three generalized potentials, in two and three dimensions, respectively.
\end{remark}

\subsubsection{Compressible isotropic Cauchy elasticity}

In the case of isotropic Cauchy elastic solids, the dependence of $\boldsymbol{\Omega}$ on $\mathbf{C}$ is reduced to dependence on the three principal invariants $I_1$, $I_2$, and $I_3$. 
Thus, in this case, $\boldsymbol{\Omega}$  is a $1$-form on a three-dimensional space, which has the following generic representation: 
\begin{equation} 
	\boldsymbol{\Omega}(I_1,I_2,I_3)=f_1(I_1,I_2,I_3)\,dI_1+f_2(I_1,I_2,I_3)\,dI_2+f_3(I_1,I_2,I_3)\,dI_3\,.
\end{equation}
However, Darboux theorem tells us that it takes one of the following three normal forms \citep{Edelen1977}
\begin{empheq}[left={\empheqlbrace }]{align}
	\label{Isotropic-Cauchy-1} 
	\boldsymbol{\Omega}(I_1,I_2,I_3) & =  d\psi_1(I_1,I_2,I_3)\,, \\
	\label{Isotropic-Cauchy-2} 
	\boldsymbol{\Omega}(I_1,I_2,I_3) & =  \phi_1(I_1,I_2,I_3) \,d\psi_1(I_1,I_2,I_3)\,, \\
	\label{Isotropic-Cauchy-3} 
	\boldsymbol{\Omega}(I_1,I_2,I_3) & =  \phi_1(I_1,I_2,I_3) \,d\psi_1(I_1,I_2,I_3)+d\psi_2(I_1,I_2,I_3) 
	\,.
\end{empheq}
The first two cases, correspond to hyperelastic and Ericksen elastic solids, respectively. Note that the integral manifolds of $\boldsymbol{\Omega}=0$ for either a hyperelastic or an Ericksen elastic solid are the surfaces $\psi_1(I_1,I_2,I_3)=\text{constant}$. An isotopic Cauchy elastic solid that is neither Green nor Ericksen is called an \emph{Edelen isotropic elastic solid}.

\begin{remark}
The constitutive equation of a \textit{Cauchy elastic fluid} depends explicitly only on its mass density \citep{TruesdellNoll2004, Gurtin2010} or equivalently on $J = \sqrt{I_3}$ \cite[p.~198]{WangTruesdell1973}. We observe that Cauchy elastic fluids are inherently hyperelastic.
\end{remark}

\paragraph{Constitutive equations in terms of Edelen-Darboux potentials.}
The constitutive equations for Green, Ericksen and Edelen isotropic elastic solids are, respectively, written as
\begin{equation}
\begin{dcases}
	\mathbf{S} = 2\frac{\partial \psi_1}{\partial I_1}\frac{\partial I_1}{\partial \mathbf{C}^\flat}
	+2\frac{\partial \psi_1}{\partial I_2}\frac{\partial I_2}{\partial \mathbf{C}^\flat}
	+2\frac{\partial \psi_1}{\partial I_3}\frac{\partial I_3}{\partial \mathbf{C}^\flat} \,,\\
	\mathbf{S} = 2\phi_1\left[\frac{\partial \psi_1}{\partial I_1}\frac{\partial I_1}{\partial \mathbf{C}^\flat}
	+\frac{\partial \psi_1}{\partial I_2}\frac{\partial I_2}{\partial \mathbf{C}^\flat}
	+\frac{\partial \psi_1}{\partial I_3}\frac{\partial I_3}{\partial \mathbf{C}^\flat} \right]\,,\\
	\mathbf{S} = 2\phi_1\left[\frac{\partial \psi_1}{\partial I_1}\frac{\partial I_1}{\partial \mathbf{C}^\flat}
	+\frac{\partial \psi_1}{\partial I_2}\frac{\partial I_2}{\partial \mathbf{C}^\flat}
	+\frac{\partial \psi_1}{\partial I_3}\frac{\partial I_3}{\partial \mathbf{C}^\flat} \right]
	+2\frac{\partial \psi_2}{\partial I_1}\frac{\partial I_1}{\partial \mathbf{C}^\flat}
	+2\frac{\partial \psi_2}{\partial I_2}\frac{\partial I_2}{\partial \mathbf{C}^\flat}
	+2\frac{\partial \psi_2}{\partial I_3}\frac{\partial I_3}{\partial \mathbf{C}^\flat} \,.
\end{dcases}
\end{equation}
Recall that \citep{MarsdenHughes1983,SadikYavari2024}
\begin{equation}
	\frac{\partial I_1}{\partial\mathbf{C}^\flat}=\mathbf{G}^{\sharp}\,,\qquad
	\frac{\partial I_2}{\partial\mathbf{C}^{\flat}}
	=I_2\,(\mathbf{C}^{-1})^{-\sharp}-I_3\,(\mathbf{C}^{-2})^{\sharp}
	=I_2\,\mathbf{C}^{-\sharp}-I_3\,\mathbf{C}^{-2\sharp}	\,,\qquad 
	\frac{\partial I_3}{\partial\mathbf{C}^{\flat}}=I_3\mathbf{C}^{-\sharp}
	\,.
\end{equation}
Thus
\begin{equation}
\begin{dcases}
	\mathbf{S}= 2\psi_{11}\mathbf{G}^{\sharp}
	+2\left(I_2\,\psi_{12}-I_3 \psi_{13} \right)\mathbf{C}^{-\sharp}
	-2I_3\psi_{12}\,\mathbf{C}^{-2\sharp} \,,\\
	\mathbf{S}= 2\phi_1\,\psi_{11}\mathbf{G}^{\sharp}
	+2\phi_1\left(I_2 \,\psi_{12}-I_3 \,\psi_{13}\right)\mathbf{C}^{-\sharp} 	
	-2I_3\,\phi_1\,\psi_{12}\,\mathbf{C}^{-2\sharp} \,,\\
	\mathbf{S}= 2(\phi_1\,\psi_{11}+\psi_{21})\mathbf{G}^{\sharp}
	+2\left[I_2(\phi_1\,\psi_{12}+\psi_{22})-I_3(\phi_1\,\psi_{13}+\psi_{23}) \right]\mathbf{C}^{-\sharp} 
	-2I_3(\phi_1\,\psi_{12}+\psi_{22})\mathbf{C}^{-2\sharp}
	\,,
\end{dcases}
\end{equation}
where 
\begin{equation}
	\psi_{ij}=\frac{\psi_i}{\partial I_j}\,,\quad i=1,2\,,~j=1,2,3 \,.
\end{equation}
In particular, note that $\psi_{ij}\neq \psi_{ji}$.
In terms of the Cauchy stress, the constitutive equations read
\begin{equation}
\begin{dcases}
	\boldsymbol{\sigma} = \frac{2}{\sqrt{I_3}}\left[ \psi_{11}\mathbf{b}^{\sharp}
	+\left(I_2\,\psi_{12}-I_3 \psi_{13} \right)\mathbf{g}^{\sharp}
	-I_3\psi_{12}\,\mathbf{c}^{\sharp} \right] \,,\\
	\boldsymbol{\sigma} = \frac{2}{\sqrt{I_3}}\left[ \phi_1\,\psi_{11}\mathbf{b}^{\sharp}
	+\phi_1\left(I_2 \,\psi_{12}-I_3 \,\psi_{13}\right)\mathbf{g}^{\sharp} 	
	-I_3\,\phi_1\,\psi_{12}\,\mathbf{c}^{\sharp} \right] \,,\\
	\boldsymbol{\sigma} = \frac{2}{\sqrt{I_3}}\left[ (\phi_1\,\psi_{11}+\psi_{21})\mathbf{b}^{\sharp}
	+\left[I_2(\phi_1\,\psi_{12}+\psi_{22})-I_3(\phi_1\,\psi_{13}+\psi_{23}) \right]\mathbf{g}^{\sharp} 
	-I_3(\phi_1\,\psi_{12}+\psi_{22})\mathbf{c}^{\sharp} \right]
	\,.
\end{dcases}
\end{equation}

\paragraph{Cyclic deformations.}
A cyclic deformation of an isotropic Cauchy elastic solid is a closed path in the three-dimensional space $\mathcal{I}=\{(I_1,I_2,I_3)|I_1,I_2,I_3>0\}\subset \mathbb{R}^3$. \citet{Edelen1977} defined\textit{ a path of zero work} $\Gamma$ in $\mathcal{I}$ to be a curve for which $\int_{\Gamma}\boldsymbol{\Omega}=0$. Let us consider a point $(\mathring{I}_1,\mathring{I}_2,\mathring{I}_3)\in\mathcal{I}$. For each of the classes \eqref{Isotropic-Cauchy-1}-\eqref{Isotropic-Cauchy-3}, \citet{Edelen1977} characterized all those points in $\mathcal{I}$ that can be connected to $(\mathring{I}_1,\mathring{I}_2,\mathring{I}_3)\in\mathcal{I}$ by a path of zero work. In the case of a hyperelastic solid \eqref{Isotropic-Cauchy-1}
\begin{equation}
	0=\int_{\Gamma}\boldsymbol{\Omega}=\int_{\Gamma}d\psi_1(I_1,I_2,I_3)
	=\psi_1(I_1,I_2,I_3)-\psi_1(\mathring{I}_1,\mathring{I}_2,\mathring{I}_3)
	\,,
\end{equation}
and hence such points satisfy the following relation
\begin{equation} \label{Accessibility-Hyperelastic}
	\psi_1(I_1,I_2,I_3)=\psi_1(\mathring{I}_1,\mathring{I}_2,\mathring{I}_3)
	\,,
\end{equation}
which is the equation of a surface. This means that there are points close to $(\mathring{I}_1,\mathring{I}_2,\mathring{I}_3)$ that cannot be connected to it via a path of zero work---the Carath\'eodory inaccessibility condition for $\boldsymbol{\Omega}$ \citep{Caratheodory1909}. Ericksen solids have the inaccessibility property while Edelen solids do not.

\paragraph{Paths of zero stress-work  for an isotropic Edelen solid.}
We consider a compressible isotropic Cauchy elastic solid with the stress-work $1$-form \begin{align}
    \boldsymbol{\Omega}(I_1,I_2,I_3)  =  \phi(I_1,I_2,I_3) \,d\psi(I_1,I_2,I_3)+d\chi(I_1,I_2,I_3),\quad d\phi\wedge d\psi\neq 0,
\end{align} i.e., an Edelen isotropic solid. We assume that a homogenous body made of this material is in a given state of deformation denoted by $x$. We assume that $x$ is the origin with coordinates $(0,0,0)$ in the $(\phi,\psi,\chi)$-space. Let us consider an arbitrary neighboring deformed state $y$ with coordinates $(\phi_0,\psi_0,\chi_0)$. We follow \citet{Bryant2013} and construct a path $\Gamma$ that connects $x$ to $y$ such that $\boldsymbol{\Omega} = 0$ along the entire path.
In the $(\phi,\psi)$-plane let $\mathcal{c}$ be a curve $(\phi(t),\psi(t)),~t\in [0,1]$ such that $(\phi(0),\psi(0))=(0,0)$, and $(\phi(1),\psi(1))=(\phi_0,\psi_0)$ (obviously, there are infinitely many such curves). Let us define
\begin{equation} 
	\chi(t) = -\int_0^t \phi(\tau) \,d\psi(\tau) \,.
\end{equation}
Note that $-\int_0^1 \phi(\tau) \,d\psi(\tau)$ is the signed area of the region bounded by $\mathcal{c}$ and the cord that joins $(0,0)$ and $(\phi_0,\psi_0)$. One can choose a curve $\mathcal{c}:~(\phi(t),\psi(t)),~t\in [0,1]$  for which this signed area is exactly equal to $\chi_0$. 
The desired zero-stress-work path $\Gamma$ that connects $x$ and $y$ is defined as
\begin{equation} 
	\Gamma:~ (\phi(t),\psi(t),\chi(t))\,,\quad t\in [0,1]
	\,.
\end{equation}

\subsubsection{Incompressible isotropic Cauchy elasticity}\label{Sec:Incomp-Cauchy}

In incompressible isotropic Cauchy elasticity, $I_3=1$, and hence, the dependence of $\boldsymbol{\Omega}$ on $\mathbf{C}$ is reduced to a dependence on the two principal invariants $I_1$ and $I_2$. Notice that the (constitutively indeterminate) pressure field $p$ does not contribute to stress work, and hence, the stress-work $1$-form is not affected by the pressure field. 
Thus, $\boldsymbol{\Omega}$  is a $1$-form on a two-dimensional space, which has the following generic representation:
\begin{equation} 
	\boldsymbol{\Omega}(I_1,I_2)=f_1(I_1,I_2)\,dI_1+f_2(I_1,I_2)\,dI_2	\,.
\end{equation}
However, Darboux theorem tells us that it takes one of the following two normal forms
\begin{empheq}[left={\empheqlbrace }]{align}
	\label{Incomp-Isotropic-Cauchy-1} 
	\boldsymbol{\Omega}(I_1,I_2) & =  d\chi(I_1,I_2)\,, \\
	\label{Incomp-Isotropic-Cauchy-2} 
	\boldsymbol{\Omega}(I_1,I_2) & =  \phi(I_1,I_2) \,d\psi(I_1,I_2)  
	\,.
\end{empheq}
This, in particular, implies that an incompressible isotropic Cauchy elastic solid is either hyperelastic or Ericksen elastic. 
We also observe that all incompressible isotropic Cauchy elastic solids have the stress-work inaccessibility property.

\begin{remark}
Another interesting observation is that incompressible isotropic Cauchy elastic solids with constitutive equations depending only on either $I_1$ or $I_2$ are inherently hyperelastic.
\end{remark}

\paragraph{Constitutive equations in terms of Edelen-Darboux potentials.}
The constitutive equations for incompressible isotropic Green and Ericksen elastic solids are, respectively, written as
\begin{equation}
\begin{dcases}
	\mathbf{S} = -p\,\mathbf{C}^{-\sharp}
	+2\frac{\partial \chi}{\partial I_1}\frac{\partial I_1}{\partial \mathbf{C}^\flat}
	+2\frac{\partial \chi}{\partial I_2}\frac{\partial I_2}{\partial \mathbf{C}^\flat}
	\,,\\
	\mathbf{S} = -p\,\mathbf{C}^{-\sharp}
	+2\phi\left[\frac{\partial \psi}{\partial I_1}\frac{\partial I_1}{\partial \mathbf{C}^\flat}
	+\frac{\partial \psi}{\partial I_2}\frac{\partial I_2}{\partial \mathbf{C}^\flat}  \right]\,.
\end{dcases}
\end{equation}
Using the relations ($I_3=1$)
\begin{equation}
	\frac{\partial I_1}{\partial\mathbf{C}^\flat}=\mathbf{G}^{\sharp}\,,\qquad
	\frac{\partial I_2}{\partial\mathbf{C}^{\flat}}
	=I_2\,\mathbf{C}^{-\sharp}-\mathbf{C}^{-2\sharp}	\,,
\end{equation}
in the above constitutive equations, one obtains (note that as $p$ is an indeterminate scalar field at this stage one can replace $-p+I_2\,\psi_{12}$ by $-p$)
\begin{equation}
\begin{dcases}
	\mathbf{S} = -p\,\mathbf{C}^{-\sharp}
	+2\chi_{1}\,\mathbf{G}^{\sharp}
	-2\chi_{2}\,\mathbf{C}^{-2\sharp}
	\,,\\
	\mathbf{S} = -p\,\mathbf{C}^{-\sharp}
	+2\phi\, \psi_{1}\,\mathbf{G}^\sharp
	-2\phi \,\psi_{2} \mathbf{C}^{-2\sharp}\,.
\end{dcases}
\end{equation}
In terms of Cauchy stress these read
\begin{equation}
\begin{dcases}
	\boldsymbol{\sigma} = -p\,\mathbf{g}^{\sharp}
	+2\chi_{1}\,\mathbf{b}^{\sharp}
	-2\chi_{2}\,\mathbf{c}^{\sharp}
	\,,\\
	\boldsymbol{\sigma} = -p\,\mathbf{g}^{\sharp}
	+2\phi\, \psi_{1}\,\mathbf{b}^\sharp
	-2\phi\, \psi_{2}\, \mathbf{c}^{\sharp}\,.
\end{dcases}
\end{equation}

\subsubsection{$2$D compressible and incompressible isotropic Cauchy elasticity}

In compressible two-dimensional isotropic Cauchy elasticity, the dependence of the stress-work $1$-form on strain is through the two invariants $I_1=\operatorname{tr}\mathbf{C}$ and $I_2=\det\mathbf{C}$.
From \eqref{Darboux-2D}, we see that the canonical forms of the stress-work $1$-form are $\boldsymbol{\Omega}=d\psi_1$ and $\boldsymbol{\Omega}=\phi^1 d\psi_1$. In incompressible $2$D Cauchy elasticity $I_2=1$, and hence $\boldsymbol{\Omega}$ depends on strain via $I_1$ (the Lagrange multiplier $p$ does not contribute to the stress-work $1$-form). This implies that $\boldsymbol{\Omega}=d\psi_1$, i.e., incompressible $2D$ Cauchy elastic solids are hyperelastic.

\subsubsection{Compressible and incompressible anisotropic Cauchy elastic solids}

As an example of anisotropic materials, we consider a transversely isotropic Cauchy elastic solid.
Thus, $\mathbf{S} = \mathbf{S}(X, I_1,I_2,I_3,I_4,I_5)$. This implies that the dependence of the stress-work $1$-form $\boldsymbol{\Omega}$ on $\mathbf{C}$ is reduced to a dependence on the five invariants. Hence, $\boldsymbol{\Omega}$ is a $1$-form defined on a five-dimensional manifold.
On a five-dimensional manifold, the rank of a $1$-form can take any of the values $0,1,2$. Thus, one has the following possibilities (note that $d\boldsymbol{\Omega}\wedge d\boldsymbol{\Omega}\wedge d\boldsymbol{\Omega} $ is a $6$-form and identically vanishes on any $5$-manifold):
\begin{equation}
\begin{aligned}
    k=0:& \quad d\boldsymbol{\Omega}=0\,,~ \boldsymbol{\Omega}\wedge 	
    (d\boldsymbol{\Omega})^0=\boldsymbol{\Omega}\neq 0 ~ \Rightarrow ~ \boldsymbol{\Omega}=d\psi_1\,, \\
    k=1:& \quad d\boldsymbol{\Omega}\wedge d\boldsymbol{\Omega}=0, ~ 
    \begin{dcases}
    \boldsymbol{\Omega}\wedge d\boldsymbol{\Omega} =0    ~ \Rightarrow ~ \boldsymbol{\Omega}
    =\phi_1 d\psi_1\,, \\
    \boldsymbol{\Omega}\wedge d\boldsymbol{\Omega} \neq 0   ~ \Rightarrow ~ 
    \boldsymbol{\Omega}=\phi_1 d\psi_1+d\psi_2\,,
    \end{dcases}\\
    k=2:& \quad d\boldsymbol{\Omega}\wedge d\boldsymbol{\Omega}\wedge d\boldsymbol{\Omega}=0, ~ 
    \begin{dcases}
    \boldsymbol{\Omega}\wedge d\boldsymbol{\Omega}\wedge d\boldsymbol{\Omega} =0    
    ~ \Rightarrow ~ \boldsymbol{\Omega}=\phi_1 d\psi_1+\phi_2 d\psi_2\,,\\
    \boldsymbol{\Omega}\wedge d\boldsymbol{\Omega}\wedge d\boldsymbol{\Omega} \neq 0    
    ~ \Rightarrow ~ \boldsymbol{\Omega}=\phi_1 d\psi_1+\phi_2 d\psi_2+d\psi_3\,,
    \end{dcases}
\end{aligned}
\end{equation}
where $\phi_i=\phi_i(I_1,\hdots,I_5)$, $i=1,2$ and $\psi_j=\psi_j(I_1,\hdots,I_5)$, $j=1,2,3$. Therefore, a compressible transversely isotropic Cauchy elastic solid has at most five generalized energy functions.

For incompressible solids, $I_3=1$, and hence \citep{MarsdenHughes1983,SadikYavari2024}
\begin{equation}
	dI_3=\frac{\partial I_3}{\partial \mathbf{C}^\flat}\!:\!d\mathbf{C}^\flat
	=I_3\mathbf{C}^{-\sharp}\!:\!d\mathbf{C}^\flat=0	\,.
\end{equation}
Therefore, for incompressible (anisotropic) solids we have the constraint $\mathbf{C}^{-\sharp}\!:\!d\mathbf{C}^\flat=C^{-AB}\,dC_{AB}=0$. 
Note that stress has a reactive part and a constitutive part: $\mathbf{S}=-p \mathbf{C}^{-\sharp}+\hat{\mathbf{S}}(\mathbf{C},\mathbf{G})=-p \mathbf{C}^{-\sharp}+\hat{\mathbf{S}}(I_1,I_2,I_4,I_5)$. It is also noted that only the constitutive part of stress contributes to stress work $1$-form. Therefore, $\boldsymbol{\Omega}$ is a $1$-form defined on a four-dimensional manifold.
On a four-dimensional manifold, the rank of a $1$-form can take any of the values $0,1,2$. Thus, one has the following possibilities (note that $\boldsymbol{\Omega}\wedge d\boldsymbol{\Omega}\wedge d\boldsymbol{\Omega} $ is a $5$-form and identically vanishes on any $4$-manifold):
\begin{equation}
\begin{aligned}
    k=0:& \quad d\boldsymbol{\Omega}=0\,,~ \boldsymbol{\Omega}\wedge 
    (d\boldsymbol{\Omega})^0=\boldsymbol{\Omega}\neq 0 ~ \Rightarrow ~ \boldsymbol{\Omega}=d\psi_1\,, \\
    k=1:& \quad d\boldsymbol{\Omega}\wedge d\boldsymbol{\Omega}=0, ~ 
    \begin{dcases}
    \boldsymbol{\Omega}\wedge d\boldsymbol{\Omega} =0    ~ \Rightarrow ~ \boldsymbol{\Omega}
    =\phi_1 d\psi_1\,, \\
    \boldsymbol{\Omega}\wedge d\boldsymbol{\Omega} \neq 0    ~ \Rightarrow ~ \boldsymbol{\Omega}
    =\phi_1 d\psi_1+d\psi_2\,,
    \end{dcases}\\
    k=2:& \quad d\boldsymbol{\Omega}\wedge d\boldsymbol{\Omega}\wedge d\boldsymbol{\Omega}=0, ~ 
    \boldsymbol{\Omega}\wedge d\boldsymbol{\Omega}\wedge d\boldsymbol{\Omega} =0   
     ~ \Rightarrow ~ \boldsymbol{\Omega}=\phi_1 d\psi_1+\phi_2 d\psi_2\,,
\end{aligned}
\end{equation}
where $\phi_i=\phi_i(I_1,I_2,I_4,I_5)$ and $\psi_i=\psi_i(I_1,I_2,I_4,I_5)$, $i=1,2$. Therefore, an incompressible transversely isotropic Cauchy elastic solid has at most four generalized energy functions.

\subsubsection{The natural additive decomposition of stress into conservative and nonconservative parts} \label{Sec:AdditiveDecomposition}

It should be noted that Darboux's classification of stress-work $1$-form implies that stress can always be additively decomposed into a conservative part (corresponding to $d\psi_1$, $d\psi_2$, and $d\psi_3$ in \eqref{Cauchy-1}, \eqref{Cauchy-3}, and \eqref{Cauchy-5}, respectively) and a non-conservative part, i.e., 
\begin{equation} \label{Stress-c-nc}
	\mathbf{P}=\mathbf{P}_{\text{c}}+\mathbf{P}_{\text{nc}} \,.
\end{equation}
Let us consider a Cauchy elastic solid in equilibrium in the absence of body forces, i.e., $\operatorname{Div}\mathbf{P}=\mathbf{0}$, or $\operatorname{Div}(\mathbf{P}_{\text{c}}+\mathbf{P}_{\text{nc}})=\mathbf{0}$. This can be rewritten as $\operatorname{Div}\mathbf{P}_{\text{c}}+\rho_0\mathbf{B}_{\text{a}}=\mathbf{0}$, where $\rho_0\mathbf{B}_{\text{a}}=\operatorname{Div}\mathbf{P}_{\text{nc}}$ can be thought of as a body force.
In other words, a non-hyperelastic Cauchy elastic system can be viewed as an underlying hyperelastic system subjected to a particular type of body force that depends on the divergence of a function of the deformation gradient. The non-zero work of stress in a cyclic deformations is written as
\begin{equation}
	W_{\text{s}}(\mathcal{U},[t_1,t_2])
	= \int_{t_1}^{t_2} \int_{\partial\mathcal{U}} \mathbf{T}_{\text{nc}}\cdot \mathbf{V}\,dA \,dt
	-\int_{t_1}^{t_2} \int_{\mathcal{U}} \rho_0\mathbf{B}_{\text{a}}\cdot \mathbf{V}\,dV \,dt\,,
\end{equation}
where $\mathbf{T}_{\text{nc}}=\mathbf{P}_{\text{nc}}\cdot\mathbf{N}$ is the traction corresponding to the non-conservative stress.

\begin{remark}
We note that the decomposition \eqref{Stress-c-nc} is not unique. Adding a conservative stress to a non-conservative stress gives a stress that remains non-conservative. Two decompositions, $\mathbf{P} = \tilde{\mathbf{P}}_{\text{c}} + \tilde{\mathbf{P}}_{\text{nc}} = \mathbf{P}_{\text{c}} + \mathbf{P}_{\text{nc}}$, are equivalent if $\tilde{\mathbf{P}}_{\text{nc}} - \mathbf{P}_{\text{nc}} = -(\tilde{\mathbf{P}}_{\text{c}} - \mathbf{P}_{\text{c}})$ is a conservative stress.
\end{remark}

\subsubsection{A comparison of Cauchy elastic response and pseudoelasticity}

\citet{Fung1980} referred to the response of arteries as \emph{pseudo-elastic} due to the presence of hysteresis during cyclic deformations.
Similarly, in shape memory alloys, the term \emph{pseudoelasticity} is used to describe the fact that  the material returns to its original shape after unloading. At first sight, it appears that pseudo-elastic materials have some features of Cauchy elastic material. However, the loading and unloading responses differ, resulting in a hysteresis loop, indicating that the behavior is always dissipative \citep{HuoMuller1993}.
For modeling Mullins effect in filled rubber \citep{Mullins1948,Mullins1969}, \citet{Ogden1999} assumed that the material has a \emph{pseudo-energy function} $W=W(\mathbf{F},\mathbf{g},\mathbf{G},\eta)=\hat{W}(\mathbf{C}^\flat,\mathbf{G},\eta)$, where $0\leq \eta \leq 1$ is an internal variable. In loading this internal variable is inactive, i.e., it takes a constant value ($d\eta=0$), while in unloading it is activated and has the following implicit relationship with $\mathbf{F}$:
\begin{equation}
	\frac{\partial W}{\partial \eta}(\mathbf{F},\eta)=0	\,,
\end{equation}
and hence, one may write $\eta=\bar{\eta}(\mathbf{F})$.
The first Piola-Kirchhoff stress is still written as $\mathbf{P}=\mathbf{g}^\sharp\frac{\partial W}{\partial\mathbf{F}}$.
For such materials, the stress-work $1$-form is written as
\begin{equation}
	\boldsymbol{\Omega}=\mathbf{P}\!:\!d\mathbf{F}=dW-\xi \,d\eta\,,
\end{equation}
where $\xi=\frac{\partial W}{\partial \eta}$. 
Note that on a loading path $\eta=\eta_0$, i.e., $d\eta=0$, and hence, $\boldsymbol{\Omega}=dW(\mathbf{F},\mathbf{g},\mathbf{G},\eta_0)$. On an unloading path $\xi(\mathbf{F},\mathbf{g},\mathbf{G},\bar{\eta}(\mathbf{F}))=0$, and hence, $\boldsymbol{\Omega}=dW(\mathbf{F},\mathbf{g},\mathbf{G},\bar{\eta}(\mathbf{F}))$. Let us consider a closed path $\Gamma=\Gamma^+\cup \Gamma^-$ in the strain space, where $\Gamma^+$ and $\Gamma^-$ are the loading and unloading paths, respectively. Let us assume that deformation gradient has the values $\mathbf{F}_1$ and $\mathbf{F}_2$ at the starting and the end points of $\Gamma^+$, respectively. On the loading path the internal variable has the constant value $\eta_0$, and hence
\begin{equation}
	\int_{\Gamma^+}\boldsymbol{\Omega} = \int_{\Gamma^+} dW=
	W(\mathbf{F}_2,\mathbf{g},\mathbf{G},\eta_0)-W(\mathbf{F}_1,\mathbf{g},\mathbf{G},\eta_0)
	\,.
\end{equation}
On the unloading path 
\begin{equation}
	\int_{\Gamma^-}\boldsymbol{\Omega} = \int_{\Gamma^-} dW=
	W(\mathbf{F}_1,\mathbf{g},\mathbf{G},\bar{\eta}(\mathbf{F}_1))
	-W(\mathbf{F}_2,\mathbf{g},\mathbf{G},\bar{\eta}(\mathbf{F}_2))
	\,.
\end{equation}
Therefore, on the closed path
\begin{equation}
	\int_{\Gamma}\boldsymbol{\Omega} =
	W(\mathbf{F}_2,\mathbf{g},\mathbf{G},\eta_0)-W(\mathbf{F}_1,\mathbf{g},\mathbf{G},\eta_0)
	+W(\mathbf{F}_1,\mathbf{g},\mathbf{G},\bar{\eta}(\mathbf{F}_1))
	-W(\mathbf{F}_2,\mathbf{g},\mathbf{G},\bar{\eta}(\mathbf{F}_2))
	\,,
\end{equation}
which is generally nonzero. Now let us consider the reverse closed path $-\Gamma=(-\Gamma^-)\cup (-\Gamma^+)$, where this time $-\Gamma^-$ is the loading path while $-\Gamma^+$ is the unloading path. Note that the loading path still starts from $\mathbf{F}_1$, and hence, on the new loading path one still has $\eta=\eta_0$. Thus, on the loading path
\begin{equation}
	\int_{-\Gamma^-}\boldsymbol{\Omega} = \int_{-\Gamma^-} dW=
	W(\mathbf{F}_2,\mathbf{g},\mathbf{G},\eta_0) - W(\mathbf{F}_1,\mathbf{g},\mathbf{G},\eta_0)
	\,.
\end{equation}
Similarly, on the unloading path
\begin{equation}
	\int_{-\Gamma^+}\boldsymbol{\Omega} = \int_{-\Gamma^+} dW=
	W(\mathbf{F}_1,\mathbf{g},\mathbf{G},\bar{\eta}(\mathbf{F}_1))
	-W(\mathbf{F}_2,\mathbf{g},\mathbf{G},\bar{\eta}(\mathbf{F}_2))
	\,.
\end{equation}
Therefore,
\begin{equation}
\begin{aligned}
	\int_{-\Gamma}\boldsymbol{\Omega}  &=
	W(\mathbf{F}_2,\mathbf{g},\mathbf{G},\eta_0) - W(\mathbf{F}_1,\mathbf{g},\mathbf{G},\eta_0)
	+W(\mathbf{F}_1,\mathbf{g},\mathbf{G},\bar{\eta}(\mathbf{F}_1))
	-W(\mathbf{F}_2,\mathbf{g},\mathbf{G},\bar{\eta}(\mathbf{F}_2)) \\
	& =\int_{\Gamma}\boldsymbol{\Omega}
	\,,
\end{aligned}
\end{equation}
which is a characteristic of dissipative systems.

\subsubsection{The necessary and sufficient conditions for a Cauchy elastic solid to be hyperelastic}

Given the constitutive equations of a Cauchy elastic solid, the stress-work $1$-form can be used to determine the existence of a corresponding energy function.
In terms of the second Piola-Kirchhoff stress one can write
\begin{equation}
    d\boldsymbol{\Omega}=\frac{1}{2} dS^{AB}\wedge dC_{AB}
    =\frac{1}{2} \frac{\partial S^{AB}}{\partial C_{CD}}\,dC_{CD}\wedge dC_{AB}\,.
\end{equation}
Thus, $d\boldsymbol{\Omega}=0$ if and only if
\begin{equation}\label{Hyperelastic-Potential}
    \frac{\partial S^{AB}}{\partial C_{CD}}=\frac{\partial S^{CD}}{\partial C_{AB}}\,.
\end{equation}
These are the necessary and sufficient conditions for the existence of a hyperelastic potential.\footnote{This assertion presumes that the stress space is simply connected. In the absence of simple connectedness, the condition remains necessary but is no longer sufficient. In such cases, sufficiency requires the satisfaction of additional integrability conditions, which are contingent on the topological properties---specifically, the de Rham cohomology---of the underlying space \citep{Yavari2013}.}
In terms of the first Piola-Kirchhoff stress
\begin{equation} 
	d\boldsymbol{\Omega} = dP_a{}^{A} \wedge dF^a{}_{A}
	= \frac{\partial P_a{}^{A}}{\partial F^b{}_{B}} \,dF^b{}_{B} \wedge dF^a{}_{A}
	\,.
\end{equation}
Hence, $d\boldsymbol{\Omega}=0$ if and only if
\begin{equation} 
	\frac{\partial P_a{}^{A}}{\partial F^b{}_{B}} = \frac{\partial P_b{}^{B}}{\partial F^a{}_{A}}
	\,.
\end{equation}
In terms of the Kirchhoff stress, the necessary and sufficient conditions for hyperelasticity are the following symmetry conditions:
\begin{equation} \label{Hyperelastic-Potential-Kirchhoff}
	\frac{\partial \tau^{ab}}{\partial h_{cd}} = \frac{\partial \tau^{cd}}{\partial h_{ab}}
	\,,
\end{equation}
where $h_{ab}$ are components of $\mathbf{h}^\flat=\mathbf{g}\mathbf{h}$.

\begin{remark}
There is a connection between Cauchy elasticity and symplectic geometry \citep{Cardin1991,Cardin1995}.\footnote{There is also a connection with contact geometry \citep{Mrugala1978,Geiges2001}  that we will not discuss it here.}
One can think of the stress-work $1$-form \eqref{Liouville-1-form} as a Liouville-$1$-form. Thus
\begin{equation} \label{Symplectic-form-F}
	\boldsymbol{\Theta}=d\boldsymbol{\Omega}(\mathbf{F}(X,t),\mathbf{G}(X,t),\mathbf{g}(\varphi(X,t)))
	=dP_a{}^{A}(X,t)\wedge dF^a{}_{A}(X,t)\,,
\end{equation}
is a symplectic $2$-form on the cotangent bundle of deformation gradients, i.e., $T^*(\Omega^{1}_{\varphi}(\mathcal{B};T\mathcal{S}))$. Recall that deformation gradient can be thought of as a covector-valued $1$-form \citep{Kanso2007,Yavari2008,Angoshtari2015}.
One can instead use the right Cauchy-Green strain, and in this case
\begin{equation} \label{Symplectic-form-C}
	\boldsymbol{\Theta}(X,t)=d\boldsymbol{\Omega}(\mathbf{C}^\flat(X,t),\mathbf{G}(X,t))
	=\frac{1}{2}dS^{AB}(X,t)\wedge dC_{AB}(X,t)\,,
\end{equation}
is a symplectic $2$-form on the cotangent bundle of right Cauchy-Green strains, i.e., $T^*(S^{2}T^{\ast}\mathcal{B})$.
This makes $(T^*\mathcal{B},\boldsymbol{\Theta})$ a symplectic manifold.
Hyperelasticity corresponds to Lagrangian submanifolds on which the restriction of the symplectic form vanishes.
\end{remark}

\subsubsection{Examples of non-hyperelastic Cauchy elastic solids} \label{Examples-NonHyperelastic}

In this section, we examine the two constitutive equations mentioned in Footnote \ref{Becker-Hencky} and demonstrate that they are non-hyperelastic models.

\vskip 0.1 in \noindent
\textbf{Example 1: Hencky's model.} \citet{Hencky1928} considered the following constitutive equation \citep[Footnote~17]{Neff2014}. 
\begin{equation}
	\boldsymbol{\sigma}= 2 \mu \log \mathbf{V}^\sharp
	+\lambda [\operatorname{tr}(\log \mathbf{V}^\sharp)]\, \mathbf{g}^\sharp
	\,.
\end{equation}
In terms of Kirchhoff stress this is written as $\boldsymbol{\tau} = 2J \mu \log \mathbf{V}^\sharp+J\lambda [\operatorname{tr}(\log \mathbf{V}^\sharp)] \mathbf{g}^\sharp$. Note that $\operatorname{tr}(\log \mathbf{V}^\sharp)=\log \lambda_1+\log \lambda_2+\log \lambda_3=\log (\lambda_1\lambda_2\lambda_3)=\log J$. Thus,
\begin{equation}
    \boldsymbol{\tau} = 2\mu J \,\mathbf{h}^\sharp+\lambda J\log J \,\mathbf{g}^\sharp
	\,.
\end{equation}
Hence
\begin{equation}
\begin{aligned}
	\frac{\tau^{ab}}{\partial h_{cd}} & = 
    2\mu \frac{\partial J}{\partial h_{cd}} h^{ab}
    +2\mu J \frac{\partial h^{ab}}{\partial h_{cd}}+\lambda (1+\log J)\frac{\partial J}{\partial h_{cd}} g^{ab}\\
    & = 
    2\mu \frac{\partial J}{\partial h_{cd}} h^{ab}
    +\mu J \left(g^{ac}g^{bd}+g^{ad}g^{bc}\right)
    +\lambda (1+\log J)\frac{\partial J}{\partial h_{cd}} g^{ab}
    \,.
\end{aligned}
\end{equation}
The derivative of $J$ with respect to $\mathbf{h}$ is calculated as follows. Note that $J=\det\mathbf{V}=\det e^{\mathbf{h}}$ and recall the following two identities
\begin{equation}
	\det (\mathbf{A}+\epsilon \mathbf{B}) 
	= \det\mathbf{A}+ \epsilon\, (\det\mathbf{A})\, \mathbf{A}^{-1}\!:\! \mathbf{B} +o(\epsilon)\,,\qquad
	e^{\mathbf{A}+\epsilon \mathbf{B}}=e^{\mathbf{A}} +\epsilon\, \mathbf{B} \,e^{\mathbf{A}}
	+o(\epsilon)
	\,.
\end{equation}
Therefore
\begin{equation}
\begin{aligned}
	\det e^{\mathbf{h}+\epsilon \mathbf{k}} 
	& = \det \left(e^{\mathbf{h}} +\epsilon\, \mathbf{k} \,e^{\mathbf{h}}	+o(\epsilon)\right) \\
	& =\det e^{\mathbf{h}}+ \epsilon\,(\det e^{\mathbf{h}})(e^{\mathbf{h}})^{-1}\!:\!\mathbf{k} \,e^{\mathbf{h}}
	+o(\epsilon) \\
	& =\det e^{\mathbf{h}}+ \epsilon\,(\det e^{\mathbf{h}}) \,e^{-\mathbf{h}}\!:\!\mathbf{k} \,e^{\mathbf{h}}
	+o(\epsilon)
	\,.
\end{aligned}
\end{equation}
This means that
\begin{equation}
\begin{aligned}
	\frac{\partial J}{\partial \mathbf{h}}\!:\!\mathbf{k}=
	(D\det e^{\mathbf{h}}) [\mathbf{k}]
	=(\det e^{\mathbf{h}}) \,e^{-\mathbf{h}}\!:\!\mathbf{k} \,e^{\mathbf{h}}
	= J\, \mathbf{V}^{-1}\!:\!\mathbf{k} \,\mathbf{V}
	\,.
\end{aligned}
\end{equation}
In components this is written as
\begin{equation}
	\frac{\partial J}{\partial h_{ab}} k_{ab} = J V^{-ab} k_{am}V^m{}_b= J g^{am} k_{am}
	= J g^{ab} k_{ab} \,,
\end{equation}
which implies that
\begin{equation}
	\frac{\partial J}{\partial h_{ab}}= J g^{ab}
	\,.
\end{equation}
Therefore
\begin{equation}
	\frac{\partial \tau^{ab}}{\partial h_{cd}}  =  2\mu J g^{cd} h^{ab} +\mu J \left(g^{ac}g^{bd}+g^{ad}g^{bc}\right)
	+\lambda J(1+\log J) g^{cd} g^{ab}    \,.
\end{equation}
Note that
\begin{equation}
	\frac{\tau^{ab}}{\partial h_{cd}} - \frac{\tau^{cd}}{\partial h_{ab}}
	= 2\mu J (g^{cd} h^{ab}-h^{cd} g^{ab}) \neq 0
	\,,
\end{equation}
which indicates that this constitutive equation is not hyperelastic, see also \citep[Footnote~17]{Neff2014}.
A similar constitutive equation is the following \citep{NeffEidelMartin2016}
\begin{equation}
    \boldsymbol{\tau} = 2\mu \,\mathbf{h}^\sharp+\lambda \log J \,\mathbf{g}^\sharp
	\,.
\end{equation}
In this case
\begin{equation}
	\frac{\partial \tau^{ab}}{\partial h_{cd}} = 
  	2\mu \frac{\partial h^{ab}}{\partial h_{cd}}+ \frac{\lambda}{J}\frac{\partial J}{\partial h_{cd}} g^{ab}
	= \mu J \left(g^{ac}g^{bd}+g^{ad}g^{bc}\right)+ \frac{\lambda}{J} g^{ab} g^{cd}    \,.
\end{equation}
Therefore, for this constitutive equation the symmetry conditions \eqref{Hyperelastic-Potential-Kirchhoff} are satisfied, and hence, this is a hyperealstic model. 
\citet{NeffEidelMartin2016} aptly referred to this as the ``hyperelastic Hencky model".

\vskip 0.1 in \noindent
\textbf{Example 2: Becker's model.} \citet{Becker1893} proposed the following constitutive equation
\begin{equation} \label{Becker-Model}
	\mathbf{T} = 2\mu \log \mathbf{U}^\sharp
	+ \lambda \operatorname{tr}( \log \mathbf{U}) \,\mathbf{G}^\sharp
	= 2\mu \log \mathbf{U}^\sharp
	+ \lambda ( \log J)\, \mathbf{G}^\sharp
	\,,
\end{equation}
where $\mathbf{T}=\mathbf{U}\mathbf{S}$ is the Biot stress, which is work conjugate to $\mathbf{U}$ and $\mu$ and $\lambda$ are elastic constants. 
It is straightforward to show that the stress-work $1$-form in terms of Biot stress reads
\begin{equation}
	\boldsymbol{\Omega}
	=\frac{1}{2} \mathbf{S}\!:\!d\mathbf{C}^\flat = \mathbf{T}\!:\!d\mathbf{U}^\flat
	\,.
\end{equation}
Therefore, in terms of the Biot stress, the necessary and sufficient conditions for hyperelasticity are the following symmetry conditions:
\begin{equation} \label{Hyperelasticity-Biot}
    \frac{\partial T^{AB}}{\partial U_{CD}}=\frac{\partial T^{CD}}{\partial U_{AB}}\,.
\end{equation}
Since the elastic constants $\mu$ and $\lambda$ are independent, the symmetry conditions \eqref{Hyperelasticity-Biot} must be satisfied for each term individually for this constitutive equation to be hyperelastic. We demonstrate that these conditions are not satisfied for the second term, and therefore, \eqref{Becker-Model} is a non-hyperelastic model when $\lambda\neq 0$. 
Specifically, note that
\begin{equation} 
    \frac{\partial \log J}{\partial U_{AB}}
    = \frac{1}{2I_3} \frac{\partial I_3}{\partial U_{AB}}
    = \frac{1}{2I_3} \frac{\partial I_3}{\partial C_{MN}}\frac{\partial C_{MN}}{\partial U_{AB}}
    = \frac{1}{2} C^{-MN} \frac{\partial C_{MN}}{\partial U_{AB}}
    =U^{-AB}
    \,,
\end{equation}
where $C^{-AB}$ and $U^{-AB}$ are the components of $\mathbf{C}^{-\sharp}$ and $\mathbf{U}^{-\sharp}$, respectively. Therefore,
\begin{equation} 
    \frac{\partial }{\partial U_{CD}}   \left[ \lambda ( \log J)\,G^{AB} \right]
    = \lambda \,G^{AB}\,U^{-CD}
    \,.
\end{equation}
Obviously, $\lambda \,G^{AB}\,U^{-CD}\neq \lambda \,G^{CD}\,U^{-AB}$ and the symmetry conditions \eqref{Hyperelasticity-Biot} do not hold.
Therefore, for $\lambda\neq 0$, the constitutive equation \eqref{Becker-Model} is non-hyperelastic, which is consistent with \citep[Proposition~5.1]{NeffMunchMartin2016}.

\section{The Work Line Bundle of Cauchy Elasticity and the Geometric Hysteresis\footnote{The term ``geometric hysteresis" was suggested to us by Francesco Fedele.}}
\label{Sec:Geometric-Phase}

In this section, we demonstrate that the stress $1$-form serves as a connection $1$-form on the manifold of strains---\textit{the stress-work connection}---which can be interpreted as the base manifold of a vector bundle, specifically the work line bundle of Cauchy elasticity.
Put simply, a bundle is characterized by a mapping that projects a larger space (total space) onto a smaller space (base space) \citep{Nakahara2018}. 
We will show that the work line bundle of Cauchy elasticity is closely related to the known observation that the work of stress in cyclic deformations may be nonzero. Curvature of the stress-work connection is responsible for this non-vanishing work. This curvature identically vanishes for hyperelastic solids. Interestingly, the stress-work connection has a non-vanishing torsion, even for hyperelastic solids.

\subsection{Geometric hysteresis in incompressible isotropic Cauchy elasticity}

In incompressible isotropic Cauchy elastic solids, the dependence of stress on strain is reduced to that on the two principle in invariants $I_1$ and $I_2$.
The stress-work $1$-form $\boldsymbol{\Omega}$ is a $1$-form on a $2$-dimensional base manifold $\mathcal{M}$, which has local coordinates $\{I_1,I_2\}$. This can be considered as the (only) connection form $\omega^1{}_2=\boldsymbol{\Omega}$ on $\mathcal{M}$. We assume that this connection is metric compatible. 
Choosing the coordinate frame and coframe fields $\{\mathbf{e}_1=\frac{\partial}{\partial I_1},\mathbf{e}_2=\frac{\partial}{\partial I_2}\}$ and $\{\vartheta^1=dI_1,\vartheta^2=dI_2\}$ metric has the simple form $\mathbf{G}=\delta_{\alpha\beta}\vartheta^{\alpha}\otimes\vartheta^{\beta}=\vartheta^1\otimes\vartheta^1+\vartheta^2\otimes\vartheta^2=dI_1\otimes dI_1+dI_2\otimes dI_2$. This is the natural metric for measuring distance between different deformed local configurations.
Recall that $\nabla\mathbf{e}_{\alpha}=\mathbf{e}_{\gamma}\otimes \omega^{\gamma}{}_{\alpha}$ and $\nabla_{\mathbf{e}_{\beta}}\mathbf{e}_{\alpha}= \langle\omega^{\gamma}{}_{\alpha},\mathbf{e}_{\beta}\rangle \,\mathbf{e}_{\gamma}$.
Thus, $\nabla\mathbf{e}_1=-\mathbf{e}_2\otimes\boldsymbol{\Omega}$ and $\nabla\mathbf{e}_2=\mathbf{e}_1\otimes\boldsymbol{\Omega}$.
It is straightforward to show that 
\begin{equation}
	\nabla_{\mathbf{e}_1}\mathbf{e}_1=-\phi \psi_1 \mathbf{e}_2\,,\qquad
	\nabla_{\mathbf{e}_2}\mathbf{e}_1=-\phi \psi_2 \mathbf{e}_2\,,\qquad
	\nabla_{\mathbf{e}_1}\mathbf{e}_2=\phi \psi_1 \mathbf{e}_1\,,\qquad		
	\nabla_{\mathbf{e}_2}\mathbf{e}_2=\phi \psi_2 \mathbf{e}_1
	\,,
\end{equation}
where  $\psi_i=\frac{\partial \psi}{\partial I_i}$, $i=1,2$.

Recall that on an $n$-manifold Cartan's first and second structural equations read \citep{Sternberg2013,Hehl2003}: 
\begin{equation} \label{Structural-Equations}
	\mathcal{T}^{\alpha}
	=d\vartheta^{\alpha}+\omega^{\alpha}{}_{\gamma}\wedge\,\vartheta^{\gamma}\,,
	\qquad
	\mathcal{R}^{\alpha}{}_{\beta}
	=d\omega^{\alpha}{}_{\beta}+\omega^{\alpha}{}_{\gamma}\wedge\,\omega^{\gamma}{}_{\beta} \,.
\end{equation}
The first structural equations \eqref{Structural-Equations}$_1$ give the torsion $2$-forms:
\begin{equation} 
	\mathcal{T}^1=\omega^{1}{}_{2}\wedge\,\vartheta^{2}=\boldsymbol{\Omega} \wedge dI_2
	=\phi \psi_1 \,dI_1\wedge dI_2 \,,\qquad
	\mathcal{T}^2=\omega^{2}{}_{1}\wedge\,\vartheta^{1}=-\boldsymbol{\Omega} \wedge dI_1
	=\phi \psi_2 \,dI_1\wedge dI_2
	\,.
\end{equation}
Recall that torsion tensor is defined as
\begin{equation} \label{Torsion-Isotropic-Incompressible}
	\mathbf{T}(\mathbf{e}_1,\mathbf{e}_2)
	=\nabla_{\mathbf{e}_1}\mathbf{e}_2-\nabla_{\mathbf{e}_2}\mathbf{e}_1
	-[\mathbf{e}_1,\mathbf{e}_2]
	=\phi(\psi_1 \mathbf{e}_1+\psi_2 \mathbf{e}_2)
	=\phi \left(\psi_1 \frac{\partial}{\partial I_1}+\psi_2 \frac{\partial}{\partial I_2} \right)
	\,.
\end{equation}
The second structural equations \eqref{Structural-Equations}$_2$ imply that the only non-zero curvature $2$-form is
\begin{equation} 
	\mathcal{R}^{1}{}_{2}=d\omega^{1}{}_{2}=d\boldsymbol{\Omega} 
	= (\phi_1 \psi_2-\phi_2 \psi_1) \,dI_1\wedge dI_2	\,.
\end{equation}
When this curvature is non-zero (non-hyperelastic solids) its integral over a region enclosed by the trajectory of a cyclic motion---the net work of stress, which is a fiber variable---is a geometric hysteresis. It is geometric in the sense that even if one considers inertial effects this geometric hysteresis is independent of how fast or slow the body has experienced the cyclic  deformation. 
We observe that the torsion tensor is always non-vanishing, even for hyperelastic solids. However, the curvature tensor is non-vanishing only in the case of non-hyperelastic solids.

It should be noted that the pressure field does not contribute to the stress-work $1$-form, and hence, does not affect the net work in any cyclic motion. From \eqref{Incomp-Isotropic-Cauchy-2}, note that 
\begin{equation}
	d\boldsymbol{\Omega}(I_1,I_2)  =  d\phi(I_1,I_2) \wedge d\psi(I_1,I_2)  
	=\left[\frac{\partial \phi}{\partial I_1}\frac{\partial \psi}{\partial I_2}
	-\frac{\partial \phi}{\partial I_2}\frac{\partial \psi}{\partial I_1}\right] dI_1\wedge dI_2
	=\left(\phi_{,1}\psi_{,2}-\phi_{,2}\psi_{,1} \right)\,dI_1\wedge dI_2
	\,.
\end{equation}
For example, if $\psi$ is only a function of $I_1$, $\phi$ must be a function of $I_2$ (or both $I_1$ and $I_2$) for the material to be non-hyperelastic. It should also be noted that not every cyclic deformation in a non-hyperelastic solid corresponds to a non-zero net work. Suppose $\Gamma$ is a closed curve in the $(I_1,I_2)$ space. It encloses a region $\mathcal{D}$ in $(I_1,I_2)$ space. Using Stokes' theorem the geometric hysteresis is calculated as
\begin{equation}
	\mathcal{H}(\Gamma)=\int_{\Gamma}\boldsymbol{\Omega}=\int_{\mathcal{D}}d\boldsymbol{\Omega}
	=\int_{\mathcal{D}}\left[\frac{\partial \phi}{\partial I_1}\frac{\partial \psi}{\partial I_2}
	-\frac{\partial \phi}{\partial I_2}\frac{\partial \psi}{\partial I_1}\right] dI_1\wedge dI_2\,.
\end{equation}
For a non-hyperelastic solid for which $d\boldsymbol{\Omega}\neq \mathbf{0}$, the net work is non-zero only if $\mathcal{D}$ has non-vanishing area.

One can equivalently assume that $\phi$ and $\psi$ explicitly depend on the principal stretches instead, i.e., $\phi=\phi(\lambda_1,\lambda_2)$ and $\psi=\psi(\lambda_1,\lambda_2)$. The net work in a cyclic motion is written as
\begin{equation}
	\mathcal{H}(\Gamma)=
	\int_{\mathcal{D}}\left[\frac{\partial \phi}{\partial \lambda_1}\frac{\partial \psi}{\partial \lambda_2}
	-\frac{\partial \phi}{\partial \lambda_2}\frac{\partial \psi}{\partial \lambda_1}\right] 
	d\lambda_1\wedge d\lambda_2\,.
\end{equation}
This, in particular, shows that a one-dimensional cyclic motion of a Cauchy elastic solid, e.g., uniaxial deformation of a bar, always results in a zero net work. We will look at some concrete examples of geometric phase in \S\ref{Sec:Examples}.

\begin{figure}[t!]
\centering
\includegraphics[width=0.4\textwidth]{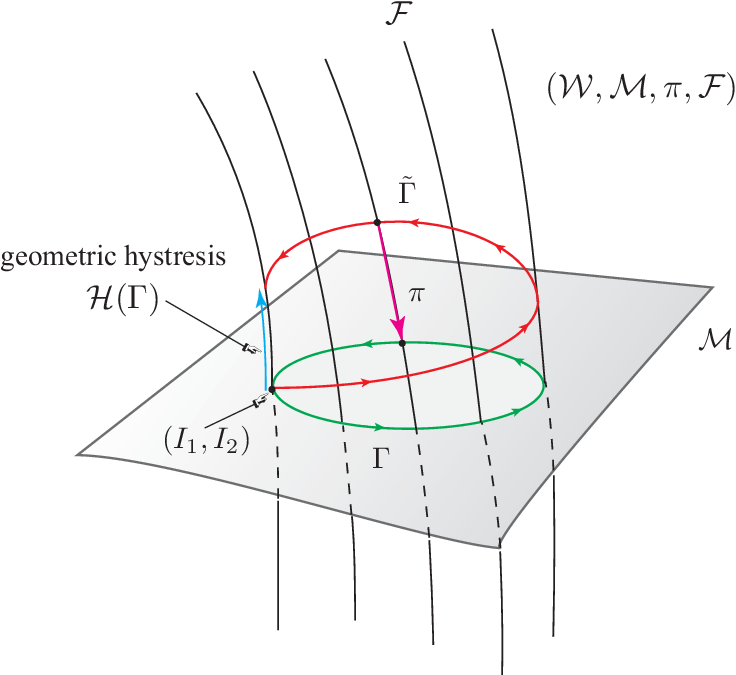}
\vspace*{0.10in}
\caption{Geometric hystresis in incompressible isotropic Cauchy elasticity. The fiber (line) bundle structure of Cauchy elasticity is schematically shown. A fiber $\mathcal{F}$ is attached to each point $(I_1,I_2)$ of the base manifold $\mathcal{M}$. A cyclic motion is a closed curve $\Gamma$ in the base manifold $\mathcal{M}$. $\mathcal{M}$ is the base (shape) manifold and $\mathcal{F}$ is a generic fiber. The corresponding curve in the line bundle is not closed; the lack of closure is related to the geometric hysteresis $\mathcal{H}(\Gamma)$. Note that $\mathcal{H}(-\Gamma)=-\mathcal{H}(\Gamma)$. In compressible isotropic elasticity $\mathcal{M}$ is a $3$-manifold while in compressible anisotropic elasticity its dimension depends on the symmetry class. } 
\label{Geometric-Hystresis}
\end{figure}

\paragraph{The work line bundle.}
One can define a fiber bundle (and more specifically, a line bundle) over $\mathcal{M}$ by attaching a line (a one-dimensional vector space) to each point $m=(I_1,I_2)$ of the base manifold; this is the fiber at $m$, which we denote by $\mathcal{F}$. 
Each fiber is an additive group of real numbers, and the work of stress is an element of this group.
The line bundle is denoted by $(\mathcal{W},\mathcal{M},\pi,\mathcal{F})$, where $\mathcal{W}$ is the total space and $\pi:\mathcal{W}\to\mathcal{M}$ is the projection map. $\mathcal{W}$ is a trivial bundle in the sense that $\mathcal{W}=\mathcal{M}\times \mathcal{F}$.
A deformation is a curve in the base space, and a cyclic deformation is a closed curve in the base manifold, see Fig.~\ref{Geometric-Hystresis}. 
Let us consider a closed curve $\Gamma$ in $\mathcal{M}$ and suppose that it has the parametrization $\Gamma:[0,t_0]\to\mathcal{M}$, $t \mapsto \Gamma(t)=(I_1(t),I_2(t))$. 
We lift this curve to the total space and denote it by $\tilde{\Gamma}: [0,t_0]\to\mathcal{W}$ such that $\tilde{\Gamma}(t)=(\Gamma(t),W(t))=(I_1(t),I_2(t),W(t))$, where
\begin{equation}
\begin{aligned}
	W(t) & = \int_0^{t} \boldsymbol{\Omega} (\Gamma(t))
	= \int_0^{t} \phi(\Gamma(s)) \,d\psi(\Gamma(s)) \\
	& = \int_0^{t} \phi(I_1(s),I_2(s))
	\left[\frac{\partial \psi}{\partial I_1} I'_1(s)+\frac{\partial \psi}{\partial I_2} I'_2(s)\right] ds
	\,.
\end{aligned}
\end{equation}
In other words, the fiber coordinate at $\Gamma(t)$ is the work done by the stress over the time interval $[0,t]$. It should be emphasized that the fiber coordinate $W(t)$ explicitly depends on the initial point $\Gamma(0)=(I_1(0),I_2(0))$, which does not necessarily correspond to a stress-free configuration, as well as on the path $\Gamma$.
We observe that for any closed curve $\Gamma \in \mathcal{M}$, its lift $\tilde{\Gamma}$ to $\mathcal{W}$ is unique but not necessarily a closed curve.

\begin{remark}
The fiber coordinate of the lifted curve $\tilde{\Gamma}$ is related to the torsion tensor of the connection $\boldsymbol{\Omega}$.
Using \eqref{Torsion-Isotropic-Incompressible} one can write $\mathbf{T}(\mathbf{e}_1,\mathbf{e}_2)\cdot (I'_1\mathbf{e}_1+I'_1\mathbf{e}_2)=\phi(\psi_1 I'_1+\psi_2 I'_2)$, and hence,
\begin{equation}
	W(t) =  \int_0^{t} 
	\mathbf{T}\left(\frac{\partial}{\partial I_1},\frac{\partial}{\partial I_2}\right)\cdot  \Gamma'(s)\, ds
	\,.
\end{equation}
\end{remark}

\subsection{Geometric hysteresis in compressible isotropic Cauchy elasticity}

For compressible isotropic Cauchy elasticity, the base (shape) manifold is parametrized by $\{I_1,I_2,I_3\}$.
Let us choose the coordinate frame and coframe fields $\{\mathbf{e}_1=\frac{\partial}{\partial I_1},\mathbf{e}_2=\frac{\partial}{\partial I_2},\mathbf{e}_3=\frac{\partial}{\partial I_3}\}$ and $\{\vartheta^1=dI_1,\vartheta^2=dI_2,,\vartheta^3=dI_3\}$. Metric has the simple form $\mathbf{G}=\delta_{\alpha\beta}\vartheta^{\alpha}\otimes\vartheta^{\beta}=\vartheta^1\otimes\vartheta^1+\vartheta^2\otimes\vartheta^2+\vartheta^3\otimes\vartheta^3=dI_1\otimes dI_1+dI_2\otimes dI_2+dI_3\otimes dI_3$. 
Let us pick \eqref{Isotropic-Cauchy-3} to be the stress-work connection form, i.e., $\boldsymbol{\Omega}=\phi d\psi+d\chi$. In a three-dimensional manifold there are three connection $1$-forms. Without loss of generality, let us assume that $\omega^1{}_2=\boldsymbol{\Omega}$ and $\omega^2{}_3=\omega^3{}_1=0$. 
Assigning these three $1$-forms is equivalent to equipping $\mathcal{M}$ with a connection $\nabla$. This connection always has non-vanishing torsion and, in the case of non-hyperelastic solids, also has non-vanishing curvature.

Let us consider a closed curve $\Gamma \in \mathcal{M}$ with the parametrization $\Gamma: [0,t_0] \to \mathcal{M}, t \mapsto \Gamma(t) = (I_1(t), I_2(t), I_3(t))$. We lift this curve to the total space and denote it by $\tilde{\Gamma}: [0,t_0] \to \mathcal{W}$, such that $\tilde{\Gamma}(t) = (\Gamma(t), W(t)) = (I_1(t), I_2(t),  I_3(t), W(t))$, where
\begin{equation}
\begin{aligned}
	W(t) & = \int_0^{t} \boldsymbol{\Omega} (\Gamma(t))
	= \int_0^{t} \left[\phi(\Gamma(s)) \,d\psi(\Gamma(s))+d\chi(\Gamma(s))\right] \\
	& = \int_0^{t} \left[\phi(\Gamma(s)) \, \frac{\partial \psi}{\partial \Gamma(s)}
	+\frac{\partial \chi}{\partial \Gamma(s)} \right]\cdot \Gamma'(s) \,ds \\
	& = \int_0^{t} \Big[(\phi \psi_1+\chi_1)I'_1(s)+(\phi \psi_2+\chi_2)I'_2(s)+(\phi \psi_3+\chi_3)I'_3(s)
	\Big]\,ds
	\,.
\end{aligned}
\end{equation}
Note that
\begin{equation}
\begin{aligned}
	d\boldsymbol{\Omega}(I_1,I_2,I_3)  
	& = \left(\phi_{,1}\psi_{,2}-\phi_{,2}\psi_{,1} \right)\,dI_1\wedge dI_2
	+\left(\phi_{,2}\psi_{,3}-\phi_{,3}\psi_{,2} \right)\,dI_2\wedge dI_3 \\
	& \quad +\left(\phi_{,1}\psi_{,3}-\phi_{,3}\psi_{,1} \right)\,dI_1\wedge dI_3
	\,.
\end{aligned}
\end{equation}
We observe that the above expression is the only non-vanishing curvature $2$-form of the shape manifold: $\mathcal{R}^1{}_2=d\omega^1{}_2=d\boldsymbol{\Omega}$, $\mathcal{R}^2{}_3=\mathcal{R}^3{}_2=0$. 
Geometric hystresis is the integral of the non-vanishing connection form over a closed path $\Gamma$ in the shape manifold, or integral of any surface that has this closed path as its boundary, i.e.,
\begin{equation}
	\mathcal{H}(\Gamma)=\int_{\mathcal{D}}\mathcal{R}^1{}_2
	=\int_{\mathcal{D}}d\boldsymbol{\Omega}\,.
\end{equation}

\begin{remark}
Consider the following choices for the connection $1$-forms
\begin{equation}
    \omega^1{}_2=a\,\boldsymbol{\Omega}\,,\qquad
    \omega^2{}_3=b\,\boldsymbol{\Omega}\,,\qquad
    \omega^3{}_1=c\,\boldsymbol{\Omega}\,,\qquad  a+b+c=1 \,.
\end{equation}
In this case, the second structural tensors give us $\mathcal{R}^1{}_2=d\omega^1{}_2$, $\mathcal{R}^2{}_3=d\omega^2{}_3$, and $\mathcal{R}^3{}_1=d\omega^3{}_1$ (note that $\boldsymbol{\Omega}\wedge \boldsymbol{\Omega}=0$). Thus
\begin{equation}
	\mathcal{H}(\Gamma)=\int_{\mathcal{D}}\left(\mathcal{R}^1{}_2+\mathcal{R}^2{}_3+\mathcal{R}^3{}_1\right)
	=\int_{\mathcal{D}}d\boldsymbol{\Omega}\,.
\end{equation}
\end{remark}

\subsection{Geometric hysteresis in  anisotropic Cauchy elasticity}

Geometric hysteresis is defined similarly in anisotropic Cauchy elasticity. Suppose a compressible anisotropic Cauchy elastic solid has an integrity basis $\{I_1,\hdots,I_m\}$. In this case the base manifold $\mathcal{M}$ is $m$-dimensional. Assuming that $\omega^1{}_2$ is the only non-vanishing connection $1$-form on $\mathcal{M}$, the only non-vanishing curvature $2$-form is $\mathcal{R}^1{}_2=d\omega^1{}_2$. The geometric hysteresis is defined as $W=\int_{\Gamma}\boldsymbol{\Omega}$, where $\Gamma$ is a closed curve in $\mathcal{M}$. If the solid is incompressible, the geometric hysteresis is defined similarly, but in this case, $\Gamma$ is a closed curve in an $(m-1)$-dimensional submanifold of $\mathcal{M}$ defined by the constraint $I_3=1$.

\section{Active Solids and Cauchy Elasticity} \label{Sec:Active-Matter}

There are many examples of active solids in nature and engineering. These active solids can change their microscopic structure to remodel, grow, change shape, or exert work on their environment. The active component typically requires external sources of energy that is brought into the system through chemical, mechanical, or electrmagnetic fields. Therefore, this additional component in the solid may depend on many different fields as active solids respond to either a pre-programmed course of action or to achieve a certain response to stimuli \citep{goriely17}. 

In the particular case when the response \textit{only} depends on strain, these active solids can be modeled as Cauchy elastic materials, which provides a new area of application for the theory of Cauchy elasticity.
For instance, in biomechanics, the activity of muscles is sometimes modeled by assuming an additive split of stress into elastic and active stresses. Another approach is to assume a multiplicative decomposition of the deformation gradient into elastic and active parts \citep{Kondaurov1987,Ambrosi2012,Goriely2017ActiveSolids,Giantesio2019}.
\citet{Ambrosi2012} pointed out that an active stress tensor does not necessarily have a corresponding energy function. A direct way to model active solids is to include an extra \textit{active} component to the stress, in which case, one writes the second Piola-Kirchhoff stress as \citep{Goriely2017ActiveSolids}
\begin{equation} \label{Stress-Additive}
	\mathbf{S} = 2\frac{\partial \psi}{\partial \mathbf{C}^\flat}+\mathbf{S}_{\text{a}} \,,
\end{equation}
where $\mathbf{S}_{\text{a}}=\mathbf{S}_{\text{a}}(X,\mathbf{C}^\flat,\mathbf{G})$ is the active stress. In the case where the  active stress explicitly depends on strain, this system is clearly a Cauchy elastic material, and in general, in a cyclic motion the net work done by active stress is nonzero. 
The following two observations are nothworthy: i) Eq.~\eqref{Stress-Additive} suggests that some biological systems like muscular response can be modeled, in the simplest case, as Cauchy materials. ii) The additive split of stress \eqref{Stress-Additive} is not an assumption; it is a consequence of Darboux classification of Cauchy elastic solids, see \S\ref{Sec:AdditiveDecomposition}.
We next examine the stress representation \eqref{Stress-Additive} in a few examples of constitutive assumptions for active stress in the literature.

\subsection{Hydrostatic active stress} 

\citet{Panfilov2005} suggested the following constitutive equation for active stress
\begin{equation}
	\mathbf{S}_{\text{a}}(X,\mathbf{C}^\flat,\mathbf{G})=\tilde{S}(X)\,\mathbf{C}^{-\sharp}\,.
\end{equation}
One can easily show that the conditions \eqref{Hyperelastic-Potential} are satisfied, and hence, this active stress has an active potential $\psi_{\text{a}}(\mathbf{C}^\flat,\mathbf{G})=\frac{1}{2}\tilde{S}\,\operatorname{tr}\mathbf{C}^2=\frac{1}{2}\tilde{S}\,\mathbf{C}^{-\sharp}\!:\!\mathbf{C}^\flat$ such that
\begin{equation}
	\mathbf{S}_{\text{a}}(X,\mathbf{C}^\flat,\mathbf{G})=\frac{\partial\psi_{\text{a}}}{\partial\mathbf{C}^\flat}\,.
\end{equation}
Therefore, this active system can be written with a hyperelastic strain energy function and the net work of stress in any cyclic deformation is zero.

\subsection{Active stress in fiber-reinforced solids} 

Consider an isotropic solid that is reinforced by a family of fibers. At $X\in\mathcal{B}$ let the $\mathbf{G}$-unit tangent to the fiber be $\mathbf{N}(X)$. Such a solid is effectively transversely isotropic. In addition to the principal invariants one has the following two extra invariants: $I_4=\mathbf{N}\cdot\mathbf{C}\cdot\mathbf{N}$, and $I_5=\mathbf{N}\cdot\mathbf{C}^2\cdot\mathbf{N}$ \citep{DoyleEricksen1956,spencer1982formulation}. An example of active stress constitutive equation that has been widely used in the literature is the following \citep{Ambrosi2012,Giantesio2019}:
\begin{equation}\label{Active-Stress-Fiber}
    \mathbf{S}_{\text{a}}(X,\mathbf{C}^\flat,\mathbf{G})=\tilde{S}(X,I_4)\,\mathbf{N}\otimes\mathbf{N}\,.
\end{equation}
For an example of $\tilde{S}(X,I_4)$ given in \citep{Pathmanathan2010}, \citet{Ambrosi2012} pointed out that active stress is the derivative of a potential (that \citet{Pathmanathan2010} call ``active strain energy"). Then they write ``Such a scalar function should not be understood as a strain energy, as no conservation applies...". However, we showed that  as soon as the stress is derived from a potential, the response of the material is hyperelastic and the net work of stress in any cyclic motion is zero. In other words, such a solid is conservative.\footnote{What they probably mean is that despite the fact that the stress derives from a potential component, the active component is not an intrinsic property of the elastic material but an extra component.}
For the active stress constitutive equation \eqref{Active-Stress-Fiber} one has
\begin{equation}
    \frac{\partial \mathbf{S}_{\text{a}}}{\partial\mathbf{C}^\flat}
    =\frac{\partial\tilde{S}}{\partial I_4}\,\frac{\partial I_4}{\partial \mathbf{C}^\flat}\otimes\mathbf{N}\otimes\mathbf{N}
    =\frac{\partial\tilde{S}}{\partial I_4}\,\mathbf{N}\otimes\mathbf{N}\otimes\mathbf{N}\otimes\mathbf{N}
    \,.
\end{equation}
Clearly,  conditions \eqref{Hyperelastic-Potential} are satisfied for any choice of $\tilde{S}(X,I_4)$. More specifically, the active energy function is written as $\psi_{\text{a}}(\mathbf{C}^\flat,\mathbf{G})=\int \tilde{S}(X,I_4)\,dI_4$.

\begin{remark}
If one assumes that $\tilde{S}=\tilde{S}(X,I_1,I_4)$ then
\begin{equation}
    \frac{\partial \mathbf{S}_{\text{a}}}{\partial\mathbf{C}^\flat}
    =\frac{\partial\tilde{S}}{\partial I_1}\,\mathbf{G}^\sharp\otimes\mathbf{N}\otimes\mathbf{N}
    +\frac{\partial\tilde{S}}{\partial I_4}\,\mathbf{N}\otimes\mathbf{N}\otimes\mathbf{N}\otimes\mathbf{N}
    \,.
\end{equation}
Note that $\mathbf{G}^\sharp\otimes\mathbf{N}\otimes\mathbf{N}\neq \mathbf{N}\otimes\mathbf{N}\otimes\mathbf{G}^\sharp$, and hence if $\partial{\tilde{S}}/\partial I_1\neq 0$, the active stress \eqref{Active-Stress-Fiber} is not derived from a potential and this system is a non-hyperelastic Cauchy material.
\end{remark}

\subsection{Active stress in arteries} 

The response of an artery under load and internal muscular activation was modeled by \citet{Rachev1999} as a thick-walled tube made of an incompressible orthotropic  elastic solid.
In addition to the constitutively determined (up to an unknown pressure field) elastic (passive) stresses, they included an active circumferential stress arising from muscle contraction and relaxation. 
Suppose that in its undeformed configuration the artery has inner and outer radii $R_i$ and $R_o$, respectively. With respect to the cylindrical coordinates $(R,\Theta,Z)$ and $(r,\theta,z)$ in the reference and current configurations, respectively, consider the following family of deformations
\begin{equation}
    r(R,\Theta,Z)=r(R)\,,\qquad
    \theta(R,\Theta,Z)=\Theta\,,\qquad
    z(R,\Theta,Z)=\lambda\,Z
    \,,
\end{equation}
where $\lambda$ is an unknown axial stretch. Here, we will ignore eigenstrains (``residual strains" according to \citet{Rachev1999}). The deformation gradient and the right Cauchy-Green strain have the following representations: $\mathbf{F}=\operatorname{diag}\left\{r'(R),1,\lambda\right\}$ and $\mathbf{C}^\flat=\operatorname{diag}\left\{{r'}^2(R),r^2(R),\lambda^2\right\}$.
The corresponding tensors in terms of physical components are:\footnote{The physical components read $\bar{F}^a{}_A=\sqrt{g_{aa}}\,\sqrt{G^{AA}}\,F^a{}_A$ and $\bar{C}_{AB}=\sqrt{G^{AA}}\,\sqrt{G^{BB}}\,C_{AB}$ (no summation) \citep{Truesdell1953}.}
\begin{equation}
    \bar{\mathbf{F}}=\begin{bmatrix}
    r'(R) & 0 & 0 \\
    0 & \frac{r(R)}{R} & 0 \\
    0 & 0 & \lambda 
    \end{bmatrix}
    \,, \qquad 
    \bar{\mathbf{C}}^\flat=\begin{bmatrix}
    {r'}^2(R) & 0 & 0 \\
    0 & \frac{r^2(R)}{R^2} & 0 \\
    0 & 0 & \lambda^2 
    \end{bmatrix}
    \,.
\end{equation}
Thus, the principal stretches are $\lambda_r=r'(R)$, $\lambda_{\theta}=\frac{r(R)}{R}$, and $\lambda_z=\lambda$.
Assuming an incompressible solid $J=r^2(R)\,{r'}^2(R)\,\lambda^2\,R^{-2}=1$, and hence
\begin{equation}
    r(R)=\sqrt{r_i^2+\frac{R^2-R_i^2}{\lambda}}
    \,,
\end{equation}
where $r_i=r(R_i)$.
\citet{Rachev1999} assumed that the only non-zero active stress is $\hat{\sigma}^{\theta\theta}=S \lambda_{\theta} f(\lambda_{\theta})$. From the incompressibility constraint, $r'(R)=\frac{1}{\lambda\lambda_{\theta}}$.
Note that $S^{\Theta\Theta}=\sigma^{\theta\theta}$ is the only nonzero active stress. The stress-work $1$-form is written as
\begin{equation}
    \boldsymbol{\Omega}=\frac{1}{2}S^{\Theta\Theta}dC_{\Theta\Theta}
    =\frac{R^2}{2r^2(R)}\,\hat{S}^{\Theta\Theta}d\hat{C}_{\Theta\Theta}
    =\frac{1}{2\lambda^2_{\theta}}\,S \lambda_{\theta} f(\lambda_{\theta})\,d(\lambda^2_{\theta})
    =Sf(\lambda_{\theta})\,d\lambda_{\theta}\,.
\end{equation}
Therefore
\begin{equation}
    d\boldsymbol{\Omega}=f(\lambda_{\theta})\,\frac{\partial S}{\partial \lambda}\,d\lambda \wedge d\lambda_{\theta}\,.
\end{equation}
We observe that if $S$ is not a functions of the axial stretch $\lambda$, then the assumed active stress has a potential, and hence, is conservative.

\section{Examples of Cyclic Universal Deformations} \label{Sec:Examples}

For a given class of materials, \emph{universal deformations} are defined as those deformations that can be sustained without the need for body forces, relying solely on the application of boundary tractions, applicable to all materials within the class. For homogeneous compressible isotropic hyperelastic solids, \citet{Ericksen1955} showed that the set of universal deformations coincides with the set of homogeneous deformations. For homogeneous incompressible isotropic solids, \citet{Ericksen1954} found four families of universal deformations (other than isochoric homogeneous deformations). A fifth family was discovered more than a decade later \citep{SinghPipkin1965,KlingbeilShield1966}.
Universal deformations are significant for several reasons. In particular, they provide valuable insights for designing experiments aimed at determining the constitutive relations of a specific material \citep{Rivlin1951,DoyleEricksen1956,Saccomandi2001}.
Recently, \citet{Yavari2024} showed that the set of universal deformations of isotropic Cauchy elasticity is identical to that of isotropic hyperelasticity, for both compressible and incompressible cases. In this section, we will analyze three examples of cyclic deformations for incompressible materials that remain universal at all times.

\subsection{Biaxial extension of an incompressible block}

We consider a homogeneous block, made of an incompressible Ericksen elastic solid, with dimensions $a_0$, $b_0$, and $c_0$ in its initial undeformed configuration. We use  a Cartesian coordinate system $\{X^1,X^2,X^3\}$ with coordinate lines parallel to the edges of the undeformed block and use Cartesian Coordinates $\{x^1,x^2,x^3\}$ in the current configuration following a family of (biaxial tension) deformations:
\begin{equation} \label{Biaxial-Extension}
    x^1(X,t)=\lambda_1(t)\,X^1\,,\qquad
    x^2(X,t)=\lambda_2(t)\,X^2\,,\qquad
    x^3(X,t)=\frac{1}{\lambda_1(t)\,\lambda_2(t)}\,X^3
    \,.
\end{equation}
The velocity field reads 
\begin{equation}
\begin{aligned}
    V^1(X,t)=\lambda_1'(t)\,X^1\,,\qquad
    V^2(X,t)=\lambda_2'(t)\,X^2\,,\qquad
    V^3(X,t) =  -\left[\frac{\lambda_1'(t)}{\lambda_1^2(t) \lambda_2(t)}
    +\frac{\lambda_2'(t)}{\lambda_1(t) \lambda_2^2(t)} \right]X^3
    \,.
\end{aligned}
\end{equation}
Thus
\begin{equation}
	\int_{\mathcal{B}}\mathcal{K} \,dV= \frac{a_0b_0c_0\rho_0}{6}
	\left\{a_0^2\frac{\left[\lambda_2(t) \lambda_1'(t)
	+\lambda_1(t) \lambda_2'(t)\right]^2}{\lambda_1(t)^4 \lambda_2(t)^4}
	+b_0^2 {\lambda_2'}^2(t)+c_0^2 {\lambda_1'}^2(t) \right\}
    \,.
\end{equation}
The principal invariants read ($I_3=1$)
\begin{equation}
    I_1=\frac{1}{\lambda_1^2(t)\, \lambda_2^2(t)}+\lambda_1^2(t)+\lambda_2^2(t)\,,\qquad
    I_2=\lambda_1^2(t)\, \lambda_2^2(t)+\frac{1}{\lambda_1^2(t)}+\frac{1}{\lambda_2^2(t)}
    \,.
\end{equation}

\subsubsection{Biaxial extension of an incompressible Ericksen elastic block}

\begin{figure}[t!]
\centering
\includegraphics[width=0.70\textwidth]{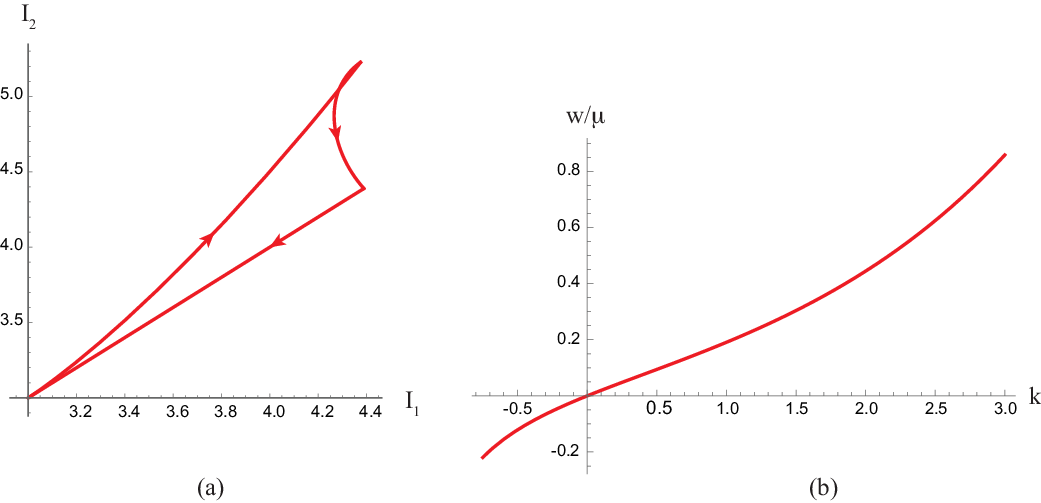}
\vspace*{0.10in}
\caption{(a) A closed curve in the $(I_1,I_2)$ plane corresponds to a cyclic deformation. (b) Non-dimensionalized volume density of the non-dimensionalized net work of stress $\frac{w}{\mu}=\frac{W}{a_0 b_0 c_0\mu}$ as a function of the parameter $k$ in $\phi(I_1,I_2)=(I_2-3)^k$.} 
\label{Homogeneous-Cyclic}
\end{figure}

For an Ericksen elastic solid, the stress-work $1$-form can be written as
\begin{equation}
\begin{aligned}
	\boldsymbol{\Omega}(\lambda_1,\lambda_2) 
	& = \phi\Bigg[\frac{\left[\lambda_1^4(t) \lambda_2^2(t)-1\right] \left[\psi_1
	+\psi_2 \lambda_2^2(t)\right]}{\lambda_1^3(t) \lambda_2^2(t)}\, d\lambda_1(t) \\
	&\qquad\qquad +\frac{\left[\lambda_1^2(t) \lambda_2^4(t)-1\right] \left[\psi_1
	+\psi_2 \lambda_1^2(t)\right]}{\lambda_1^2(t) \lambda_2^3(t)}\,d\lambda_2(t)\Bigg]	\,.
\end{aligned}
\end{equation}
We have two potentials to specify. For $\psi$, we use a neo-Hookean Cauchy elastic material  $\psi(I_1,I_2)=\frac{\mu}{2}(I_1-3)$ and choose the second potential as $\phi(I_1,I_2)=(I_2-3)^k$, where $k$ is a material parameter. Thus, $\psi_1=\frac{\mu}{2}$ and $\psi_2=0$.
Let us consider the following piecewise linear cyclic deformation ($i=1,2$)
\begin{equation} \label{Biaxial-Loading-Example}
	\lambda_i(t)=
	\begin{dcases}
	1+\left(\mathring{\lambda}_i-1\right)\frac{t}{\mathring{t}_i}\,, & 0 \leq t \leq \mathring{t}_i\,, \\
	2\mathring{\lambda}_i-1+\left(1-\mathring{\lambda}_i\right)\frac{t}{\mathring{t}_i}\,, 
	& \mathring{t}_i \leq t \leq 2\mathring{t}_i\,, \\
	1\,, & t > 2\mathring{t}_i\,,
	\end{dcases}
\end{equation}
for some given constants $\mathring{\lambda}_1$ and $\mathring{\lambda}_2$. To visualize the dynamics, we use the  following numerical values for this cyclic loading: $\mathring{\lambda}_1=1.5$, $\mathring{\lambda}_2=1.75$, $\mathring{t}_1=1.0$, and $\mathring{t}_2=2.0$ (the cyclic deformation is within the time interval $[0,2\mathring{t}_2]=[0,4]$). Fig.~\ref{Homogeneous-Cyclic}a shows a closed curve in the $(I_1,I_2)$ plane that corresponds to this cyclic deformation. The density of the net work of stress is plotted in Fig.~\ref{Homogeneous-Cyclic}b as a function of the parameter $k$ that specifies the Ericksen potential of this material. It shows clearly that depending on the material parameter and the direction of the cycle, energy is either needed to perform the work or is gained.

It is also important to note that  this cyclic deformation  is \textit{not} a cyclic motion. As a consequence, the work of stress is not the total work done on the block. Indeed, a direct computation of the kinetic energy gives
\begin{equation}
    \int_{\mathcal{B}}(\mathcal{K}_2-\mathcal{K}_1) \,dV=-a_0b_0c_0\rho_0\left(0.104167 a_0^2+0.0416667 c_0^2 \right).
\end{equation}

\begin{remark}
An active material may have access to a time-dependent source/sink of energy. This can be modeled by assuming that $\phi$ has an explicit time dependence, e.g., in the above example $\phi(t,I_1,I_2)=f(t)(I_2-3)^k$, where $f(t)$ is some function of time.
\end{remark}

\subsubsection{Biaxial extension of an incompressible neo-Hookean block reinforced by active fibers}

Consider again the block from the previous example but assume that it is made of a homogeneous neo-Hookean solid reinforced with active fibers parallel to one of the coordinate axes in the undeformed configuration. We still consider the time-dependent biaxial extension \eqref{Biaxial-Extension} and assume that the active stress has the following form
\begin{equation}
    \mathbf{S}_a(X,\mathbf{C}^\flat,\mathbf{G})=\tilde{S}(I_1)\,\mathbf{N}\otimes\mathbf{N}\,,
\end{equation}
where for this example and with respect to the Cartesian coordinates $(X^1,X^2,X^3)$, we consider the following three cases: $\mathbf{N}=\hat{\mathbf{E}}_1=\{1,0,0\}$, $\mathbf{N}=\hat{\mathbf{E}}_2=\{0,1,0\}$, and $\mathbf{N}=\hat{\mathbf{E}}_3=\{0,0,1\}$.

\begin{figure}[t!]
\centering
\includegraphics[width=0.45\textwidth]{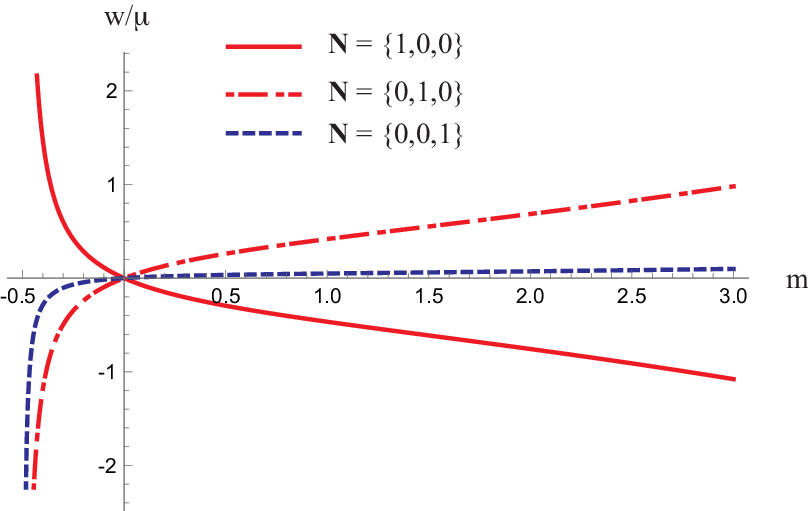}
\vspace*{0.10in}
\caption{Non-dimensionalized volume density of the non-dimensionalized net work of stress $w/\mu$ as a function of the parameter $m$ in the active stress expression for three different fiber distributions and under the cyclic deformation \eqref{Biaxial-Loading-Example}.} 
\label{Active-Fibers-Cyclic}
\end{figure}

The stress-work $1$-form is written as
\begin{equation}
	\boldsymbol{\Omega}(\lambda_1,\lambda_2) 
	= \begin{dcases} 
	\left[(\tilde{S}+\mu)\lambda_1(t)- \frac{\mu}{\lambda_1^3(t) \lambda_2^2(t)}\right] d\lambda_1(t)	
	+\mu \left[\lambda_2(t)- \frac{1}{\lambda_1^2(t) \lambda_2^3(t)}\right] d\lambda_2(t)\,,&
	\mathbf{N}=\hat{\mathbf{E}}_1\,,\\
	\mu \left[\lambda_1(t)- \frac{1}{\lambda_1^3(t) \lambda_2^2(t)}\right] d\lambda_1(t)	
	+\left[(\tilde{S}+\mu )\lambda_2(t)- \frac{\mu}{\lambda_1^2(t) \lambda_2^3(t)}\right] d\lambda_2(t)\,,&
	\mathbf{N}=\hat{\mathbf{E}}_2\,,\\
	\left[\mu \lambda_1(t)- \frac{\tilde{S}+\mu}{\lambda_1^3(t) \lambda_2^2(t)}\right] d\lambda_1(t)	
	+\left[\mu \lambda_2(t)- \frac{\tilde{S}+\mu}{\lambda_1^2(t) \lambda_2^3(t)}\right] d\lambda_2(t)\,,&
	\mathbf{N}=\hat{\mathbf{E}}_3\,.
	\end{dcases}
\end{equation}
Let us assume that $\tilde{S}(I_1)=(I_1-3)^m$ and consider the loading \eqref{Biaxial-Loading-Example}.
The density of the net work of stress for all the three cases is plotted in Fig.~\ref{Active-Fibers-Cyclic} as a function of the parameter $m$. 

\subsection{Finite torsion and extension of an Ericksen cylinder}

Consider a circular cylindrical bar that has length $L$ and radius $R_0$, respectively, in its initial undeformed configuration and is made of a homogeneous incompressible isotropic Ericksen material.
Let us use cylindrical coordinates $(R,\Theta,Z)$ and $(r,\theta,z)$ in the reference and current configurations, respectively. 
The metric of the Euclidean ambient space and the induced material metric have the following diagonal matrix representations: $\mathbf{g}=\operatorname{diag}\{1,r^2,1\}$, and $\mathbf{G}=\operatorname{diag}\{1,R^2,1\}$.
This bar is under a time-dependent torsion slow enough for the inertial effects to be negligible. The kinematics is described by the following family of deformations
\begin{equation} \label{Torsion-Deformation}
	r=r(R,t)\,,\qquad \theta=\Theta+\psi(t) Z\,,\qquad z=\lambda(t) Z\,,
\end{equation}
where $\psi(t)$ is twist per unit length, and $\lambda(t)$ is the axial stretch.
This is a subset of Family $3$ universal deformations.
There are four possible loading scenarios: either $\psi(t)$ or the applied torque is given, and either $\lambda(t)$ or the applied force is given.
The deformation gradient is written as
\begin{equation}
   \mathbf{F}=\mathbf{F}(R,t)=\begin{bmatrix}
  r'(R,t) & 0  & 0  \\
  0 & 1  & \psi(t)  \\
  0 & 0  & \lambda^2(t)
\end{bmatrix}\,,
\end{equation}
where $r'(R,t)=\frac{\partial r(R,t)}{\partial R}$.
The incompressibility condition is expressed as
\begin{equation}
	J=\sqrt{\frac{\det\mathbf{g}}{\det\mathbf{G}}}\det\mathbf{F}=\frac{\lambda(t)\,r(R,t)\,r'(R,t)}{R}=1\,.
\end{equation}
This condition, along with $r(0,t)=0$, yields
\begin{equation} \label{r_0}
	r(R,t)=\frac{R}{\sqrt{\lambda(t)}}\,,\qquad 0\leq R \leq R_0\,.
\end{equation}
The velocity field is written as
\begin{equation} 
	V^r(R,t)=-\frac{1}{2}\lambda^{-\frac{3}{2}}\dot{\lambda}(t)\,R \,,\qquad 
	V^{\theta}(Z,t)=\dot{\psi}(t) Z\,,\qquad 
	V^z(Z,t)=\dot{\lambda}(t) Z\,.
\end{equation}
The kinetic energy density is written as 
\begin{equation} 
	\frac{1}{ |\mathcal{B}|}\int_{\mathcal{B}}\mathcal{K}(X,t)\,dV
	=\frac{\rho_0}{36 R_0} 
	\left[ 3{\lambda'}^2(t) \left(4 L^2+\frac{R_o^2}{\lambda^3(t)}\right)+4 L^2 R_o^2 {\psi'}^2(t)\right]\,.
\end{equation}
The right Cauchy-Green strain and its rate have the following representations
\begin{equation} 
\begin{aligned}
	\mathbf{C}^\flat(R,t) &=\begin{bmatrix}
	 \frac{1}{\lambda(t)} & 0 & 0 \\
	 0 & \frac{R^2}{\lambda (t)} & \frac{R^2 \psi (t)}{\lambda (t)} \\
	 0 & \frac{R^2 \psi (t)}{\lambda (t)} & \frac{R^2 \psi^2(t)}{\lambda (t)}+\lambda^2(t)
	\end{bmatrix}\,,\\
	\dot{\mathbf{C}}^\flat(R,t) &=\begin{bmatrix}
	 -\frac{\dot{\lambda}(t)}{\lambda^2(t)} & 0 & 0 \\
	 0 & -\frac{R^2 \dot{\lambda}(t)}{\lambda^2(t)} 
	 & \frac{R^2 \left[\lambda(t) \dot{\psi}(t)-\psi(t) \dot{\lambda}(t)\right]}{\lambda^2(t)} \\
	 0 & \frac{R^2 \left[\lambda(t) \dot{\psi}(t)-\psi(t) \dot{\lambda}(t)\right]}{\lambda^2(t)}
	   & \frac{\dot{\lambda}(t) \left[2 \lambda^3(t)-R^2 \psi^2(t)\right]
	   +2 R^2 \lambda(t) \psi(t) \dot{\psi}(t)}{\lambda^2(t)} \\
	\end{bmatrix}\,.
\end{aligned}
\end{equation}
The principal invariants are 
\begin{equation}
   I_1(R,t)=\frac{2+R^2 \psi^2(t)+\lambda^3(t)}{\lambda(t)}  \,,\qquad
   I_2(R,t)=\frac{1+R^2 \psi^2(t)+2 \lambda^3(t)}{\lambda^2(t)} \,.
\end{equation}
The non-zero components of the Cauchy stress read
\begin{equation}
\begin{dcases}
   \sigma^{rr}(R,t) = -p(R,t)+\frac{\alpha(R,t)}{\lambda(t)}-\beta(R,t)\,\lambda(t)  \,,  \\
   \sigma^{\theta\theta}(R,t) = -p(R,t)\,\frac{\lambda(t)}{R^2}
   +\alpha(R,t)\left[\frac{1}{R^2}+\psi^2(t) \right] -\frac{\beta(R,t)\,\lambda^2(t)}{R^2}  \,, \\
   \sigma^{zz}(R,t) = -p(R,t)+\alpha(R,t)\, \lambda^2(t)
   -\beta(R,t)\,\frac{1 +R^2 \psi^2(t)}{\lambda^2(t)} \,, \\
   \sigma^{\theta z}(R,t) = \psi(t) \left[\alpha(R,t) \lambda(t)+\beta(R,t)\right]    \,,
\end{dcases}
\end{equation}
where $\alpha=2\phi\, \psi_{1}$ and $\beta=2\phi\, \psi_{2}$.
Utilizing the circumferential and axial equilibrium equations, one concludes that $p=p(R,t)$.
The radial equilibrium equation $\frac{\partial \sigma^{rr}}{\partial r}+\frac{1}{r}\sigma^{rr}-r\sigma^{\theta\theta}=0$, in terms of the referential coordinates reads
\begin{equation}
	\frac{\partial \sigma^{rr}(R,t)}{\partial R}- \frac{\psi^2(t)}{\lambda(t)}\, R\, \alpha(R,t)=0\,.
\end{equation}
Therefore, using the boundary condition $\sigma^{rr}(R_0,t)=0$, one obtains
\begin{equation}
	\sigma^{rr}(R,t)=-\frac{\psi^2(t)}{\lambda(t)}\int_{R}^{R_0} \xi\,\alpha(\xi,t) \,d\xi\,.
\end{equation}
Thus, the pressure field is written as
\begin{equation} 
	-p(R,t)=-\frac{\psi^2(t)}{\lambda(t)}\int_{R}^{R_0} \xi\,\alpha(\xi,t) \,d\xi
	-\frac{\alpha(R,t)}{\lambda(t)}+\beta(R,t)\, \lambda(t) \,.
\end{equation}
Hence, the non-zero physical components of the Cauchy stress read\footnote{The physical components of the Cauchy stress are expressed as ${\bar{\sigma}}^{ab}=\sigma^{ab}\sqrt{g_{aa}\,g_{bb}}$ (no summation) \citep{Truesdell1953}.}
\begin{equation}
\begin{dcases}
   \bar{\sigma}^{rr}(R,t) = -\frac{\psi^2(t)}{\lambda(t)}\int_{R}^{R_0} \xi\,\alpha(\xi,t) \,d\xi   \,,  \\
   \bar{\sigma}^{\theta\theta}(R,t) = -\frac{\psi^2(t)}{\lambda(t)}\int_{R}^{R_0} \xi\,\alpha(\xi,t) \,d\xi 
   +\alpha(R,t)\,\frac{R^2\,\psi^2(t)}{\lambda(t)}   \,, \\
   \bar{\sigma}^{zz}(R,t) = -\frac{\psi^2(t)}{\lambda(t)}\int_{R}^{R_0} \xi\,\alpha(\xi,t) \,d\xi 
	+\alpha(R,t)\left[\lambda^2(t)-\frac{1}{\lambda(t)}\right] 
   +\beta(R,t)\left[ \lambda(t)-\frac{1 +R^2 \psi^2(t)}{\lambda^2(t)}\right] \,, \\
   \bar{\sigma}^{\theta z}(R,t) = \frac{R\,\psi(t)}{\sqrt{\lambda(t)}} \left[\alpha(R,t) \lambda(t)+\beta(R,t)\right]    
   \,.
\end{dcases}
\end{equation}
The work of stress in a time interval $[0,t_0]$ is calculated as
\begin{equation}
\begin{aligned}
	W = \pi L \int_{0}^{t_0} \int_{0}^{R_0} 
	& \Bigg\{\frac{2 \left[\lambda^3(t)-1\right] \left[\beta +\alpha \lambda (t)\right]
	-R^2 \psi^2(t) \left[2 \beta +\alpha \lambda(t) \right]}{\lambda (t)^3}\,\dot{\lambda}(t) \\
	& \quad
	+\frac{2 R^2 \psi(t) \left[\beta+\alpha \lambda(t)\right]}{\lambda^2(t)}\,\dot{\psi}(t)\Bigg\} R\,dR\,dt 
	\,.
\end{aligned}
\end{equation}

\begin{figure}[t!]
\centering
\includegraphics[width=0.70\textwidth]{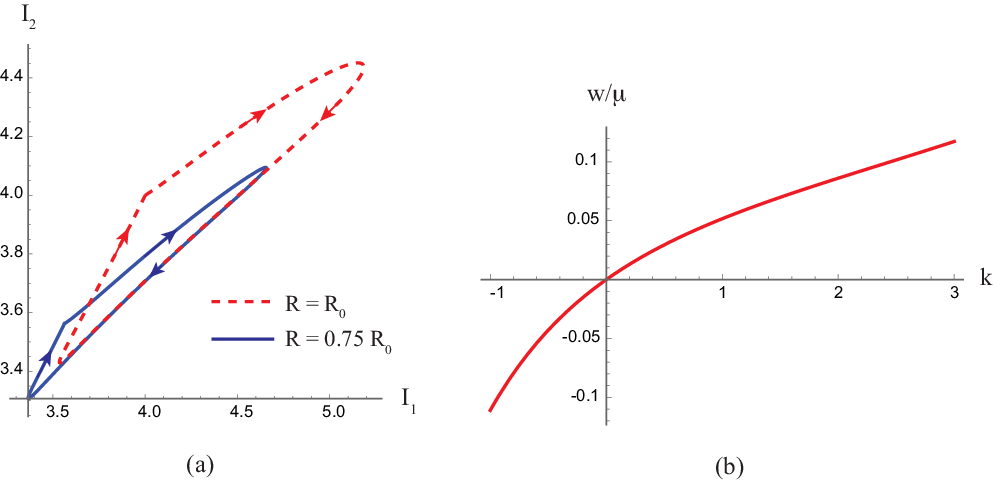}
\vspace*{0.10in}
\caption{(a) Two closed curves in the $(I_1,I_2)$ plane for two values of $R$ that correspond to a cyclic deformation. (b) Non-dimensionalized volume density of the net work of stress $w/\mu$ as a function of the parameter $k$ in $\phi(I_1,I_2)=(I_2-3)^k$.} 
\label{Torsion-Cyclic}
\end{figure}

Let us consider an Ericksen material with a primary neo-Hookean potential $\psi(I_1,I_2)=\frac{\mu}{2}(I_1-3)$ and secondary potential $\phi(I_1,I_2)=(I_2-3)^k$.
As an example of a cyclic motion, we consider the following applied axial stretch and twist per unit length 
\begin{equation} \label{Torsion-Cyclic-Deformation}
	\lambda(t)=1+\lambda_0\sin\frac{k_1\pi\,t}{t_0}\,,\qquad 
	\psi(t)=1+\psi_0\sin\frac{k_2\pi\,t}{t_0}\,,\qquad t\in [0,t_0]\,,
\end{equation}
for some given constants $\lambda_0$ and $\psi_0$. We use the following numerical values for this cyclic loading: $\lambda_0=0.75$, $\psi_0=0.5$, $t_0=1.0$, $k_1=1$, and $k_2=2.0$. Fig.~\ref{Torsion-Cyclic}a shows two closed curves in the $(I_1,I_2)$ plane for $R=R_0$ and $R=0.75 R_0$ that correspond to this cyclic deformation. 

The non-dimensionalized density of the net work of stress is plotted in Fig.~\ref{Torsion-Cyclic}b as a function of the parameter $k$ that specifies the Ericksen potential of this material.
It should be noted that the cyclic deformation \eqref{Torsion-Cyclic-Deformation} is not a cyclic motion. However, the initial and final kinetic energies in a cycle of deformation are identical, and hence, the total work done on the body is equal to the work of stress, see Remark \ref{Cyclic-Deformation-Motion}.

\section{Linear Cauchy Elasticity} \label{Sec:Linear-Cauchy-elasticity}

Now that we have a full nonlinear theory, it is of interest to consider the linearized version of this theory to better understand the impact of non-hyperelasticity on the behaviour of elastic materials.
Linear elasticity is defined through the general stress-strain relationship at any material point:
\begin{equation} 
	\sigma_{ab}=\mathsf{C}_{abcd}\,\epsilon_{cd} \,,\qquad a,b=1,2,3\,,
\end{equation}
where  $\sigma_{ab}$ is the component of the Cauchy stress tensor in Cartesian coordinates, $\boldsymbol{\mathsf{C}}$ denotes the \textit{elasticity tensor}, $\epsilon_{ab}=\frac{1}{2}(u_{a,b}+u_{b,a})$ is the \textit{linear elastic strain},  and  summation over repeated indices is implied. 

The balance of angular momentum, $\sigma_{ab}=\sigma_{ba}$, along with the symmetry of the linearized strain, implies the following \textit{minor symmetries}:
\begin{equation} 
	\mathsf{C}_{abcd}=\mathsf{C}_{bacd}=\mathsf{C}_{abdc}
	 \,.
\end{equation}
In Green elasticity, the existence of a quadratic strain energy function in the linear strains implies the \textit{major symmetries}: $\mathsf{C}_{abcd} = \mathsf{C}_{cdab}$. However, in linear Cauchy elasticity, we have, in general $\mathsf{C}_{abcd}\neq \mathsf{C}_{cdab}$.\footnote{It has been pointed out by several researchers that relaxation functions in linear viscoelasticity do not have to possess the major symmetries and this does not violate the second law of thermodynamics \citep{RogersPipkin1963,Shunat1967,Day1971}.} Nevertheless, the elasticity tensor can be decomposed into symmetric and antisymmetric parts as follows
\begin{equation} 
	\mathsf{C}_{abcd}=\cCs_{abcd}+\cCa_{abcd} \,,
\end{equation}
where
\begin{equation} 
	\cCs_{abcd}=\frac{1}{2}\left[\mathsf{C}_{abcd}+\mathsf{C}_{cdab}\right]\,,\qquad
	\cCa_{abcd}=\frac{1}{2}\left[\mathsf{C}_{abcd}-\mathsf{C}_{cdab}\right]\,.
\end{equation}
It is important to note that there have been extensive works on characterizing anisotropic elasticity (including viscoelasticity) tensors that lack major symmetries \citep{RogersPipkin1963, Podio1987, Yong1991, He1996, Ostrosablin2017}. An important observation is that in dimension three, isotropic linear Cauchy elastic solids do not have any antisymmetric elastic constants \citep{RogersPipkin1963, Truesdell1964, Ostrosablin2017}. This observation has recently been re-discovered in the context of ``odd elasticity", which pertains to linear non-hyperelastic Cauchy elasticity; it has been observed that in $3$D isotropic elasticity, there are no ``odd elastic constants" \citep{Scheibner2020, Fruchart2023}.
Similarly, cubic Cauchy elastic solids also lack any antisymmetric elastic constants \citep{RogersPipkin1963, Ostrosablin2017}. 

Linear Cauchy elasticity encompasses eight distinct symmetry classes: triclinic, monoclinic, tetragonal, trigonal, orthotropic, transversely isotropic, cubic, and isotropic \citep{cowin1995anisotropic, Chadwick2001, ting2003generalized, cowin2007tissue,Ostrosablin2017}.
Since it is often easier to work with matrices, we can introduce the \textit{Voigt notation} through the bijection 
\begin{equation} 
	(11,22,33,23,31,12)\leftrightarrow(1,2,3,4,5,6)\,.
\end{equation}
Then the constitutive equations in this notation are given by 
\begin{equation} 
	\sigma_{\alpha}=\sum_{\beta=1}^6 C_{\alpha\beta}\,\epsilon_{\beta}
	=\sum_{\beta=1}^6(c_{\alpha\beta}+b_{\alpha\beta})\,\epsilon_{\beta} \,,\qquad 
	\alpha=1,\hdots,6\,.
\end{equation}
Here, $c_{\alpha\beta}=c_{\beta\alpha}$ and $b_{\alpha\beta}=-b_{\beta\alpha}$ represent the symmetric and antisymmetric $6\times 6$ stiffness matrices, respectively. Consequently, the elasticity matrix, as described in \citep{Ostrosablin2017}, has the following form:
\begin{equation} \label{elasticity-matrix}
    \mathbf{c}=\begin{bmatrix}
	c_{11} & c_{12}-b_{12} & c_{13}-b_{13} & c_{14}-b_{14} & c_{15}-b_{15} & c_{16}-b_{16} \\
	c_{12}+b_{12} & c_{22} & c_{23}-b_{23} & c_{24}-b_{24} & c_{25}-b_{25} & c_{26}-b_{26}\\
	c_{13}+b_{13} & c_{23}+b_{23} & c_{33} & c_{34}-b_{34} & c_{35}-b_{35} & c_{36}-b_{36}  \\
	c_{14}+b_{14} & c_{24}+b_{24} & c_{34}+b_{34} & c_{44} & c_{45}-b_{45} & c_{46}-b_{46}  \\
	c_{15}+b_{15} & c_{25}+b_{25} & c_{35}+b_{35} & c_{45}+b_{45} & c_{55} & c_{56}-b_{56}  \\
	c_{16}+b_{16} & c_{26}+b_{26} & c_{36}+b_{36} & c_{46}+b_{46} & c_{56}+b_{56} & c_{66}  
 \end{bmatrix}\,.
\end{equation}

\begin{remark}
A classical linear elastic solid is considered stable if its energy density is a positive-definite quadratic form, i.e., $\sum_{\alpha,\beta=1}^6 C_{\alpha\beta}\,\epsilon_{\alpha}\,\epsilon_{\beta}>0$, whenever at least one strain component is nonzero. This condition guarantees uniqueness of solutions in both linear hyperelastic elasticity and linear Cauchy elasticity \citep{Ogden1984}. 
However, note that $\sum_{\alpha,\beta=1}^6 C_{\alpha\beta}\,\epsilon_{\alpha}\,\epsilon_{\beta}=\sum_{\alpha,\beta=1}^6 c_{\alpha\beta}\,\epsilon_{\alpha}\,\epsilon_{\beta}>0$, which only depends on the symmetric constants. Hence, stability in this sense is determined by only the symmetric part of the elastic tensor.
\end{remark}

\subsection{The geometric hysteresis in linear Cauchy elasticity} \label{Berry-Phase}

The stress work $1$-form of linear elasticity is defined as 
\begin{equation} 
	\boldsymbol{\alpha}
	=\boldsymbol{\sigma}\!:\!\dot{\boldsymbol{\epsilon}}\,dt
	=\mathsf{C}_{abcd}\,\epsilon_{cd}\,\dot{\epsilon}_{ab}\,dt
	=C_{\alpha\beta}\,\epsilon_{\beta}\,\dot{\epsilon}_{\alpha}\,dt
	=\left[c_{\alpha\beta}+b_{\alpha\beta}\right] \epsilon_{\beta}\,\dot{\epsilon}_{\alpha}\,dt\
	=\left[c_{\alpha\beta}+b_{\alpha\beta}\right] \epsilon_{\beta}\,d\epsilon_{\alpha}\,.
\end{equation}
We consider a cyclic deformation in a time interval $[t_1,t_2]$ such that $\boldsymbol{\epsilon}(x,t_1)=\boldsymbol{\epsilon}(x,t_2)$ \citep{Sternberg1979}. A cyclic deformation is a path $\Gamma$ in the space of linearized strains. Let $\mathcal{U}\subset\mathcal{B}$ be a sub-body. Then, under a cyclic deformation, the net work done by the stress in the sub-body is
\begin{equation}
	W(\mathcal{U},\Gamma)
	= \int_{\mathcal{U}}\int_{\Gamma} \left[c_{\alpha\beta}+b_{\alpha\beta}\right] 
	\epsilon_{\beta}\,d\epsilon_{\alpha} \,dV\,.
\end{equation}
Let us assume that the closed curve $\Gamma$ is the boundary of a surface $\Omega$. Using Stokes' theorem and recalling that the elastic constants are homogeneous, one writes
\begin{equation}
	\int_{\Gamma} \left[c_{\alpha\beta}+b_{\alpha\beta}\right] \epsilon_{\beta}\,d\epsilon_{\alpha}
	=\int_{\Omega} \left[c_{\alpha\beta}+b_{\alpha\beta}\right] 
	d\epsilon_{\beta}\wedge d\epsilon_{\alpha}
	\,,
\end{equation}
where $\wedge$ is the wedge product of differential forms and summation  over repeated Greek indices $\beta<\alpha$ is assumed. Noting that $d\epsilon_{\beta}\wedge d\epsilon_{\alpha}=-d\epsilon_{\alpha}\wedge d\epsilon_{\beta}$ \citep{Flanders1963,Abraham2012} the above integral can be written as
\begin{equation}
	\int_{\Gamma} \left[c_{\alpha\beta}+b_{\alpha\beta}\right] \epsilon_{\beta}\,d\epsilon_{\alpha}
	=\int_{\Omega} b_{\alpha\beta}\, d\epsilon_{\beta}\wedge d\epsilon_{\alpha}
	\,,
\end{equation}
and hence
\begin{equation}
	W(\mathcal{U},\Gamma)
	= \int_{\mathcal{U}} \int_{\Omega} b_{\alpha\beta}\, d\epsilon_{\beta}\wedge d\epsilon_{\alpha}
	\,dV
	= \int_{\mathcal{U}}\int_{\Gamma} b_{\alpha\beta}\,\epsilon_{\beta}\,d\epsilon_{\alpha} \,dV
	= \int_{\mathcal{U}}\int_{t_1}^{t_2} b_{\alpha\beta}\,\epsilon_{\beta}\,\dot{\epsilon}_{\alpha}\,dt \,dV\,.
\end{equation}
We observe that only the antisymmetric elastic constants contribute to the net work of stress \citep{GreenNaghdi1971,Scheibner2020}. It is worth noting that a non-vanishing net work requires the area of $\Omega$ to be non-zero.

For isotropic and cubic Cauchy elastic solids, the net work always vanishes as there are no non-vanishing antisymmetric elastic constants.
Recently, it was proved that homogeneous displacements are universal for all linear Cauchy elastic solids \citep{YavariSfyris2024}. Denoting the volume of the sub-body by $|\mathcal{U}|$, for a homogeneous cyclic deformation we have 
\begin{equation}
	W(\mathcal{U},\Gamma)
	= |\mathcal{U}|\,\int_{t_1}^{t_2} b_{\alpha\beta}\,\epsilon_{\beta}\,\dot{\epsilon}_{\alpha}\,dt
	= |\mathcal{U}|\,\int_{\Gamma} b_{\alpha\beta}\,\epsilon_{\beta}\,d\epsilon_{\alpha}
	= |\mathcal{U}|\,\int_{\Omega} b_{\alpha\beta}\, d\epsilon_{\beta}\wedge d\epsilon_{\alpha}\,.
\end{equation}
Next we write the net stress-work density $w(\Gamma)=|\mathcal{U}|^{-1}\,W(\mathcal{U},\Gamma)$ explicitly for the remaining six symmetry classes.

\subsection{Experimental determination of antisymmetric elastic constants} \label{Experiments}

Since the net work does not vanish in  cyclic deformations, we can use it to extract information on the antisymmetric elastic constants and characterize them  experimentally. Here, we suggest a set of displacement-control loadings  to determine all the antisymmetric elastic constants for each symmetry class.

\begin{itemize}[topsep=2pt,noitemsep, leftmargin=10pt]
\item \emph{Triclinic Cauchy linear elastic solids}: For triclinic solids, the only symmetry transformations are the identity and its opposite. As a result, there are no constraints, apart from positive-definiteness, on the elastic constants in the elasticity matrix. Thus, there are a total of $36$ independent elastic constants. Recall that
\begin{equation}
	w(\Gamma)
	= b_{\alpha\beta}\,\int_{t_1}^{t_2} \epsilon_{\beta}(t)\,\dot{\epsilon}_{\alpha}(t)\,dt\,.
\end{equation}
Let us denote the set of even and odd positive natural numbers by $\mathbb{N}^e$ and $\mathbb{N}^o$, respectively.
Consider a displacement-control loading in which $\epsilon_{\xi}(t)$ and $\epsilon_{\eta}(t)$ ($\xi\neq \eta$) are the only non-zero strains.
We consider the following closed path in the strain space.
\begin{equation}\label{Cyclic-Deformation}
\begin{aligned}
	\Gamma_{\xi\eta} & :~ \left(\epsilon_{\xi}(t)\,, \epsilon_{\eta}(t)\right)
	= \left(\mathring{\epsilon}_{\xi}\,\sin\frac{\pi k_{\xi} t}{t_0}\,,
	\mathring{\epsilon}_{\eta}\,\sin\frac{\pi k_{\eta} t}{t_0}\right)\,, \qquad t\in[0,t_0]\,,
\end{aligned}
\end{equation}
where $\mathring{\epsilon}_{\xi}$ and $\mathring{\epsilon}_{\eta}$ are the constant strain amplitudes, $k_{\xi}, k_{\eta}\in\mathbb{N}$, and $k_{\xi}+ k_{\eta}\in \mathbb{N}^o$. Thus
\begin{equation}
	w(\Gamma_{\xi\eta})
	= b_{\xi\eta}\,\int_{0}^{t_0} \left[\epsilon_{\eta}(t)\,\dot{\epsilon}_{\xi}(t)
	-\epsilon_{\xi}(t)\,\dot{\epsilon}_{\eta}(t)\right] dt
	=\frac{4 k_{\xi} k_{\eta} \,\mathring{\epsilon}_{\xi}\, \mathring{\epsilon}_{\eta}}
	{k_{\eta}^2-k_{\xi}^2}\,b_{\xi\eta}\,.
\end{equation}
This implies that the fifteen antisymmetric elastic constants can be calculated using the net work of stress in the following fifteen different displacement-control cyclic deformations:
\begin{equation}
	b_{\xi\eta}	=\frac{k_{\eta}^2-k_{\xi}^2}
	{4 k_{\xi} k_{\eta} \,\mathring{\epsilon}_{\xi}\, \mathring{\epsilon}_{\eta}}
	\,w(\Gamma_{\xi\eta})\,,\qquad \xi,\eta=1,\hdots,6\,,~\xi<\eta\,.
\end{equation}

\begin{remark}
The external source/sink of energy can be time dependent. One can assume that the symmetric elastic constants are fixed, i.e., time-independent elastic response, while the antisymmetric elastic constants are time-dependent. This would be an example of a material evolution governed by active sources.
\end{remark}

\item \emph{Monoclinic Cauchy linear elastic solids}: In monoclinic solids, a plane of reflection symmetry exists. Without loss of generality, one can assume that the $x_3$-axis is perpendicular to the plane of material symmetry. A monoclinic solid has $20$ independent elastic constants, comprising $13$ symmetric and $7$ antisymmetric ones. In Cartesian coordinates $(x_1,x_2,x_3)$, the seven non-zero antisymmetric elastic constants are $b_{12}, b_{13}, b_{23}, b_{16}, b_{26}, b_{36}$, and $b_{45}$, and the elasticity matrix has the following form \citep{Ostrosablin2017}:\footnote{The elasticity matrices for the trigonal, hexagonal, and transversely isotropic cases in \citep{Ostrosablin2017} differ slightly from the more commonly used versions in the literature. Specifically, in \citet{Ostrosablin2017}’s matrices, the $(6,6)$ component for both the trigonal and transversely isotropic cases lacks a factor of $\frac{1}{2}$, and in the trigonal case, the $(4,6)$ and $(6,4)$ components include a factor of $\sqrt{2}$. These minor differences arise because \citet{Ostrosablin2017} adopts the Kelvin notation rather than the Voigt notation.}
\begin{equation}
    \mathbf{c}=\begin{bmatrix}
    c_{11} & c_{12}-b_{12} & c_{13}-b_{13} & 0 & 0 & c_{16}-b_{16} \\
    c_{12}+b_{12} & c_{22} & c_{23}-b_{23} & 0 & 0 & c_{26}-b_{26}  \\
    c_{13}+b_{13} & c_{23}+b_{23} & c_{33} & 0 & 0 & c_{36}-b_{36}  \\
    0 & 0 & 0 & c_{44} & c_{45}-b_{45} & 0  \\
    0 & 0 & 0 & c_{45}+b_{45} & c_{55} & 0  \\
    c_{16}+b_{16} & c_{26}+b_{26} & c_{36}+b_{36} & 0 & 0 & c_{66}  
    \end{bmatrix}\,.
\end{equation}
Thus
\begin{equation}
\begin{aligned}
	w(\Gamma) &= \int_{t_1}^{t_2} \Big [
	b_{12}\left(\epsilon_{22}\dot{\epsilon}_{11}-\epsilon_{11}\dot{\epsilon}_{22} \right)+
	b_{13}\left(\epsilon_{33}\dot{\epsilon}_{11}-\epsilon_{11}\dot{\epsilon}_{33} \right)+
	b_{23}\left(\epsilon_{33}\dot{\epsilon}_{22}-\epsilon_{22}\dot{\epsilon}_{33} \right)\\
	& \qquad\qquad+  b_{16}\left(\epsilon_{12}\dot{\epsilon}_{11}-\epsilon_{11}\dot{\epsilon}_{12} \right) 
	+b_{26}\left(\epsilon_{12}\dot{\epsilon}_{22}-\epsilon_{22}\dot{\epsilon}_{12} \right) \\
	& \qquad\qquad+b_{36}\left(\epsilon_{12}\dot{\epsilon}_{33}-\epsilon_{33}\dot{\epsilon}_{12} \right)+
	b_{45}\left(\epsilon_{31}\dot{\epsilon}_{23}-\epsilon_{23}\dot{\epsilon}_{31} \right)
	\Big] dt \,.
\end{aligned}
\end{equation}
The seven antisymmetric elastic constants can again  be calculated using the cyclic deformations given in \eqref{Cyclic-Deformation}. \\

\item \emph{Tetragonal Cauchy linear elastic solids}: In a tetragonal solid, there are five planes of symmetry. Four of these planes have coplanar normals, while the fifth plane is perpendicular to the other four. One can assume, without loss of generality, that in the Cartesian coordinate system $(x_1,x_2,x_3)$, the normal corresponding to the fifth plane of symmetry aligns with the $x_3$-axis, with its corresponding plane of symmetry parallel to the $x_1x_2$-plane. The first two planes of symmetry are parallel to the $x_1x_3$ and $x_2x_3$-planes, respectively. The remaining two planes of symmetry are related to those parallel to the $x_1x_3$-plane through rotations of $\frac{\pi}{4}$ and $\frac{3\pi}{4}$ about the $x_3$-axis. A tetragonal solid possesses $7$ independent elastic constants, comprising $6$ symmetric and $1$ antisymmetric constants. The only antisymmetric elastic constant is $b_{13}$ and $b_{23}=b_{13}$ \citep{Ostrosablin2017}, and  the elasticity matrix has the following form:
\begin{equation}
    \mathbf{c}=\begin{bmatrix}
    c_{11} & c_{12} & c_{13}-b_{13} & 0 & 0 & 0 \\
    c_{12} & c_{11} & c_{13}-b_{13} & 0 & 0 & 0  \\
    c_{13}+b_{13} & c_{13}+b_{13} & c_{33} & 0 & 0 & 0  \\
    0 & 0 & 0 & c_{44} & 0 & 0  \\
    0 & 0 & 0 & 0 & c_{44} & 0  \\
    0 & 0 & 0 & 0 & 0 & c_{66}  
    \end{bmatrix}\,.
\end{equation}
Hence
\begin{equation}
	w(\Gamma)
	= b_{13} \int_{t_1}^{t_2} \left[\epsilon_{33}\left(\dot{\epsilon}_{11}+\dot{\epsilon}_{22}\right) 
	+ \dot{\epsilon}_{33}\left(\epsilon_{22}-\epsilon_{11}\right)  \right] dt \,.
\end{equation}
The elastic constant $b_{13}$ can be calculated using the cyclic deformation $\Gamma_{13}$ corresponding to $\xi=1$ and $\eta=3$ in \eqref{Cyclic-Deformation}. \\

\item \emph{Trigonal Cauchy linear elastic solids}: In a trigonal solid, there are three planes of symmetry, where the normals of these planes lie in the same plane and are related via $\frac{\pi}{3}$ rotations. Essentially, two of the planes of symmetry are obtained from the third one through rotations about a fixed axis by $\frac{\pi}{3}$ and $-\frac{\pi}{3}$. In a Cartesian coordinate system $(x_1,x_2,x_3)$, let us assume that the $x_3$-axis is the trigonal axis. A trigonal solid has $8$ independent elastic constants: $6$ are symmetric and $2$ are antisymmetric. In the Cartesian coordinates $(x_1,x_2,x_3)$, the non-zero antisymmetric elastic constants are $b_{13}, b_{15}$, $b_{25}=-b_{15}$, and $b_{46}=b_{15}$, and the elasticity matrix has the following form \citep{Ostrosablin2017}:
\begin{equation}
    \mathbf{c}=\begin{bmatrix}
    c_{11} & c_{12} & c_{13}-b_{13} & 0 & c_{15}-b_{15} & 0 \\
    c_{12} & c_{11} & c_{13}-b_{13} & 0 & -c_{15}+b_{15} & 0  \\
    c_{13}+b_{13} & c_{13}+b_{13} & c_{33} & 0 & 0 & 0  \\
    0 & 0 & 0 & c_{44} & 0 & -c_{15}-b_{15}  \\
    c_{15}+b_{15} & -c_{15}-b_{15} & 0 & 0 & c_{44} & 0  \\
    0 & 0 & 0 & -c_{15}+b_{15} & 0 & \frac{1}{2}(c_{11}-c_{12})  
    \end{bmatrix}\,.
\end{equation}
Thus
\begin{equation}
\begin{aligned}
	w(\Gamma)= \int_{t_1}^{t_2} \left\{
	b_{13}\left(\epsilon_{33}\dot{\epsilon}_{11}-\epsilon_{11}\dot{\epsilon}_{33} \right)
	+b_{15}\left[  \epsilon_{31}\left(\dot{\epsilon}_{11}-\dot{\epsilon}_{22}\right)
	+\dot{\epsilon}_{31}\left(\epsilon_{22}-\epsilon_{11}\right)
	+\epsilon_{12}\dot{\epsilon}_{23}-\epsilon_{23}\dot{\epsilon}_{12}
	\right]  \right\} dt 
	\,.
\end{aligned}
\end{equation}
The elastic constant $b_{13}$ can be calculated using the cyclic deformation $\Gamma_{13}$ corresponding to $\xi=1$ and $\eta=3$ in \eqref{Cyclic-Deformation}. 
For calculating $b_{15}$, the following cyclic deformations are used:
\begin{equation}
\begin{aligned}
	\Gamma: &\quad  \epsilon_{11}(t)=\mathring{\epsilon}_{11}\,\sin\frac{\pi k_1 t}{t_0}\,,\qquad
	\epsilon_{22}(t)=\mathring{\epsilon}_{22}\,\sin\frac{\pi k_1 t}{t_0}\,,\qquad
	\epsilon_{33}(t)=0\,,\\
	& \quad \epsilon_{23}(t)=\mathring{\epsilon}_{23}\,\sin\frac{\pi k_4 t}{t_0}\,,\qquad
	\epsilon_{13}(t)=\mathring{\epsilon}_{13}\,\sin\frac{\pi k_4 t}{t_0}\,,\qquad
	\epsilon_{12}(t)=\mathring{\epsilon}_{12}\,\sin\frac{\pi k_4 t}{t_0}\,,\qquad t\in[0,t_0]
	\,,
\end{aligned}
\end{equation}
where $\mathring{\epsilon}_{11}$, $\mathring{\epsilon}_{22}$, $\mathring{\epsilon}_{23}$, $\mathring{\epsilon}_{13}$, and $\mathring{\epsilon}_{12}$ are constant strain amplitudes, $k_1, k_4\in\mathbb{N}$, and $k_1+ k_4\in \mathbb{N}^o$. Thus
\begin{equation}
	b_{15} =
	\frac{k_4^2-k_1^2}
	{4 k_1 k_4 \,\mathring{\epsilon}_{13} (\mathring{\epsilon}_{11}-\mathring{\epsilon}_{22})}
	\,w(\Gamma)
	\,.
\end{equation}\\

\item \emph{Orthotropic Cauchy linear elastic solids}: An orthotropic solid possesses three mutually orthogonal symmetry planes. We assume that these coincide with the coordinate planes in the Cartesian coordinates $(x_1,x_2,x_3)$.
An orthotropic solid has $12$ independent elastic constants, consisting of $9$ symmetric and $3$ antisymmetric ones. In the Cartesian coordinates $(x_1,x_2,x_3)$, the three non-zero antisymmetric elastic constants are $b_{12}, b_{13}$, and $b_{23}$, and the elasticity matrix has the following representation \citep{Ostrosablin2017}:
\begin{equation}
    \mathbf{c}=\begin{bmatrix}
    c_{11} & c_{12}-b_{12} & c_{13}-b_{13} & 0 & 0 & 0 \\
    c_{12}+b_{12} & c_{22} & c_{23}-b_{23} & 0 & 0 & 0  \\
    c_{13}+b_{13} & c_{23}+b_{23} & c_{33} & 0 & 0 & 0  \\
    0 & 0 & 0 & c_{44} & 0 & 0  \\
    0 & 0 & 0 & 0 & c_{55} & 0  \\
    0 & 0 & 0 & 0 & 0 & c_{66}  
    \end{bmatrix}\,.
\end{equation}
Hence
\begin{equation}
\begin{aligned}
	w(\Gamma)= \int_{t_1}^{t_2} \left [
	b_{12}\left(\epsilon_{22}\dot{\epsilon}_{11}-\epsilon_{11}\dot{\epsilon}_{22} \right)+
	b_{13}\left(\epsilon_{33}\dot{\epsilon}_{11}-\epsilon_{11}\dot{\epsilon}_{33} \right)+
	b_{23}\left(\epsilon_{33}\dot{\epsilon}_{22}-\epsilon_{22}\dot{\epsilon}_{33} \right)
	\right]   dt 
	\,.
\end{aligned}
\end{equation}
Consider a homogeneous body made of an orthotropic Cauchy linear elastic solid undeformed at time $t=0$ on which we apply the following family of triaxial displacement-control loadings:
\begin{equation}
\begin{aligned}
	\Gamma:~ \left(\epsilon_{11}(t)\,, \epsilon_{22}(t)\,, \epsilon_{33}(t)\right)
	= \left(\mathring{\epsilon}_{11}\,\sin\frac{\pi k_1 t}{t_0}\,,
	\mathring{\epsilon}_{22}\,\sin\frac{\pi k_2 t}{t_0}\,,
	\mathring{\epsilon}_{33}\,\sin\frac{\pi k_3 t}{t_0}\right)\,,\qquad t\in[0,t_0]
	\,,
\end{aligned}
\end{equation}
where $\mathring{\epsilon}_{11}$, $\mathring{\epsilon}_{22}$, and $\mathring{\epsilon}_{33}$ are constant strain amplitudes.
For $k_1, k_2,k_3\in\mathbb{N}$, these are cyclic loadings. For $k_1=k_2=k_3$, the net work of stress is zero. We assume that $k_1, k_2,k_3$ are three distinct positive integers. Thus
\begin{equation}
\begin{aligned}
	w(\Gamma) &=
	\frac{2 k_1 k_2 \,\mathring{\epsilon}_{11} \mathring{\epsilon}_{22} 
	\left[(-1)^{k_1+k_2}-1\right]}{k_1^2-k_2^2}\,b_{12}
	 +\frac{2 k_1 k_3 \,\mathring{\epsilon}_{11} \mathring{\epsilon}_{33}
	\left[(-1)^{k_1+k_3}-1\right]}{k_1^2-k_3^2}\,b_{13}\\
	&\quad
	+\frac{2 k_2 k_3 \,\mathring{\epsilon}_{22} \mathring{\epsilon}_{33}
	\left[(-1)^{k_2+k_3}-1\right]}{k_2^2-k_3^2}\,b_{23}
	\,.
\end{aligned}
\end{equation}
The above expression suggests that one can calculate the three antisymmetric elastic constants using the following three loading paths:
\begin{equation}
\begin{aligned}
	\Gamma_1 & :~ \left(\epsilon_{11}(t)\,, \epsilon_{22}(t)\,, \epsilon_{33}(t)\right)
	= \left(\mathring{\epsilon}_{11}\,\sin\frac{\pi k_1 t}{t_0}\,,
	\mathring{\epsilon}_{22}\,\sin\frac{\pi k_2 t}{t_0}\,,0\right)\,,\quad k_1+k_2\in \mathbb{N}^o\,, \\
	\Gamma_2 &:~ \left(\epsilon_{11}(t)\,, \epsilon_{22}(t)\,, \epsilon_{33}(t)\right)
	= \left(\mathring{\epsilon}_{11}\,\sin\frac{\pi k_1 t}{t_0}\,,0\,,
	\mathring{\epsilon}_{33}\,\sin\frac{\pi k_3 t}{t_0}\right)\,,\quad k_1+k_3\in \mathbb{N}^o\,,\\
	\Gamma_3 &:~ \left(\epsilon_{11}(t)\,, \epsilon_{22}(t)\,, \epsilon_{33}(t)\right)
	= \left(0\,,	\mathring{\epsilon}_{22}\,\sin\frac{\pi k_2 t}{t_0}\,,
	\mathring{\epsilon}_{33}\,\sin\frac{\pi k_3 t}{t_0}\right)\,,\quad k_2+k_3\in \mathbb{N}^o\,.
\end{aligned}
\end{equation}
Using the above cyclic deformations one obtains
\begin{equation}
\begin{aligned}
	b_{12} = \frac{k_2^2-k_1^2}
	{4 k_1 k_2 \,\mathring{\epsilon}_{11} \mathring{\epsilon}_{22} }\,w(\Gamma_1)\,,\qquad
	b_{13} = \frac{k_3^2-k_1^2}
	{4 k_1 k_3 \,\mathring{\epsilon}_{11} \mathring{\epsilon}_{33}}\,w(\Gamma_2)\,,\qquad
	b_{23} = \frac{k_2^2-k_3^2}
	{4 k_2 k_3\, \mathring{\epsilon}_{22} \mathring{\epsilon}_{33}}\,w(\Gamma_3)
	\,.
\end{aligned}
\end{equation}

\item \emph{Transversely isotropic Cauchy linear elastic solids}: A transversely isotropic solid has an axis of symmetry such that planes perpendicular to it are isotropy planes. Let us assume that the axis of transverse isotropy corresponds to the $x_3$-axis in the Cartesian coordinates $(x_1, x_2, x_3)$. A transversely isotropic solid is characterized by $8$ independent elastic constants ($5$ symmetric and $3$ antisymmetric). In the Cartesian coordinates $(x_1, x_2, x_3)$, the three non-zero antisymmetric elastic constants are  $b_{13}$, $b_{16}$, $b_{26}=-b_{16}$, and $b_{45}$. The elasticity matrix is represented as \citep{RogersPipkin1963,Ostrosablin2017}:
\begin{equation}
    \mathbf{c}=\begin{bmatrix}
    c_{11} & c_{12} & c_{13}-b_{13} & 0 & 0 & -b_{16} \\
    c_{12} & c_{11} & c_{13}-b_{13} & 0 & 0 & b_{16}  \\
    c_{13}+b_{13} & c_{13}+b_{13} & c_{33} & 0 & 0 & 0  \\
    0 & 0 & 0 & c_{44} & -b_{45} & 0  \\
    -b_{16} & 0 & 0 & b_{45} & c_{44} & 0  \\
    b_{16} & 0 & 0 & 0 & 0 & \frac{1}{2}(c_{11}-c_{12})  
    \end{bmatrix}\,.
\end{equation}
Thus
\begin{equation}
\begin{aligned}
	w(\Gamma)= \int_{t_1}^{t_2} \left\{
b_{13}\left(\epsilon_{33}\dot{\epsilon}_{11}-\epsilon_{11}\dot{\epsilon}_{33} \right)+
	b_{16}\left[ \epsilon_{12}\left(\dot{\epsilon}_{11}-\dot{\epsilon}_{22}\right)
	+\dot{\epsilon}_{12}\left(\epsilon_{22}-\epsilon_{11}\right)	\right]
	+b_{45}\left(\epsilon_{31}\dot{\epsilon}_{23}-\epsilon_{23}\dot{\epsilon}_{31} \right)
	  \right\} dt 
	\,.
\end{aligned}
\end{equation}
Consider a homogeneous body made of a transversely isotropic Cauchy linear elastic solid undeformed at time $t=0$.
Let us consider the following family of triaxial displacement-control loadings:
\begin{equation}
\begin{aligned}
	\Gamma: &\quad  \epsilon_{11}(t)=\mathring{\epsilon}_{11}\,\sin\frac{\pi k_1 t}{t_0}\,,\qquad
	\epsilon_{22}(t)=\mathring{\epsilon}_{22}\,\sin\frac{\pi k_2 t}{t_0}\,,\qquad
	\epsilon_{33}(t)=\mathring{\epsilon}_{33}\,\sin\frac{\pi k_3 t}{t_0}\,,\\
	& \quad \epsilon_{23}(t)=\mathring{\epsilon}_{23}\,\sin\frac{\pi k_4 t}{t_0}\,,\qquad
	\epsilon_{13}(t)=\mathring{\epsilon}_{13}\,\sin\frac{\pi k_5 t}{t_0}\,,\qquad
	\epsilon_{12}(t)=\mathring{\epsilon}_{12}\,\sin\frac{\pi k_6 t}{t_0}\,,\qquad t\in[0,t_0]
	\,,
\end{aligned}
\end{equation}
where $\mathring{\epsilon}_{11}$, $\mathring{\epsilon}_{22}$, $\mathring{\epsilon}_{33}$, $\mathring{\epsilon}_{23}$, $\mathring{\epsilon}_{13}$, and $\mathring{\epsilon}_{12}$ are constant strain amplitudes.
Let us consider the following three loading paths:
\begin{equation}
\begin{aligned}
	\Gamma_1 &:~ \left(\epsilon_{11}(t)\,,  \epsilon_{33}(t)\right)
	= \left(\mathring{\epsilon}_{11}\,\sin\frac{\pi k_1 t}{t_0}\,,
	\mathring{\epsilon}_{33}\,\sin\frac{\pi k_3 t}{t_0}\right)\,,\quad k_1+k_3\in \mathbb{N}^o\,,\\
	& \qquad \epsilon_{22}(t)=\epsilon_{23}(t)=\epsilon_{13}(t)=\epsilon_{12}(t)=0\,,\\
	\Gamma_2 & :~ \left(\epsilon_{13}(t)\,,  \epsilon_{23}(t)\right)
	= \left(\mathring{\epsilon}_{13}\,\sin\frac{\pi k_4 t}{t_0}\,,
	\mathring{\epsilon}_{23}\,\sin\frac{\pi k_5 t}{t_0}\right)\,,\quad k_4+k_5\in \mathbb{N}^o\,,\\
	& \qquad \epsilon_{11}(t)=\epsilon_{22}(t)=\epsilon_{33}(t)=\epsilon_{12}(t)=0\,,\\
	\Gamma_3 &:~ \left(\epsilon_{11}(t)\,, \epsilon_{22}(t)\,, \epsilon_{12}(t)\right)
	= \left(\mathring{\epsilon}_{11}\,\sin\frac{\pi k_1 t}{t_0}\,,	
	\mathring{\epsilon}_{22}\,\sin\frac{\pi k_1 t}{t_0}\,,
	\mathring{\epsilon}_{12}\,\sin\frac{\pi k_6 t}{t_0}\right)\,,\quad k_1+k_6\in \mathbb{N}^o\,,\\
	& \qquad \epsilon_{33}(t)=\epsilon_{23}(t)=\epsilon_{13}(t)=0\,.
\end{aligned}
\end{equation}
Using the above cyclic deformations we find that
\begin{equation}
\begin{aligned}
	b_{13} = \frac{k_3^2-k_1^2}
	{4 k_1 k_3\, \mathring{\epsilon}_{11} \mathring{\epsilon}_{33}}\,w(\Gamma_1)\,,\qquad
	b_{45} = \frac{k_5^2-k_4^2}
	{4 k_4 k_5 \,\mathring{\epsilon}_{23} \mathring{\epsilon}_{13} }\,w(\Gamma_2)\,,\qquad
	b_{16} = \frac{k_6^2-k_1^2}
	{4 k_1 k_6 \,\mathring{\epsilon}_{12} (\mathring{\epsilon}_{22}-\mathring{\epsilon}_{11})}
	w(\Gamma_3)
	\,.
\end{aligned}
\end{equation}

\end{itemize}

\subsection{From Cauchy elasticity to odd elasticity}

Recently, there has been interest in the physics literature on Cauchy elasticity and what has been called ``\textit{odd elasticity}" \citep{Scheibner2020}. These authors are motivated by modeling of active matter, a fascinating topic with new experiments that has inspired us to revisit and revive the study of Cauchy elasticity. This new effort is welcome. Yet, we would like to clarify that odd elasticity is not a new field of elasticity; it is \textit{identical} to linear Cauchy elasticity, as first proposed by Cauchy himself. We would also like to stress that despite the possibility of new behaviors that are not captured by Green elasticity, some of the tenets of solid mechanics cannot be arbitrarily modified (and do not need to be modified). In particular, balance of energy, balance of mass, balance of linear momentum, and balance of angular momentum, are all non-negotiable. These balances exist for all materials. More fundamentally, despite some claims to the contrary, objectivity cannot be violated (for the simple reason that, without objectivity, two different observers recording the same experiment may obtain different elastic constants). Any statement to the contrary amounts to a fundamental misunderstanding of continuum mechanics.

\section{Cauchy Anelasticity} \label{Sec:CauchyAnelasticity} 

In anelasticity (in the sense of \citet{Eckart1948}\footnote{\citet{Eckart1948} realized that the natural configuration (material manifold) of an anelastic solid is a Riemannian manifold with its metric explicitly depending on the local anelastic distorsions. The connection between residually-stressed solids and Riemannian geometry was independently discovered by \citet{Kondo1949}, who later coined the term ``material manifold" \citep{kondo1950dislocation}.}), the deformation gradient is multiplicatively decomposed as $\mathbf{F}=\Fe\Fa$, where $\Fe$ and $\Fa$ are the elastic and anelastic distortions, respectively.\footnote{A one-dimensional analogue of $\mathbf{F}=\Fe\Fa$, where $\Fa$ represents the swelling part of the deformation gradient, was first introduced by \citet{Flory1944}. In finite plasticity, the multiplicative decomposition first appeared in \citep[Page 41, Eq. (12)]{Bilby1957} and \citep[Page 286, Eq. (4)]{Kroner1959}. A decade later, this decomposition was popularized in the plasticity literature by \citet{Lee1967} and \citet{Lee1969}. For more detailed discussions of the multiplicative decomposition, see \citep{,goriely17,SadikYavari2017,YavariSozio2023}.} In hyper-anelasticity, the energy function explicitly depends on the elastic distortion $W=W(\Fe,\mathring{\mathbf{G}},\mathbf{g})$, where $\mathring{\mathbf{G}}=\mathbf{g}\big|_{\mathcal{B}}$ is the metric induced from the ambient space Euclidean metric. This is the material metric in the absence of anelastic distortions. If the material is anisotropic, one instead has $W=W(\Fe,\mathring{\mathbf{G}},\mathring{\boldsymbol{\Lambda}},\mathbf{g})$, where $\mathring{\boldsymbol{\Lambda}}$ is a set of structural tensors in the absence of anelastic distortions. Adding the structural tensors to the arguments of the energy function makes it an isotropic function of its arguments. One can show that $W=W(\mathbf{F},\mathbf{G},\boldsymbol{\Lambda},\mathbf{g})$, where $\mathbf{G}=\Fa^*\mathring{\mathbf{G}}$ and $\boldsymbol{\Lambda}=\Fa^*\mathring{\boldsymbol{\Lambda}}$ \citep{YavariSozio2023}. Objectivity implies that $W=\hat{W}(\Ce^\flat,\mathring{\mathbf{G}},\mathring{\boldsymbol{\Lambda}})=\hat{W}(\mathbf{C}^\flat,\mathbf{G},\boldsymbol{\Lambda})$.

\begin{defi}
For a Cauchy anelastic body, $\mathbf{S}=\mathbf{S}(\Ce^\flat,\mathring{\mathbf{G}},\mathring{\boldsymbol{\Lambda}},\mathbf{g})$ and the stress work $1$-form is written as
\begin{equation}
	\boldsymbol{\Omega}=\frac{1}{2}\mathbf{S}\!:\!d\Ce^{\flat}
	\,.
\end{equation}
\end{defi}

The Darboux classification theorem tells us that $\boldsymbol{\Omega}$ takes one of the following six canonical forms:
\begin{empheq}[left={\empheqlbrace }]{align}
	\label{Cauchy-A1} 
	\boldsymbol{\Omega} & =  d\psi_1\,, \\
	\label{Cauchy-A2} 
	\boldsymbol{\Omega} & =  \phi_1 d\psi_1\,, \\
	\label{Cauchy-A3} 
	\boldsymbol{\Omega} & =  \phi_1 d\psi_1+d\psi_2\,, \\
	\label{Cauchy-A4} 
	\boldsymbol{\Omega} & =  \phi_1 d\psi_1+\phi_2 d\psi_2\,, \\
	\label{Cauchy-A5} 
	\boldsymbol{\Omega} & = \phi_1 d\psi_1+\phi_2 d\psi_2+d\psi_3\,,  \\
	\label{Cauchy-A6} 
	\boldsymbol{\Omega} & =  \phi_1 d\psi_1+\phi_2 d\psi_2+\phi_3 d\psi_3
	\,,
\end{empheq}
where $\psi_i=\psi_i(X,\Ce^\flat,\mathring{\boldsymbol{\Lambda}},\mathring{\mathbf{G}})$ and $\phi_i=\phi_i(X,\Ce^\flat,\mathring{\boldsymbol{\Lambda}},\mathring{\mathbf{G}})$, $i=1,2,3$. Each generalized energy function is an isotropic function of its arguments, and hence
\begin{equation}
	\psi_i=\psi_i(X,\mathbf{C}^\flat,\boldsymbol{\Lambda},\mathbf{G})\,,\qquad 
	\phi_i=\phi_i(X,\mathbf{C}^\flat,\boldsymbol{\Lambda},\mathbf{G})\,, \qquad i=1,2,3
	\,.
\end{equation}

\paragraph{Isotropic Cauchy anelasticity.} For an isotropic Cauchy anealstic solid $W=\hat{W}(\Ce^\flat,\mathring{\mathbf{G}})=\hat{W}(\mathbf{C}^\flat,\mathbf{G})=\overline{W}(I_1,I_2,I_3)$, where $I_i$, $i=1,2,3$  are the principal invariants of $\mathbf{C}^\flat$ calculated using the material metric $\mathbf{G}$. More specifically
\begin{equation} \label{Principal-Invariants-Anelasticity}
\begin{aligned}
	I_1 &=\operatorname{tr}\mathbf{C}=C^A{}_A=C_{AB}\,G^{AB}\,,\\
	I_2 &=\frac{1}{2}\left(I_1^2-\operatorname{tr}\mathbf{C}^2\right)
	=\frac{1}{2}\left(I_1^2-C^A{}_B\,C^B{}_A\right)\,,\\
	I_3 &=\det \mathbf{C}=\frac{\det \mathbf{C}^\flat}{\det \mathbf{G}}\,.
\end{aligned}
\end{equation}
Similar to isotropic Cauchy elasticity, in isotropic Cauchy anelasticity the stress work $1$-form takes one of the three canonical forms given in \eqref{Isotropic-Cauchy-1}-\eqref{Isotropic-Cauchy-3}, however with the difference that the principal invariants are given in \eqref{Principal-Invariants-Anelasticity}.

\begin{remark}
The stress work $1$-form in anisotropic Cauchy anelasticity has canonical forms similar to those in the corresponding anisotropic Cauchy anelasticity, provided that the material metric (instead of the flat Euclidean metric) is used to calculate them.
\end{remark}

\subsection{Eshelby's inclusion problem in Cauchy anelasticity}

For infinite bodies in the setting of linear elasticity, the first study of eigenstrains and the resulting stress fields was conducted by \citet{Eshelby1957}. 
The earliest three-dimensional investigation of stress fields of inclusions in nonlinear solids was carried out by \citet{DianiParks2000}, who used a multiplicative decomposition $\mathbf{F}=\mathbf{F}^e\mathbf{F}^*$. Their finite element simulations revealed a uniform hydrostatic stress within a spherical inclusion subjected to pure dilatational eigenstrains. This was later analytically proven for incompressible isotropic solids and certain classes of compressible isotropic solids in \citep{YavariGoriely2013b}, and for transversely isotropic and orthotropic solids in \citep{Golgoon2018a}. 
Several other analytical studies have further explored the stress fields of inclusions in both isotropic and anisotropic nonlinear elastic solids \citep{Yavari2015Singularity,Yavari2015twist,Golgoon2016,Yavari2021Eshelby}. 
These exact solutions do not depend on the existence of an energy function. Therefore, they can be readily extended to Cauchy elasticity.\footnote{Truesdell had a similar observation; in \citep{Truesdell1964}, he writes ``There are also a number of exact solutions of the general equations of equilibrium, solutions which yield specific predictions suitable for comparison with measurements on large strain. While the discoverers of these solutions all considered a hyperelastic material, I have examined their results and have found that every one, without exception, is easily extended to the more general theory." }

\subsection{Stress fields of line and point defects in Cauchy elastic solids}

The presence of distributed defects in solids induces anelastic distortions. Vito Volterra, in his seminal work \citep{volterra1907equilibre,Delphenich2020}, established the mathematical foundations of line defects in solids long before these defects were first experimentally observed \citep{Taylor1934,Orowan1934,Polanyi1934}. He categorized line defects into six types, three of which are now known as dislocations (translational defects), while the remaining three are referred to as disclinations (rotational defects).
In the 1950s, \citet{kondo1955geometry,kondo1955non} and \citet{bilby1955continuous} independently explored the deep connections between the mechanics of defects and non-Riemannian geometries. \citet{kondo1955geometry,kondo1955non} demonstrated that, in the presence of defects, the reference configuration of a solid may be non-Euclidean. He further recognized that the curvature and torsion of the material manifold serve as measures of incompatibility and the density of dislocations, respectively.
There are several exact solutions in the literature for line and point defects in hyperelastic solids, e.g., dislocations \citep{Gairola1979,RosakisRosakis1988,Zubov1997,Acharya2001,YavariGoriely2012a}, disclinations \citep{Zubov1997,Derezin2011,YavariGoriely2013a}, point defects \citep{YavariGoriely2012b}, dispirations \citep{Yavari2016Dispiration}, discombinations (combinations of line and point defects) \citep{YavariGoriely2014}, and distributed defects in anisotropic solids \citep{Golgoon2018b}.
These exact solutions do not depend on the existence of an energy function. Therefore, they can be readily extended to Cauchy elasticity.

\section{Cosserat-Cauchy Elasticity} \label{Sec:GeneralizedCauchyElasticity}

The first systematic formulation of generalized continua dates back to the pioneering work of the Cosserat brothers \citep{Cosserat1909}.
Cosserat elasticity remained largely unnoticed until the renewed interest in continua with microstructure during the 1950s, 1960s, and 1970s \citep{Ericksen1957, Toupin1962, Toupin1964, MindlinTiersten1962, Mindlin1964, Eringen2012}. Today, a vast body of literature exists on generalized continua.
The existing studies on generalized continua assume the existence of an energy function. In the following, we briefly discuss an extension of Cosserat elasticity when there is no underlying energy function---\emph{Cosserat-Cauchy elasticity}.

In Cosserat elasticity, it is assumed that the microstructure in the deformed configuration is characterized by three (or two in $2$D) linearly independent vectors, denoted as $\{\director{\boldsymbol{\mathsf{d}}}(x,t),~\mathfrak{a}=1,2,3\}$, referred to as director fields or simply directors \citep{Ericksen1957}.
The corresponding directors in the reference configuration are denoted by $\left\{\director{\boldsymbol{\mathsf{D}}}(X),~\mathfrak{a}=1,2,3\right\}$. 
It should be emphasized that these are not material vectors, as $\director{\boldsymbol{\mathsf{d}}}(x,t)\neq (\varphi_*\director{\boldsymbol{\mathsf{D}}})(x,t)$, in general.
The kinematics of a Cosserat elastic body is described by the pair$\left(\varphi(X,t),\director{\boldsymbol{\mathsf{d}}}(X,t)\right)$, where $\director{\boldsymbol{\mathsf{d}}}(X,t)$ is shorthand for $\director{\boldsymbol{\mathsf{d}}}(\varphi(X,t),\director{\boldsymbol{\mathsf{D}}}(X))$ \citep{Toupin1964, Stojanovic1970}.
The reciprocal of $\underaccent{\mathfrak{a}}{\boldsymbol{\mathsf{D}}}$ is denoted by $\accentset{\mathfrak{a}}{\boldsymbol{\Theta}}$ that satisfy the relations $\underaccent{\mathfrak{a}}{\mathsf{D}}^A\accentset{\mathfrak{a}}{\Theta}_B=\delta^A_B$ and $\underaccent{\mathfrak{a}}{\mathsf{D}}^A\accentset{\mathfrak{b}}{\Theta}_A=\delta^{\mathfrak{b}}_{\mathfrak{a}}$.
Similarly, the reciprocal of $\underaccent{\mathfrak{a}}{\boldsymbol{\mathsf{d}}}$ is denoted by $\accentset{\mathfrak{a}}{\boldsymbol{\vartheta}}$, satisfying the relations $\underaccent{\mathfrak{a}}{\mathsf{d}}^a\accentset{\mathfrak{a}}{\vartheta}_b=\delta^a_b$ and $\underaccent{\mathfrak{a}}{\mathsf{d}}^a\accentset{\mathfrak{b}}{\vartheta}_a=\delta^{\mathfrak{b}}_{\mathfrak{a}}$.
The referential directors can vary from point to point, and $\underaccent{\mathfrak{a}}{\mathsf{D}}^A{}_{|B}=W^A{}_{CB}\underaccent{\mathfrak{a}}{\mathsf{D}}^C$, where $W^A{}_{CB}$ is referred to as the wryness of the director field and is defined as
\begin{equation}
	W^A{}_{BC}=\underaccent{\mathfrak{a}}{\mathsf{D}}^A{}_{|C}\accentset{\mathfrak{a}}{\Theta}_B
	=-\underaccent{\mathfrak{a}}{\mathsf{D}}^A \accentset{\mathfrak{a}}{\Theta}_{B|C} \,.
\end{equation}
The director gradient is defined as $\director{\boldsymbol{\mathsf{F}}}=\nabla^{\mathbf{G}}\director{\boldsymbol{\mathsf{d}}}$. It has components $\director{\mathsf{F}}^a{}_A=\director{\mathsf{d}}^a{}_{|A}$ and is related to the relative wryness as
\begin{equation}
	\director{\mathsf{F}}^a{}_A=w^a{}_{bA}~ \director{\mathsf{d}}^b \,,
	\qquad  w^a{}_{bA}=\director{\mathsf{d}}^a{}_{|A} \accentset{\mathfrak{a}}{\vartheta}_b\,.
\end{equation}
In Cosserat hyperealsticity, it is assumed that an energy function $W=W(X,\mathbf{F},\director{\boldsymbol{\mathsf{F}}},\mathbf{G},\mathbf{g})$ exists.
Objectivity implies that $W=\hat{W}(X,\mathbf{C}^\flat,\director{\boldsymbol{\mathsf{C}}}^\flat,\mathbf{G})$, where $\director{\boldsymbol{\mathsf{C}}}^\flat=\director{\boldsymbol{\mathsf{F}}}^*\mathbf{g}=\director{\boldsymbol{\mathsf{F}}}^\star\mathbf{g}\,\director{\boldsymbol{\mathsf{F}}}$. In components, $\director{\mathsf{C}}_{AB}=\director{\mathsf{F}}^a{}_A\,\director{\mathsf{F}}^b{}_B\,g_{ab}$. 
Note that in Cosserat elasticity, there are a total of $6$ macro strains and $18$ director strains, amounting to a total of $24$ strains. The stresses corresponding to $\mathbf{C}^\flat$ and $\director{\boldsymbol{\mathsf{C}}}^\flat$ are denoted by $\mathbf{S}$ (the second Piola-Kirchhoff stress) and $\accentset{\mathfrak{a}}{\boldsymbol{\mathsf{H}}}$ (the hyperstress tensor).

A Cosserat-Cauchy elastic has the following constitutive equations
\begin{equation}
	\mathbf{S}=\hat{\mathbf{S}}(X,\mathbf{C}^\flat,\director{\boldsymbol{\mathsf{C}}}^\flat,\mathbf{G})\,,
	\qquad
	\accentset{\mathfrak{a}}{\boldsymbol{\mathsf{H}}}
	=\widehat{\accentset{\mathfrak{a}}{\boldsymbol{\mathsf{H}}}}
	(X,\mathbf{C}^\flat,\director{\boldsymbol{\mathsf{C}}}^\flat,\mathbf{G})
	\,.
\end{equation}
In Cosserat-Cauchy elasticity, in general, there is no energy function depending on strains.\footnote{Linearized Cosserat elasticity without an energy function, i.e., linerized non-hyperelastic Cosserat-Cauchy elasticity, has been called ``odd Cosserat elasticity" \citep{Surowka2023}.} The fundamental quantity of Cosserat-Cauchy elasticity is the stress work $1$-form.

\begin{defi}
In Cosserat-Cauchy elasticity the stress work $1$-form is defined as
\begin{equation}
	\boldsymbol{\Omega}(\mathbf{C}(X,t),\director{\boldsymbol{\mathsf{C}}}(X,t))
	=\frac{1}{2}S^{AB}(X,t)\,dC_{AB}(X,t)
	+\sum_{\mathfrak{a}=1}^3
	\frac{1}{2}\accentset{\mathfrak{a}}{\mathsf{H}}^{aA}(X,t)\,d\director{\mathsf{C}}_{AB}(X,t)
	\,.
\end{equation}
This is a $1$-form in a $24$-dimensional space.
\end{defi}

For a $24$-dimensional manifold, the rank of a $1$-form can take any of the values $0,1,\hdots,12$. Thus, the Darboux classification theorem tells us that there are the following possibilities (note that $\boldsymbol{\Omega}\wedge (d\boldsymbol{\Omega})^{12}$ is a $25$-form, which identically vanishes on any $24$-manifold):
\begin{equation}
\begin{dcases}
	\boldsymbol{\Omega} & =  d\psi_1\,, \\
	\boldsymbol{\Omega} & =  \phi_1 d\psi_1\,, \\
	\boldsymbol{\Omega} & =  \phi_1 d\psi_1+d\psi_2\,, \\
	\vdots \\
	\boldsymbol{\Omega} & =  \phi_1 d\psi_1+\hdots+\phi_{12} d\psi_{12} \,.
\end{dcases}
\end{equation}
We observe that an anisotropic Cosserat-Cauchy elastic solid has at most Twenty four generalized energy functions.

For a Cosserat-Cauchy elastic body, a cyclic deformation is a closed curve in $\mathbb{S}$---the space of generalized strains. This means that $\mathbf{C}(X,t_1)=\mathbf{C}(X,t_2)$ and $\director{\boldsymbol{\mathsf{C}}}^\flat(X,t_1)=\director{\boldsymbol{\mathsf{C}}}^\flat(X,t_2)$. A cyclic motion is a cyclic deformation for which $\mathbf{V}(X,t_1)=\mathbf{V}(X,t_2)$ and $\dot{\director{\boldsymbol{\mathsf{d}}}}(X,t_1)=\dot{\director{\boldsymbol{\mathsf{d}}}}(X,t_2)$, where $\dot{\director{\boldsymbol{\mathsf{d}}}}(X,t)=\frac{\partial}{\partial t}\director{\boldsymbol{\mathsf{d}}}(X,t)$ denotes the director velocity.
Given a path $\Gamma$ in the space of generalized strains, once can show that
\begin{equation} \label{Total-Work-Cosserat}
	W(\mathcal{U},\Gamma)
	=  \int_{\mathcal{U}} \left[ \mathcal{K}_2-\mathcal{K}_1
	+ \int_{\Gamma}\boldsymbol{\Omega} \right]  dV
	\,,
\end{equation}
where $\mathcal{K}=\mathcal{K}(X,t)=\frac{1}{2}\rho V^aV^bg_{ab}+\frac{1}{2} \accentset{\mathfrak{a}\mathfrak{b}}{\nu}~\dot{\director{\mathsf{d}}}^a\dot{\underaccent{\mathfrak{b}}{\mathsf{d}}}^bg_{ab}$ is the kinetic energy density ($\accentset{\mathfrak{a}\mathfrak{b}}{\nu}=\accentset{\mathfrak{b}\mathfrak{a}}{\nu}$ is the micro-mass moment of inertia), and $\mathcal{K}_i=\mathcal{K}(X,t_i)$, $i=1,2$.

\section{Discussion} \label{Discussion}

Elasticity is an old and venerable  theory that has served as a template for all field theories. Yet, it never ceases to surprise us. While the mathematical foundations are clear, the focus of most works has been on hyperelastic (Green) materials. The existence of a strain energy function from which the stresses derive has been a safety blanket for scientists and mathematicians alike. While it allows for great progress in the effective resolution of important problems, it also naturally constrains the structure of the solutions and their possible behaviors. Hyperelasticity is a perfect assumption for most passive materials until the field of applications of mechanics was enlarged to include biological and active materials. 

It is well appreciated in mechanics that the laws of thermodynamics do not preclude the validity of Cauchy elasticity. However, this leads to the puzzling possibility of a system gaining or losing energy in cyclic deformations, rendering it seemingly unphysical \citep{Coleman1962,Rivlin1986,Casey2005,Carroll2009}. If the origin of this non-zero work is due to external body forces, the puzzle can be resolved by recognizing that, in an open system, such forces may access external sources or sinks of energy. Hence, a realization of a Cauchy elastic material requires both local access to an energy source and, when operated in reverse in strain space, a local means of storing energy (or losing it—but any viscous effects would operate in both directions and are therefore unlikely to produce the desired behavior). 
For instance, \citet{Chen2011} showed that the amplitude of elastic waves can grow exponentially in some Cauchy materials. Experiments motivated by odd elastic materials exhibit this type of behavior, but only partially. These systems are typically composed of small units capable of exerting torques and forces on one another through local energy access. When properly homogenized, such structures can be modeled as Cauchy materials. However, many of these systems rely on continuous energy input rather than storing it, resulting in a useful but incomplete realization of a Cauchy material. Similarly, it is unlikely that energy can be pumped indefinitely into the system, as would be required to sustain a wave with ever-growing amplitude. Any such material would necessarily operate within a limited range of wave amplitudes.
This situation is not entirely unfamiliar. When modeling a material with an elastic model, it is understood that the model is not expected to remain valid under arbitrary deformations and motions. Materials fracture, yield, or undergo plastic deformation outside the elastic range. Cauchy elasticity is similarly restricted in its operating domain. Yet, despite this limitation, it still enables the modeling of novel behaviors in active materials and therefore warrants serious consideration.

We noted that forces can be classified as either conservative or non-conservative, and as either dissipative or non-dissipative. While all dissipative forces are non-conservative, the converse is not always true; dissipative forces form a proper subset of non-conservative forces. Similarly, while all conservative forces are non-dissipative, the converse is not necessarily true, making conservative forces a proper subset of non-dissipative forces. Consequently, the sets of non-conservative forces and non-dissipative forces have a non-empty intersection. As an introductory example, we pointed out that a particle (or a finite-dimensional mechanical system) under the influence of a non-conservative and non-dissipative force(s) serves as a zero-dimensional analogue of Cauchy elasticity.

In hyperelasticity, constitutive equations are written in terms of an energy function and its derivatives with respect to strain or invariants of strain. By contrast, in Cauchy elasticity, the fundamental quantity is the stress-work $1$-form, a generalization of the energy function, best described and studied within the theory of exterior calculus. The stress-work $1$-form is, in general,  not an exact form; it is exact only in the case of hyperelasticity. Using this formalism, it is then straightforward to apply Darboux's theorem to the classification of the stress-work $1$-forms to show that stress can always be additively decomposed into conservative and non-conservative parts. Furthermore, the work of Edelen shows that in $3$D Cauchy elasticity the stress-work $1$-form has six possible canonical forms.  In other words, \textit{Cauchy elasticity has at most six generalized energy functions in terms of the right Cauchy-Green strain.} 
The simplest non-hyperelastic Cauchy elastic materials include just two potentials and we refer to these as Ericksen solids. They have interesting properties that should be further investigated. 

The fundamental difference between hyperelasticity and Cauchy elasticity is revealed by considering the net work of stress in cyclic deformations. For a hyperleastic material, no energy is gained or released in  a cyclic deformation whereas it may not be the case for non-hyperelastic Cauchy materials. Moreover, if in a cyclic deformation  $\Gamma$ energy is released (gained), it is subsequently regained (released) in the reverse cyclic deformation $-\Gamma$. In these cycles, the geometric hysteresis serves as the distinguishing feature between non-hyperelastic and hyperelastic elasticity. 
In hindsight, this geometric hysteresis is not surprising, as it emerges naturally from the geometric structure of the problem via a material connection $1$-form and its associated curvature.
Yet, this interpretation seems to be novel to the field.

At the formal level, Cauchy elasticity has a beautiful mathematical structure that is fully compatible with the law of thermodynamics. Yet, the problem remains: Are  Cauchy solids truly ``real" solids? Our view of a material has been shaped by continua whose microscopic elastic unit is at the molecular level. Hence, we typically do not think of materials composed of small (but much larger than molecular) units that can sense and react to deformation as being governed by elasticity. Yet, from a purely observational point of view, at the macroscopic level such systems may respond to forces like elastic materials and hence can be modeled using elasticity.

Additionally, the inclusion of body forces, such as interactions with substrates, within constitutive laws expands the definition of Cauchy materials to encompass non-traditional systems. Modern research on biological materials and metamaterials has revealed numerous examples of systems connected to external energy sources or capable of dissipating energy in response to deformation. It is now clear that weakening this assumption to include Cauchy materials allows for new and relevant material behavior.
For example, it can describe the effective material behavior (constitutive equation) of a solid composed of a matrix with inclusions that are connected to an external energy source.

\section{Future work: A roadmap for Cauchy elasticity}  \label{FutureWork}

The theory that we have outlined naturally suggests a number of possible extensions.
Not surprisingly, most of elasticity theory developed for Green materials has an analog in the Cauchy world with possibly significant qualitative differences. We outline some of these extensions below.

\begin{itemize}[topsep=0pt,noitemsep, leftmargin=10pt]
\item{\textbf{Implicit elasticity.}}  Cauchy elasticity is a special case of implicit elasticity where the stress and strain tensors satisfy an implicit constitutive relation. In cases where this relation takes the form $\boldsymbol{\mathcal{F}}(\boldsymbol{\sigma}, \mathbf{b}) = \mathbf 0$, with $\mathbf{b}$ the Finger tensor and $\boldsymbol{\mathcal{F}}$ a polynomial, the set of admissible stress-strain pairs forms an algebraic variety. This observation suggests that methods from algebraic geometry may be useful in analyzing the structure and admissibility of implicit constitutive equations. Importantly, the absence of a strain-energy density in the general theory of implicit elasticity will require the same treatment with a special focus on the stress-work $1$-form.  

\item{\textbf{Dynamics.}} The dynamics of Cauchy media, and specifically wave propagation in Cauchy solids, should be investigated carefully. In contrast to hyperelastic solids, Cauchy materials lack a strain energy function and do not conserve mechanical energy. Recent work has shown that wave propagation in Cauchy solids may exhibit unphysical behavior, such as strain-dependent wave speeds and loss of strong ellipticity \citep{Chen2011}. These issues raise fundamental concerns about stability and physical admissibility and merit further investigation. Apart from a few early results (e.g., \citep{Truesdell1964,Hayes1972}), there is almost nothing in the literature on wave propagation in Cauchy elastic solids, but linear Cauchy elastic solids (also known as ``odd elastic materials") show interesting wave behaviours that have been observed experimentally \citep{Scheibner2020}. Hence, we believe that the study of dynamics in Cauchy elasticity is a fruitful field of study.

\item{\textbf{Micropolar Cauchy elasticity.}} Many of the systems for which Cauchy elasticity may be relevant includes local torque generation \citep{chen2021realization}. In the simplest cases, these local body torques can be accommodated by including a non-symmetric part to the Cauchy stress tensor, leading to an effective Cauchy material. 
However, in general when the body torques are not just a function of the strains, couple-stresss are needed to balance the angular momentum.
Hence, a general theory of micropolar Cauchy materials is needed and should reveal the most general mathematical structure of continua.

\item{\textbf{Further extensions.}} Most existing analytical works on elasticity are restricted to hyperelasticity (e.g. stability, cavitation,  existence of solutions, etc). Each of these problems can be studied within this generalized framework. We hope that our work will  motivate future efforts to extend such results to Cauchy elasticity. In particular, it would be useful to understand which phenomena or results critically depend on the existence of a strain energy function, and which are universal features of elastic materials. 
\end{itemize}

The points that we have outlined constitute a major research program with great challenges.
However, caution is still warranted before a theoretical frenzy takes place. Theory for theory's sake is facile when not properly motivated by either physical realizations, or a desire to understand the underlying mathematical structure of a theory. In the words of Truesdell, \textit{``when we start to generalize it is hard to know where to stop.''} \citep{Truesdell2013six}
Indeed, the history of elasticity has been characterized by a dynamic interplay between experimental observations, theoretical advancements, and the use of new mathematical concepts, each driving paradigm shifts in the field. We are now entering a new era of continua, one that incorporates active forces and unconventional material responses. Integrating classical foundations with contemporary innovations and advanced mathematical theories remains pivotal for our understanding of this emerging domain.

\section*{Acknowledgments}

We benefited from extensive discussions with David Steigmann throughout this research as well discussions  with Marino Arroyo, Francesco Fedele, James Hanna, and Patrizio Neff.  AY was supported by NSF -- Grant No. CMMI 1939901.
This work was also supported by an International Exchanges Scheme Grant from the Royal Society to AG. 

%

\bibliographystyle{abbrvnat}

\end{document}